\documentclass[%
 reprint,
superscriptaddress,
nofootinbib,
 amsmath,amssymb,
 aps,
]{revtex4-2}

\usepackage{amsmath}
\usepackage{graphicx}
\usepackage{dcolumn}
\usepackage{bm}
\usepackage{color}
\usepackage[dvipsnames]{xcolor}
\usepackage[normalem]{ulem}
\usepackage{placeins}
\usepackage{amssymb}
\usepackage{amsfonts}
\usepackage{orcidlink}

\newcommand{\cornell}{\affiliation{Cornell Center for Astrophysics and Planetary
		Science, Cornell University, Ithaca, New York 14853, USA}}
\newcommand{\caltech}{\affiliation{Theoretical Astrophysics, Walter Burke
		Institute for Theoretical Physics, California Institute of Technology,
		Pasadena, California 91125, USA}}
\newcommand{\aei}{\affiliation{Max Planck Institute for Gravitational Physics
		(Albert Einstein Institute), Am M{\"u}hlenberg 1, 14476 Potsdam, Germany}}
\newcommand{\fullerton}{\affiliation{Nicholas and Lee Begovich Center for
		Gravitational-Wave Physics and Astronomy, California State University
		Fullerton, Fullerton, California 92831, USA}}

\begin{document}

\preprint{APS/123-QED}

\title{
Scalarization of isolated black holes in scalar Gauss-Bonnet theory \\
in the fixing-the-equations approach
}

\author{Guillermo Lara\,\orcidlink{0000-0001-9461-6292}}
\email{glara@aei.mpg.de} \aei
\author{Harald P. Pfeiffer\,\orcidlink{0000-0001-9288-519X}} \aei
\author{Nikolas A. Wittek\, \orcidlink{0000-0001-8575-5450}} \aei
\author{Nils L.~Vu\,\orcidlink{0000-0002-5767-3949}} \caltech

\author{Kyle C.~Nelli\,\orcidlink{0000-0003-2426-8768}} \caltech
\author{Alexander Carpenter\,\orcidlink{0000-0002-9183-8006}} \fullerton
\author{Geoffrey Lovelace\,\orcidlink{0000-0002-7084-1070}} \fullerton
\author{Mark A. Scheel\,\orcidlink{0000-0001-6656-9134}} \caltech
\author{William Throwe\,\orcidlink{0000-0001-5059-4378}} \cornell

\date{\today}

\begin{abstract}
One of the most promising avenues to perform numerical evolutions in theories beyond General Relativity is the \emph{fixing-the-equations} approach, a proposal in which new ``driver'' equations are added to the evolution equations in a way that allows for stable numerical evolutions.
In this direction, we extend the numerical relativity code \texttt{SpECTRE} to evolve a ``fixed'' version of scalar Gauss-Bonnet theory in the decoupling limit, a phenomenologically interesting theory that allows for hairy black hole solutions in vacuum.
We focus on isolated black hole systems both with and without linear and angular momentum, and propose a new driver equation to improve the recovery of such stationary solutions.
We demonstrate the effectiveness of the latter by numerically evolving black holes that undergo spontaneous scalarization using different driver equations.
Finally, we evaluate the accuracy of the obtained solutions by comparing with the original unaltered theory. 
\end{abstract}

\maketitle

%%%%%%%%%%%%%%%%%%%%%%%%%%%%%%%%%%%%%%%%%%%%%%%%%%%%%%%%%%%%
%%%%%%%%%%%%%%%%%%%%%%%%%%%%%%%%%%%%%%%%%%%%%%%%%%%%%%%%%%%%
%%%%%%%%%%%%%%%%%%%%%%%%%%%%%%%%%%%%%%%%%%%%%%%%%%%%%%%%%%%%

\section{Introduction \label{sec: Introduction}}

The increasing availability of gravitational wave (GW) data,
by current (LIGO-Virgo-KAGRA~\cite{LIGOScientific:2014pky, VIRGO:2014yos, KAGRA:2020tym})
and future (LISA~\cite{LISA:2017pwj}, Einstein Telescope~\cite{Punturo:2010zz}, Cosmic Explorer~\cite{Reitze:2019iox}) ground and space-based interferometers,
promises to lead to the strictest tests yet of Einstein's general theory of relativity (GR) as the fundamental description of gravitational phenomena.
Adding on to the vast array of weak-field tests~\cite{Will:2018bme}, 
a growing suite of
model-independent tests
on the GW signal from binary compact coalescences (BCC)~\cite{LIGOScientific:2019fpa, LIGOScientific:2020tif, LIGOScientific:2021sio} show no apparent  disagreement with GR to date.
At the same time, direct observation by means of very-large-baseline interferometry show black hole (BH) images consistent with expectations based on Einstein's gravity~\cite{EventHorizonTelescope:2021dqv, EventHorizonTelescope:2022xqj}.

Nevertheless, it is conceivable that GR is not the ultimate description of gravity.
On the theoretical side,
the quantum nature of all matter and force fields in the Standard Model,
firmly established in the past half-century,
suggests 
the existence of 
a quantum theory of gravity.
From this perspective, GR arises 
as an \emph{effective field theory} (EFT)
with limited range of applicability 
and is subject to corrections relevant at shorter (higher) length (curvature) scales.
On the phenomenological side, 
modifying gravity at the largest scales
(e.g.~by introducing new light degrees of freedom)
could help explain
cosmological issues such as the observed self-accelerated expansion of the universe (or Dark Energy)~\cite{Crisostomi:2017pjs}.
Among other motivations, numerous extensions of GR have been devised.
Some of the most important classes of theories include DHOST (degenerate higher-order scalar-tensor theories)~\cite{Langlois:2015cwa, Crisostomi:2016czh, BenAchour:2016fzp} --a generalization of the Horndeski class~\cite{Horndeski:1974wa}-- and EFT expansions of gravity --see e.g.~Ref.~\cite{Endlich:2017tqa, Weinberg:2008hq}.

It is expected that GW signals will carry information about possible GR corrections.
Already, tight constraints on the speed of tensor modes 
from GW170917~\cite{Monitor:2017mdv,TheLIGOScientific:2017qsa} have cast
widely-encompassing constraints on Dark Energy models~\cite{Creminelli:2017sry, Ezquiaga:2017ekz, Baker:2017hug,Sakstein:2017xjx}. 
However, 
in order to 
fully exploit the potential of GWs
to cast the strictest bounds on beyond-GR parameters,
model-dependent tests need to be carried out for (if only a few) alternative theories of gravity. 
In this direction,
there is a need to construct full waveform models, encompassing all stages of BCC.
For GR, numerical simulations of the late-inspiral and merger phases of compact binary coalescence are essential to construct and calibrate models describing these highly dynamical and non-linear stages~\cite{Varma:2018mmi, Ossokine:2020kjp, Pratten:2020ceb, Gamba:2021ydi}.
Therefore, a key (and challenging) step
is to extend the techniques
of numerical relativity (NR)~\cite{Baumgarte:2010ndz}
beyond GR.

As it was the case for GR decades ago, one of the main difficulties is the mathematical structure of equations of motion.
These systems of partial differential equations (PDEs) need to be recast (if possible) in a way suitable for numerical evolution, such that the system admits a \emph{well-posed} initial-value problem (IVP) --a property that ensures good behaviour of the solutions: uniqueness and continuity on the space of initial data~\cite{Hadamard10030321135}. 
Failure to satisfy this property can give rise to unstable solutions and even seemingly pathological behavior --e.g.~hyperbolic PDE systems have been observed to change character to elliptic in regions outside BH horizons~\cite{East:2021bqk, Bernard:2019fjb, Ripley:2019irj, Ripley:2019aqj, Ripley:2020vpk, R:2022hlf, Corelli:2022pio, Corelli:2022phw, Thaalba:2023fmq}.
Nevertheless, recent developments have fuelled renewed interest and a steady increase in beyond-GR codes and GW-waveform examples
--see   Ref.~\cite{Ripley:2022cdh} for a review.
Indeed, some of us have obtained beyond-GR waveforms in dynamical Chern-Simons and scalar Gauss-Bonnet (sGB) theory using an order-reduction scheme to obtain perturbative corrections to the GW signal~\cite{Okounkova:2017yby, Okounkova:2019zep, Okounkova:2019zjf, Okounkova:2020rqw} --see also Refs.~\cite{Witek:2018dmd, Silva:2020omi, Elley:2022ept}.
Other studies have adopted theory-specific approaches, in some cases involving explorations of the beyond-GR parameter space~\cite{Figueras:2020dzx, Figueras:2021abd, Bezares:2020wkn, Bezares:2021yek, Bezares:2021dma,
Held:2021pht,
Held:2023aap,
Rubio:2023eva}.
More recently, novel modified Generalized Harmonic gauges applicable for a broad class of weakly-coupled Lovelock and Horndeski theories have been proposed by Kovacs and Reall~\cite{Kovacs:2020PRL, Kovacs:2020ywu}.
The latter have been numerically implemented in Refs.~\cite{East:2020hgw, East:2021bqk, Corman:2022xqg}, as well as in similar extensions to the CCZ4 system~\cite{AresteSalo:2022hua, AresteSalo:2023mmd}, to evolve binary systems in sGB EFTs.

Here we will explore a fourth avenue, sometimes referred to in the literature as the \emph{fixing-the-equations} approach~\cite{Cayuso:2017iqc}.
In this proposal (inspired by M\"uller-Israel-Stewart hydrodynamics), the equations of motion are deformed (or ``fixed'') by the introduction of auxiliary variables subject to suitable \emph{driver} evolution equations.
New timescales appearing in these driver equations control how the ``fixed'' solutions track the solutions of the original system,
thus
allowing for the evolution of the scales relevant to the problem, while ``softening'' (possibly) problematic high-frequency modes.

In this paper, we report on the first steps to implement the \emph{fixing-the-equations} approach in the numerical relativity code \texttt{SpECTRE}~\cite{deppe_2024_10619885}, developed by the SXS Collaboration.
While this method can be applied to a wide variety of theories (see e.g.~Refs.~\cite{Allwright:2018rut, Cayuso:2020lca, Bezares:2021yek, Lara:2021piy, Franchini:2022ukz, Cayuso:2023aht, Coates:2023swo}),
for concreteness and motivated by positive results in spherical symmetry~\cite{Franchini:2022ukz}, we will focus on sGB gravity in vacuum. 
The action is
\begin{align} \label{eq: action sGB}
    S\left[g_{ab}, \Psi \right] \equiv \int \, d^4 x \sqrt{-g} 
    \Big[
    \dfrac{R}{2 \kappa } 
    - \dfrac{1}{2}  (\nabla_{a} \Psi \nabla^{a} \Psi)
    + \ell^2 f(\Psi) \, \mathcal{G}
    \Big]~,
\end{align}
where \(\kappa \equiv 1/(8\pi G)\) and \(\ell\) are constant couplings, \(g = \mathrm{det}(g_{ab})\) is the determinant of the metric \(g_{ab}\),
the scalar field is \(\Psi\),
\(f(\Psi)\) is a free function describing the specific coupling to the Gauss-Bonnet scalar
\begin{align}
    \mathcal{G} \equiv R_{abcd}R^{abcd} - 4 R_{ab}R^{ab} + R^2~,
\end{align}
which is defined in terms of combinations of the Riemann tensor \(R_{abcd}\), the Ricci tensor \(R_{ab}\) and the Ricci scalar \(R\).
Scalar Gauss-Bonnet theory is of phenomenological interest  mainly because it is a model for black holes endowed with scalar \emph{hair}~\cite{Sotiriou:2013qea, Sotiriou:2014pfa}. 
As opposed their GR counterparts (described by the Kerr metric~\cite{Kerr:1963ud}), BHs in sGB theory evade common no-hair theorems~\cite{Hui:2012qt, Maselli:2015yva, 
Creminelli:2020lxn, Capuano:2023yyh} and are  characterised by a \emph{scalar charge} in addition to their mass and spin parameters.
BHs can acquire scalar hair through a variety of \emph{scalarization} mechanisms (see Ref.~\cite{Doneva:2022ewd} for a review) and it is expected that this hair will significantly impact the GW signal from binary black hole (BBH) systems.
Recently, model-dependent bounds using the inspiral phase of GW observations have been placed on shift-symmetric sGB models --giving values of \(\ell \lesssim \mathcal{O}(1) \, \mathrm{km}\)~\cite{Perkins:2021mhb, Lyu:2022gdr}.
Additional theoretical constraints on the form of the coupling function \(\ell^2 f(\Psi)\) have been derived from the assumption of consistency of sGB theory with a well-behaved (high energy) ultraviolet completion~\cite{Herrero-Valea:2021dry}.

In our implementation of the \emph{fixing-the-equations} approach, we consider the following (related) issues:
\emph{i)} Can we accurately describe the stationary solutions of the theory?,
\emph{ii)} Can we be confident that the ``fixed'' equations are able to reproduce the original physics?
We address both questions 
empirically in a case study:
the scalarization of isolated BHs
in the \emph{decoupling limit} of theory~\eqref{eq: action sGB} --where the scalar field back-reaction on the metric is neglected.
The reason why this approximation is illustrative is two-fold. First, it will help us test 
the properties of different
driver equations in a localized sector: the scalar sector. 
Second, the ``unmodified'' decoupled theory can be recast in strongly hyperbolic form and thus readily evolved so as to compare different ``fixing'' prescriptions against the ``correct'' solution.
Finally, as a main result of this paper, we
present a new driver equation that leads to an improved recovery of the stationary solutions of the theory, with respect to other equations studied in the literature --see Fig.~\ref{fig: Scalarized BH example radial profile}.

This paper is organized as follows.
In Section~\ref{sec: Theory}, we give a brief theoretical background of sGB theory and the \emph{fixing-the-equations} approach.
We describe our numerical implementation and diagnostics in Section~\ref{sec: Methodology}.
In Section~\ref{sec: Results}, we illustrate with different examples of isolated BH scalarization how stationary solutions are recovered by the `fixed'' systems and how the new driver equation improves their accuracy using a variety of diagnostics.
We summarize our results and comment on the future steps of this program in Section~\ref{sec: Conclusion}.
Further material is presented in the appendices.
In Appendix~\ref{sec: Appendix comparison with box driver} the driver equations of the main text are  compared with a third wavelike driver.
Finally, in Appendix~\ref{sec: Appendix Convergence}, convergence plots are presented.
Throughout this paper we set \(G = c = 1\) and work in \(-+++\) signature.
Early alphabet letters \(\{a, b, c, \dots\}\) represent 4-dimensional spacetime indices, whereas middle letters \(\{i, j, k, \dots\}\) represent 3-dimensional spatial indices.

%%%%%%%%%%%%%%%%%%%%%%%%%%%%%%%%%%%%%%%%%%%%%%%%%%%%%%%%%%%%
%%%%%%%%%%%%%%%%%%%%%%%%%%%%%%%%%%%%%%%%%%%%%%%%%%%%%%%%%%%%
%%%%%%%%%%%%%%%%%%%%%%%%%%%%%%%%%%%%%%%%%%%%%%%%%%%%%%%%%%%%

\section{Theory \label{sec: Theory}}

Taking the variation of action~\eqref{eq: action sGB} with respect to the scalar field \(\Psi\), we obtain the scalar equation
\begin{align} \label{eq: scalar equation sGB}
    \Box \Psi = - \ell^2 f'(\Psi) \mathcal{G}.
\end{align}
Variation with respect to the metric yields
\begin{align} \label{eq: tensor equation sGB}
    G_{ab} = \kappa \left(T^{(\Psi)}_{ab} + \ell^2 H_{ab}\right),
\end{align}
where \(G_{ab} \equiv R_{ab} - (R/2)g_{ab}\) is the Einstein tensor,
\begin{align}
    T^{(\Psi)}_{ab} \equiv \nabla_a \Psi \nabla_b \Psi - \dfrac{1}{2} g_{ab} \nabla_{c} \Psi \nabla^{c} \Psi,
\end{align}
is the canonical stress-energy tensor of the scalar,
and~\cite{Franchini:2022ukz}
\begin{align}
    H_{ab} &\equiv P_{acbd} \nabla^{c} \nabla^{d} f(\Psi)\,,
\end{align}
with
\begin{equation} \label{eq: definition of Pabcd}
    P_{abcd} \equiv R_{abcd} - 2 g_{a\left[c\right.} R_{\left.d\right] b} + 2 g_{b \left[c\right.} R_{\left.d\right] a} + g_{a\left[c\right.} g_{\left.d\right] b} R\,.
\end{equation}

In the \emph{decoupling limit} of the theory, the back-reaction of the scalar field on Eq.~\eqref{eq: tensor equation sGB} is neglected. The resulting system is equivalent to a test scalar field, described by Eq.~\eqref{eq: scalar equation sGB}, which evolves on a known GR background (see e.g.~\cite{Silva:2020omi, Elley:2022ept}): 
\begin{align} \label{eq: decoupling limit equations}
    \Box \Psi &= - \ell^2 f'(\Psi) \mathcal{G} & R_{ab} &= 0~.
\end{align}

%%%%%%%%%%%%%%%%%%%%%%%%%%%%%%%%%%%%%%%%%%%%%%%%%%%%%%%%%%%%
%%%%%%%%%%%%%%%%%%%%%%%%%%%%%%%%%%%%%%%%%%%%%%%%%%%%%%%%%%%%
%%%%%%%%%%%%%%%%%%%%%%%%%%%%%%%%%%%%%%%%%%%%%%%%%%%%%%%%%%%%

\subsection{Spontaneous scalarization}

\begin{figure}[]
  \centering
   \includegraphics[width=0.98\columnwidth,trim=40 37 10 20,clip=true]{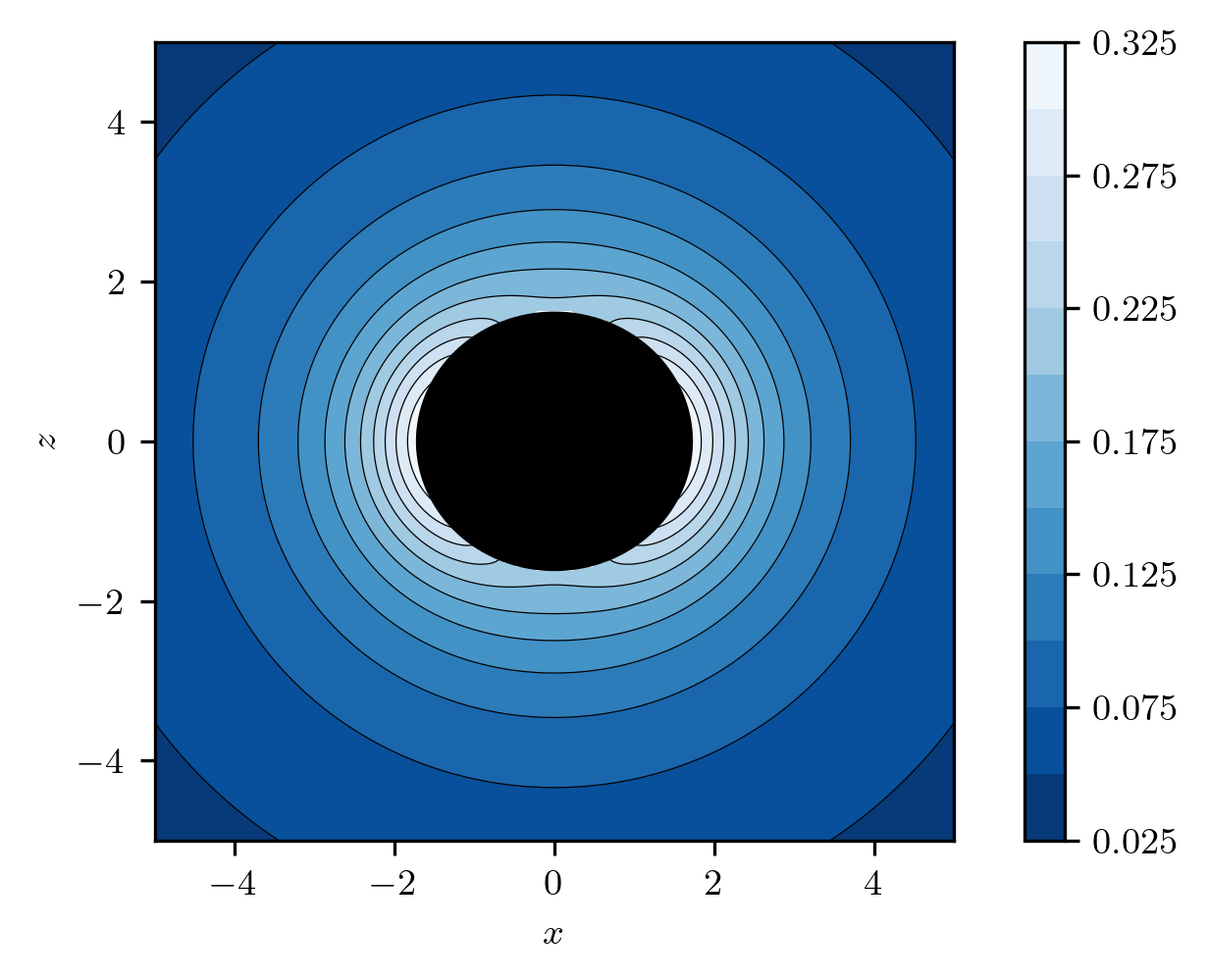}
   \caption{ \emph{Scalarized spinning BH.}
   Curvature-induced scalarization for a spinning BH with spin \(a = 0.6\) in the \(+z\) direction (top) evolved in the decoupling limit [Eqs.~\eqref{eq: decoupling limit equations}] for model~\eqref{eq: coupling function quartic model}. We have set \(\{\ell^2 / M^2 = 1, \eta = 6, \zeta = -60\}\). In colour, the absolute value of the scalar field \(\Psi\) at \(t / M  = 1000\). See Sec.~\ref{sec: Results} for more details.
   }
   \label{fig: Scalarized BH example}
\end{figure}
\begin{figure}[]
    \centering
   \includegraphics[width=3.4 in]{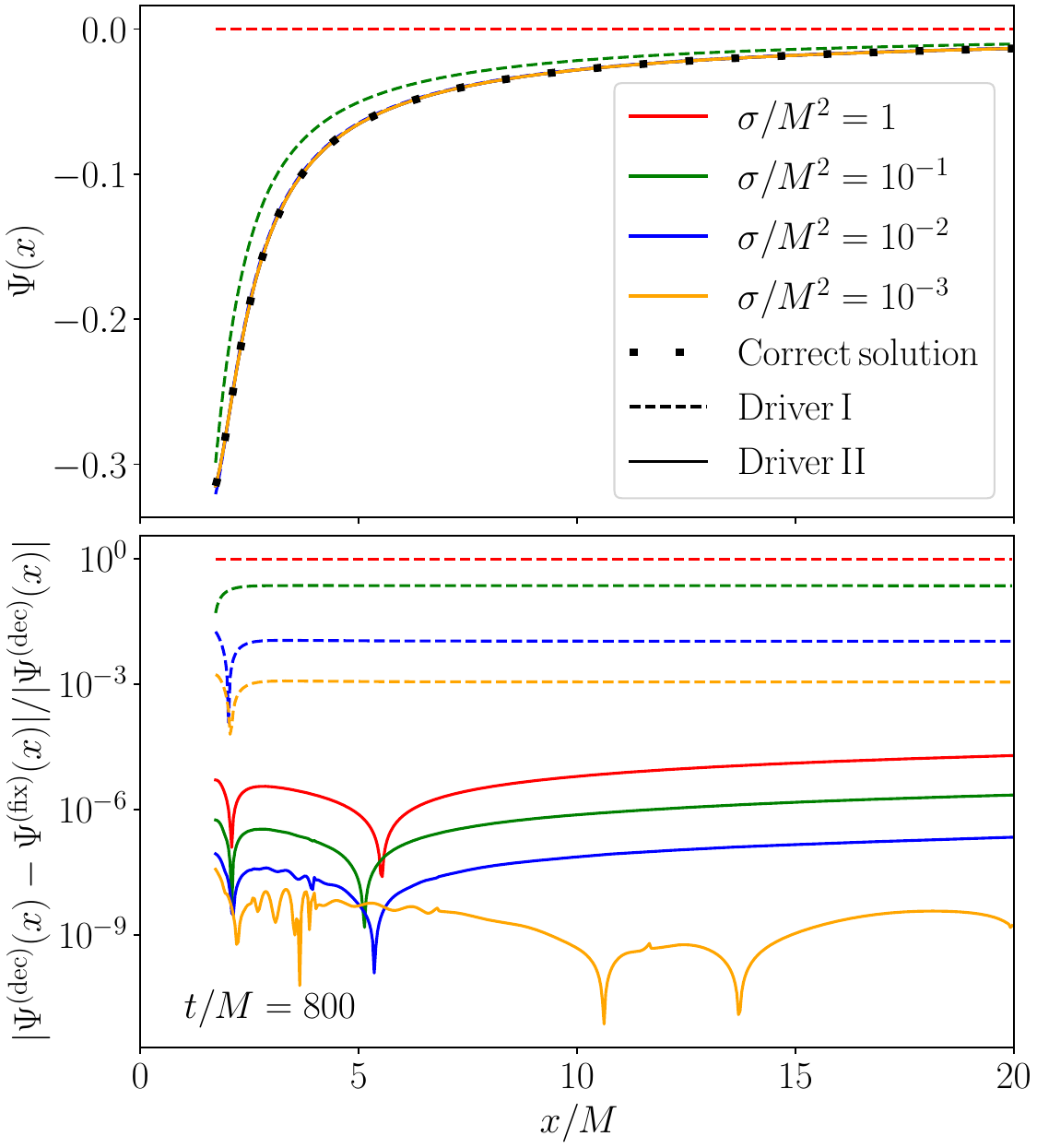}
   \caption{ \emph{Scalarized spinning BH (equatorial profile).}
   Top: Scalar profile (solid black) for the BH of Fig.~\ref{fig: Scalarized BH example}  along the equatorial \(+x\) direction at \(t / M  = 800\).
   In coloured lines, various approximations with the \emph{fixing-the-equations} approach. Dashed lines correspond to Driver I [Eq.~\eqref{eq: advection squared driver}] whereas solid lines correspond to our \emph{new} Driver II [Eq.~\eqref{eq: comving advection driver}]. We fix \(\tau = \sigma / M\). 
   The scalar profiles are plotted starting from the excision sphere, slightly inside the apparent horizon of the BH.
   Bottom: Relative error in the scalar profile with respect to the correct solution given by the decoupling limit [Eqs.~\eqref{eq: decoupling limit equations}].
   }
   \label{fig: Scalarized BH example radial profile}
\end{figure}

In order to describe BHs that undergo \emph{spontaneous scalarization}, we will consider the model of Ref.~\cite{Silva:2018qhn}
\begin{align} \label{eq: coupling function quartic model}
    f(\Psi) \equiv \dfrac{\eta}{8} \Psi^2 + \dfrac{\zeta}{16} \Psi^4~.
\end{align}
where \(\eta\) and \(\zeta\) are dimensionless parameters. 
As can be readily seen from Eq.~\eqref{eq: coupling function quartic model} and action~\eqref{eq: action sGB}, the resulting theory is \(\mathbb{Z}_2\)-symmetric and solutions related by \(\Psi \to - \Psi\) are equivalent.

A trivial solution to  Eqs.~\eqref{eq: scalar equation sGB} and \eqref{eq: tensor equation sGB} is a Kerr metric with \(\Psi \equiv 0 \). Thus, the solutions of GR are also solutions for model~\eqref{eq: coupling function quartic model} of sGB theory.
Nevertheless, a \emph{tachyonic} instability 
in the scalar sector
may render GR solutions linearly unstable~\cite{Doneva:2022ewd}.
This can be readily seen 
in the decoupling limit 
by writing the equation for a linear perturbation \(\Psi = \delta \Psi + \mathcal{O}(\delta \Psi^2)\) of the scalar, which reads
\begin{align}
    (\Box  - m^2_{\Psi, \mathrm{eff}} )\, \delta \Psi= 0
\end{align}
where \(m^2_{\Psi, \mathrm{eff}} \equiv - \ell^2 f''(\Psi)\mathcal{G}\rvert_{\Psi=0} = - \ell^2 \eta\mathcal{G}/4 \).
Depending on the combined signs of the coupling \(\eta\) and the Gauss-Bonnet scalar \(\mathcal{G}\), unstable regions where \(m^2_{\Psi, \mathrm{eff}} < 0 \) may exist in vicinity of the BH.
For a Schwarzschild BH, \(\mathcal{G} = 48M^2/r^6 > 0 \), where \(M\) is the mass of the BH, and thus \(\eta \gtrsim 0\) suffices to create an instability region leading to the growth of the scalar~\cite{Silva:2017uqg}.
More generally, when the BH is spinning, \(\mathcal{G}\) may change sign around the poles. 
For \(\eta \gtrsim 0\) a similar situation to the spinless case occurs near the equator, whereas for \(\eta \lesssim 0\), and large enough dimensionless spin parameter \(a \gtrsim 0.5 \), a \emph{spin-induced} instability can occur in the region near the poles~\cite{Dima:2020yac}. 
    
As this instability triggers 
an exponential growth of the scalar in the vicinity of the BH,
the nonlinear terms in \(f'(\Psi)\) will eventually become as large as the linear term.
Spontaneous scalarization for BHs (originally discovered for neutron stars in Damour-Esposito-Far\`ese theory~\cite{Damour:1996ke}) in sGB theory occurs
when nonlinearities are such that at a later stage the instability is quenched and the system can settle to a stable configuration with nontrivial scalar \emph{hair}~\cite{Silva:2017uqg, Doneva:2017bvd, Antoniou:2017acq} --as illustrated in Fig.~\ref{fig: Scalarized BH example}; see also Ref.~\cite{Doneva:2022ewd} for a review.
For the model~\eqref{eq: coupling function quartic model} the transition to the final configuration will depend on the parameter \(\zeta\)
since the nonlinear terms associated to it will be dominant for large \(|\Psi|\).
A back-of-the-envelope calculation suggests that the end of the tachyonic instability (where the effective mass vanishes) occurs when the scalar reaches a final (maximum) scalar amplitude of order \(\Psi \sim \pm \sqrt{-\eta/\zeta} \), for \(\zeta\) with opposite sign to \(\eta\).
For stability, the value of \(\zeta \) is further required to satisfy \(\eta/\zeta \lesssim - 0.8\)~\cite{Silva:2018qhn, Minamitsuji:2018xde}.
The parameter space for such hairy black hole solutions can be characterised in terms of the black hole mass \(M\) and corresponding \emph{scalar charge} 
--see e.g.~Ref.~\cite{Herdeiro:2020wei} for the parameter space in a similar exponential model. 

Here we define the scalar charge (per unit mass squared) as
\begin{align} \label{eq: scalar charge definition}
    q \equiv - \lim_{R \to \infty} \dfrac{1}{4\pi M^2} \oint_{S_R} dS \, \hat{s}^{i} \partial_i \Psi~,
\end{align}
where the integral is taken over a sphere of radius \(R\), and \(\hat{s}^i\) is the outward normal to the sphere.
This definition of the scalar charge coincides with the usual definition through the scalar falloff at infinity, i.e.\ \(\Psi(r \to \infty) = \Psi_\infty + q M^2 /r + \mathcal{O}(r^{-2})\) --see e.g.~Ref.~\cite{Herdeiro:2020wei}.
Notice also that, for a generic model \(f(\Psi)\), the scalar charge does not necessarily correspond to a conserved Noether charge.
    
As we will see in Sec.~\ref{sec: Methodology}, our basic set up will consist on perturbing GR solutions to obtain such hairy BHs.

%%%%%%%%%%%%%%%%%%%%%%%%%%%%%%%%%%%%%%%%%%%%%%%%%%%%%%%%%%%%
%%%%%%%%%%%%%%%%%%%%%%%%%%%%%%%%%%%%%%%%%%%%%%%%%%%%%%%%%%%%
%%%%%%%%%%%%%%%%%%%%%%%%%%%%%%%%%%%%%%%%%%%%%%%%%%%%%%%%%%%%

\subsection{Fixing the equations of scalar Gauss-Bonnet gravity}

Before using the methods of NR to obtain numerical predictions, further mathematical treatment is often needed to bring the evolution equations into the form of a strongly hyperbolic PDE system --which under appropriate initial conditions can then allow for local well-posedness of the IVP.
For GR, this procedure often makes use of auxiliary variables, constraint equations and gauge freedom --some widely-used families of formulations include the Z4 and Generalized Harmonic systems.

For the equations of full sGB theory [Eqs.~\eqref{eq: scalar equation sGB}-\eqref{eq: definition of Pabcd}]
rewriting these equations in such a way is no trivial task. 
All of the existing approaches (either perturbative~\cite{Okounkova:2017yby} or with modified gauges~\cite{Kovacs:2020PRL, Kovacs:2020ywu, AresteSalo:2022hua, AresteSalo:2023mmd}) are valid only at \emph{weak coupling} in \(\ell^2\), where the corrections with respect to GR are assumed to remain suitably small.

In the \emph{fixing-the-equations} approach~\cite{Cayuso:2017iqc}, we replace
the terms that alter the principal part
of the system 
(the terms with the highest order of derivatives) 
with new variables subject to \emph{ad hoc} \emph{driver} equations.
The aim is to have the system react in a timescale \(\tau\) such that it will have the effect of damping (high-frequency) modes \(\omega\) with \(\omega \gtrsim 1/\tau\) which are suspect of leading to instabilities during the evolution.
As of yet, there is no unique prescription for the terms to replace nor on the precise form of that the driver equations need to have.
In Sec.~\ref{sec: theory advanced drivers and recovery of stationary solutions} we will use precisely this freedom to devise a driver equation adapted to the problem of recovering stationary solutions.

For the equations of sGB [Eqs.~\eqref{eq: scalar equation sGB} and \eqref{eq: tensor equation sGB}], we replace the right-hand-side of the equations with new variables \(\{\Sigma, M_{ab}\}\) to obtain~\cite{Franchini:2022ukz}
\begin{align} \label{eq: full fixed sGB system}
    \Box \Psi &= \Sigma ~, &
    G_{ab} &= \kappa T^{(\Psi)}_{ab}  + \kappa M_{ab}~.
\end{align}
In this way, the principal part of Eqs.~\eqref{eq: full fixed sGB system} is reduced to that of the simpler Klein-Gordon and Einstein equations, for which the standard methods to rewrite them as strongly hyperbolic systems apply.

To close the system, additional evolution equations that ensure that the auxiliary variables approach their corresponding physical source terms on a specified timescale must be prescribed for \(\{\Sigma , M_{ab}\}\).
The simplest example for such \emph{driver} equations is~\cite{Allwright:2018rut}
\begin{align} \label{eq: exponential drivers}
    \tau \partial_t \Sigma &= - \left(\Sigma - \mathcal{S}\right)~, \nonumber \\
    \tau \partial_t M_{ab} &= - \left(M_{ab} - S_{ab}\right)~,
\end{align}
where 
\begin{align} \label{eq: source terms definitions}
    \mathcal{S}&\equiv  - \ell^2 f'(\Psi)\mathcal{G}~, &
    \mathcal{S}_{ab} &\equiv  \ell^2  H_{ab}~, %
\end{align}
and \(\tau\) is a relaxation
timescale (with units of mass) controlling how fast the new variables approach the original source terms. (Notice the parallel of Eqs.~\eqref{eq: exponential drivers} to exponential decay laws or to constraint damping terms in NR.)
In some sense, if the action~\eqref{eq: action sGB} is thought as an effective field theory, the damping of high-frequencies can be interpreted as a way to ``soften'' the truncation
of the theory in a way that allows one to obtain quantitative predictions.

Finally, to compute or replace the second-order time-derivative terms (\(\partial^{2}_{t} \Psi\) or \(\partial^{2}_{t} g_{ab}\)) remaining in \(\{\mathcal{S}, \mathcal{S}_{ab}\}\), additional techniques may be required, such as \emph{order-reduction}~\cite{Allwright:2018rut}, where the leading order equations of motion (in \(\ell^2\)) would be used to reduce the order of second-order time-derivative terms.

Although Eqs.~\eqref{eq: exponential drivers} may be the simplest form of driver equations, in practice, they may not be the most convenient in an actual numerical implementation since they result in standing modes for \(\{\Sigma, M_{ab}\}\), instead of propagating modes.
Indeed, in Ref.~\cite{Franchini:2022ukz}, numerical relativity simulations of the fully coupled system~\eqref{eq: full fixed sGB system} have been performed in spherical symmetry using a driver equation with wavelike properties.
We will consider other alternatives for these driver equations in Sec.~\ref*{sec: theory advanced drivers and recovery of stationary solutions}.

%%%%%%%%%%%%%%%%%%%%%%%%%%%%%%%%%%%%%%%%%%%%%%%%%%%%%%%%%%%%
%%%%%%%%%%%%%%%%%%%%%%%%%%%%%%%%%%%%%%%%%%%%%%%%%%%%%%%%%%%%
%%%%%%%%%%%%%%%%%%%%%%%%%%%%%%%%%%%%%%%%%%%%%%%%%%%%%%%%%%%%

\subsection{Case study: fixing the decoupling limit of scalar Gauss Bonnet} \label{sec: Theory fixing the decoupling limit}

As remarked in Sec.~\ref{sec: Introduction}, in this paper, we take our first steps towards an implementation in full 3-dimensional space.
Since such an undertaking is far more complex than that of Ref.~\cite{Franchini:2022ukz}, for simplicity, we begin by studying the scalar dynamics of system~\eqref{eq: full fixed sGB system}, i.e.\ we will ``fix'' the decoupling limit of scalar Gauss-Bonnet theory.
To be more precise, we will evolve
\begin{align} \label{eq: fixed decoupling limit sGB system}
    \Box \Psi &= \Sigma ~, &
    R_{ab} &= 0~,
\end{align}
where the auxiliary variable \(\Sigma\) is subject to a driver equation to be specified below.

In this particular case, the original decoupling limit equations [Eq.~\eqref{eq: scalar equation sGB} and \(R_{ab} = 0\)] can be cast as a strongly hyperbolic system
and can be readily evolved without ``fixing''. Therefore, this set up will be useful as an exercise to test the numerical implementation of the system as well as to evaluate whether it can accurately describe the dynamics of the scalar and the final stationary solutions.
Indeed, we can directly compare with the solutions of the unmodified theory, which are, in some sense the ``exact'' solution which one would like to recover with the \emph{fixing-the-equations} approach.

\subsection{Advanced drivers and recovery of stationary solutions \label{sec: theory advanced drivers and recovery of stationary solutions}}

In this section we consider more complex choices for driver equations for scalar variables.
We will begin by considering a 
driver equation 
first presented in Ref.~\cite{Cayuso:2023aht} (with some slight modifications~\footnote{
The precise difference between the driver equation here and that in Ref.~\cite{Cayuso:2023aht} lies in the terms with the lowest derivatives (e.g. terms proportional to first derivatives of the lapse and shift), as well as multiplicative factors of the lapse. 
})
\begin{multline} \label{eq: advection squared driver}
    \sigma \left(\partial_t - \beta^i \partial_i \right)^2 \Sigma
    + \tau (\partial_t - \beta^i\partial_i)\Sigma
        = - \left(\Sigma - \mathcal{S}\right)~,
\end{multline}
where \(\alpha\) is the lapse, \(\beta^i = \gamma^{ij}\beta_{j}\) is the shift, and \(\gamma_{ij}\) is the spatial metric (with inverse \(\gamma^{ij}\)) in the \((3+1)\)-decomposition of the spacetime metric
\begin{align} \label{eq: 3+1 decomposition of the metric}
    ds^2 
    &= g_{ab} \, dx^a \, dx^b \nonumber\\
    &= - \alpha^2 \, dt^2 + \gamma_{ij} (\beta^i \, dt + dx^i)(\beta^j \, dt + dx^j) ~,
\end{align}
and also where
\(\mathcal{S}\equiv - \ell^2 f'(\Psi)\mathcal{G}\) and \(\{\sigma, \tau\}\) are dimensionful non-negative parameters (with dimensions \([\mathrm{mass}]^{2}\) and \([\mathrm{mass}]\), respectively) that will control the relaxation timescale(s) of \(\Sigma\) towards the desired value given by \(\mathcal{S}\).
In the following, we will refer to Eq.~\eqref{eq: advection squared driver} as \emph{Driver I}.
Just as Eqs.~\eqref{eq: exponential drivers} are reminiscent of exponential decay laws, by looking at the time derivative terms, Eq.~\eqref{eq: advection squared driver} is reminiscent of a forced damped harmonic oscillator,
\(\ddot{x}(t) + 2 \zeta_d \omega_0 \dot{x}(t) + \omega_0^2 x(t) = \mathcal{F}\),
where
\(\omega_0 \equiv 1/\sqrt{\sigma}\) is the oscillator frequency,
\(\zeta_d \equiv \tau / (2 \sqrt{\sigma}) \) is the damping parameter,
and the source \(\mathcal{S}\) enters as a forcing term \(\mathcal{F}\equiv \mathcal{S}/\sigma\).
Moreover, it is clear that the advective operators \((\partial_t - \beta^{i}\partial_{i})\) entering Driver I are such that the propagating modes will move in the opposite direction of \(\beta^{i}\).
For BH spacetimes (depending on the specific gauge) these modes can become ingoing (into the hole) near apparent horizons,
thus allowing inaccuracies in the tracking of the source term to propagate outside of the region of interest.

While Driver I has many desirable properties, one issue associated with it (already pointed out in Refs.~\cite{Cayuso:2023aht}), as well as with related wavelike drivers of the form 
(such as the one of Ref.~\cite{Franchini:2022ukz})
\begin{align} \label{eq: box driver}
    -\sigma \Box \Sigma
    + \tau (\partial_t - \beta^{i}\partial_{i}) \Sigma
        = - \left(\Sigma - \mathcal{S}\right)~,
\end{align}
is that  stationary solutions are not recovered exactly. 
To see this, consider a stationary situation, with respect to the coordinate \(t\) associated with an approximate Killing vector \(\chi = \partial_t\). Assuming that all fields have become time independent \(\Sigma(t, x^{i}) = \Sigma (x^{i})\), we can set to zero all the time derivatives in 
Driver I.
The resulting equation, \(\sigma (\beta^i \partial_i )^2 \Sigma
- \tau  \beta^i\partial_i \Sigma
= - (\Sigma - \mathcal{S})\),
still contains spatial derivative operators.
If \(\mathcal{S}\) is spatially dependent (as for any non-trivial solution), \(\Sigma = \mathcal{S}\) cannot be a solution, and Driver I will not reproduce the correct answer.
Nevertheless, one would expect that these errors would decrease as \(\{\sigma, \tau\}\) are decreased --or as the relaxation timescales associated to these parameters are decreased. In Sec.~\ref{sec: Results} we will show with examples that this indeed seems to be the case.

To address this issue, we propose a \emph{new} Driver given by
\begin{multline} \label{eq: comving advection driver}
    \sigma (\partial_t - \beta^i \partial_i)(\partial_t + v^i \partial_i) \Sigma + \tau(\partial_t + v^i \partial_i) \Sigma = -(\Sigma - \mathcal{S})~.
\end{multline}
In the following, we will refer to this equation as \emph{Driver II}.
For a BH in uniform motion, we identify \(v^i\) as the velocity of the BH as obtained from the trajectory of the center of the apparent horizon.
The \emph{new} driver equation has now the desired limit to recover stationary solutions, which can be seen as follows: proceeding as before, for a BH with zero linear momentum (\(v^i = 0\)), setting the time derivatives to zero gives indeed \(\Sigma  = \mathcal{S}\). 
Whereas for a BH with linear momentum, the same should hold since one can identify \(D_t \equiv \partial_t +v^i \partial_i\) as the \emph{convective} time derivative.
In Sec.~\ref{sec: Results}, we will show that indeed Eq.~\eqref{eq: comving advection driver} gives an improved recovery over Eq.~\eqref{eq: advection squared driver}.

%%%%%%%%%%%%%%%%%%%%%%%%%%%%%%%%%%%%%%%%%%%%%%%%%%%%%%%%%%%%
%%%%%%%%%%%%%%%%%%%%%%%%%%%%%%%%%%%%%%%%%%%%%%%%%%%%%%%%%%%%
%%%%%%%%%%%%%%%%%%%%%%%%%%%%%%%%%%%%%%%%%%%%%%%%%%%%%%%%%%%%

\section{Methodology \label{sec: Methodology}}

%%%%%%%%%%%%%%%%%%%%%%%%%%%%%%%%%%%%%%%%%%%%%%%%%%%%%%%%%%%%
%%%%%%%%%%%%%%%%%%%%%%%%%%%%%%%%%%%%%%%%%%%%%%%%%%%%%%%%%%%%
%%%%%%%%%%%%%%%%%%%%%%%%%%%%%%%%%%%%%%%%%%%%%%%%%%%%%%%%%%%%

\subsection{First-order reduction}

All of the component subsystems of Eqs.~\eqref{eq: decoupling limit equations} and~\eqref{eq: fixed decoupling limit sGB system} are decoupled and can be written individually in first-order symmetric hyperbolic form
\begin{align} \label{eq: schematic symmetric hyperbolic system}
    \partial_t \boldsymbol{u} + \mathbb{A}^{i} \partial_{i} \boldsymbol{u} = \boldsymbol{\mathcal{S}}(\boldsymbol{u})~,
\end{align}
where  \(\boldsymbol{u}\) are first-order variables, \(\mathbb{A}= \mathbb{A}(\boldsymbol{u})\) is a square symmetric matrix, that may depend on $\boldsymbol{u}$ (but not its derivatives), and \(\boldsymbol{\mathcal{S}}(\boldsymbol{u})\) is a source term.

The principal part of the scalar equation for \(\Psi\) is that of Klein-Gordon equation. Therefore, we introduce first-order variables 
\begin{align}
    \boldsymbol{u}\equiv\{\Psi, \, \Phi^{(\Psi)}_{i} \equiv \partial_{i} \Psi, \, \Pi^{(\Psi)} \equiv -n^{c}\partial_{c}\Psi\},    
\end{align}
where \(n_c \equiv - \alpha \delta^{0}_{c}\) is the unit normal to the hypersurface,
and rewrite the scalar equation as
\begin{align} \label{eq: Klein Gordon first order form}
    \partial_t \Psi^{} - \left(1 + \gamma^{(\Psi)}_{1}\right)  \beta^i \partial_i \Psi^{}
      &\simeq 0,  \nonumber\\
    \partial_t \Pi^{(\Psi)}
      \!- \beta^k \partial_k \Pi^{(\Psi)}\! + \alpha \gamma^{ik} \partial_i \Phi^{(\Psi)}_k\! - \gamma^{(\Psi)}_{1} \gamma^{(\Psi)}_{2} \beta^i \partial_i \Psi^{}
      &\simeq 0,
      \nonumber\\
    \partial_t \Phi^{(\Psi)}_i\! - \beta^k \partial_k \Phi^{(\Psi)}_i + \alpha \partial_i \Pi^{(\Psi)} - \gamma^{(\Psi)}_{2} \alpha \partial_i \Psi^{}
      &\simeq 0.
\end{align}
Here, \(\gamma^{(\Psi)}_{1, 2}\) are constraint damping parameters and the \(\simeq 0 \) notation indicates that only the principal part terms are displayed ---see Ref.~\cite{Wittek:2023nyi} for the full expressions.

The Einstein field equations in the Generalized Harmonic gauge
have a principal part of the form
\(g^{cd}\partial_{c} \partial_{d} g_{ab} \simeq 0\),
and can therefore be written in a first-order form analogous to that of the Klein-Gordon equation.
We refer to Lindblom et al.~\cite{Lindblom:2005qh} for the full first-order equations employed here, where the first-order variables are given by
\begin{align}
    \boldsymbol{u} \equiv \{g_{ab}, \, \Phi_{iab} \equiv \partial_{i} g_{ab}, \,\Pi_{ab} \equiv -n^{c}\partial_{c}g_{ab}\}.
\end{align}

The first-order form for the driver sector depends on the particular driver used.
For Eq.~\eqref{eq: advection squared driver}
it reads
\begin{align} \label{eq: adv std first order}
    &\left(\partial_t - \beta^i \partial_i\right) \Sigma &=& -\alpha \Pi^{(\Sigma)}, \nonumber \\
    \sigma &\left(\partial_t - \beta^i \partial_i \right) \Pi^{(\Sigma)} \!\!\!\!\!\!\!\!& = &- \alpha^2 \tau   \Pi^{(\Sigma)} + \alpha \left(\Sigma - \mathcal{S}\right),
\end{align}
whereas for Eq.~\eqref{eq: comving advection driver}, it is given by
\begin{align} \label{eq: adv comov first order}
    \left(\partial_t + v^i \partial_i\right)&\Sigma &=& -\alpha \Pi^{(\Sigma)}, \nonumber \\
    \sigma \left(\partial_t - \beta^i \partial_i \right)&\Pi^{(\Sigma)} \!\!\!\!\!\!\!\!&=& -\alpha^2\tau   \Pi^{(\Sigma)} + \alpha \left(\Sigma - \mathcal{S}\right). 
\end{align}
Both systems depend on the first-order variables \(\boldsymbol{u} \equiv \{\Sigma, \, \Pi^{(\Sigma)}\}\), where \(\Pi^{(\Sigma)}\) is defined by the first equation of each system.  Notice that we do not need to introduce a first-order variable \(\Phi^{(\Sigma)}_i\) for \(\partial_i \Sigma\).

Finally, the Gauss-Bonnet scalar (in vacuum \(R_{ab}\equiv 0\)) appearing in the sources of the scalar equations for \(\{\Psi, \Sigma\}\) is computed in practice as
\begin{align}
    \mathcal{G} = C_{abcd} C^{abcd} = 8 \left(E_{ij} E^{ij} - B_{ij} B^{ij}\right),
\end{align}
where \(E_{ij} \equiv n^{a} n^{b} C_{aibj} \) and \(B_{ij} \equiv - n^{a} n^{b}  \, {^{*}C}_{a i b j}\) are the electric and magnetic parts (respectively) of the Weyl tensor \(C_{abcd}\), with left dual \({^{*}C_{abcd}\equiv \frac{1}{2}\epsilon_{abef}{C^{ef}}_{cd}}\) --their full  expressions in terms of (3+1)-decomposition quantities are given in Eqs.~(B7) and (B8) of Ref.~\cite{Okounkova:2017yby}.

%%%%%%%%%%%%%%%%%%%%%%%%%%%%%%%%%%%%%%%%%%%%%%%%%%%%%%%%%%%%
%%%%%%%%%%%%%%%%%%%%%%%%%%%%%%%%%%%%%%%%%%%%%%%%%%%%%%%%%%%%
%%%%%%%%%%%%%%%%%%%%%%%%%%%%%%%%%%%%%%%%%%%%%%%%%%%%%%%%%%%%

\subsection{Evolution code \label{sec: methodology evolution code}}

We implement systems~\eqref{eq: decoupling limit equations} and \eqref{eq: fixed decoupling limit sGB system} in \texttt{SpECTRE}~\cite{deppe_2024_10619885},
an open-source code based on task-based parallelism.

\begin{figure}[]
\centering
\includegraphics[width=3.4 in]{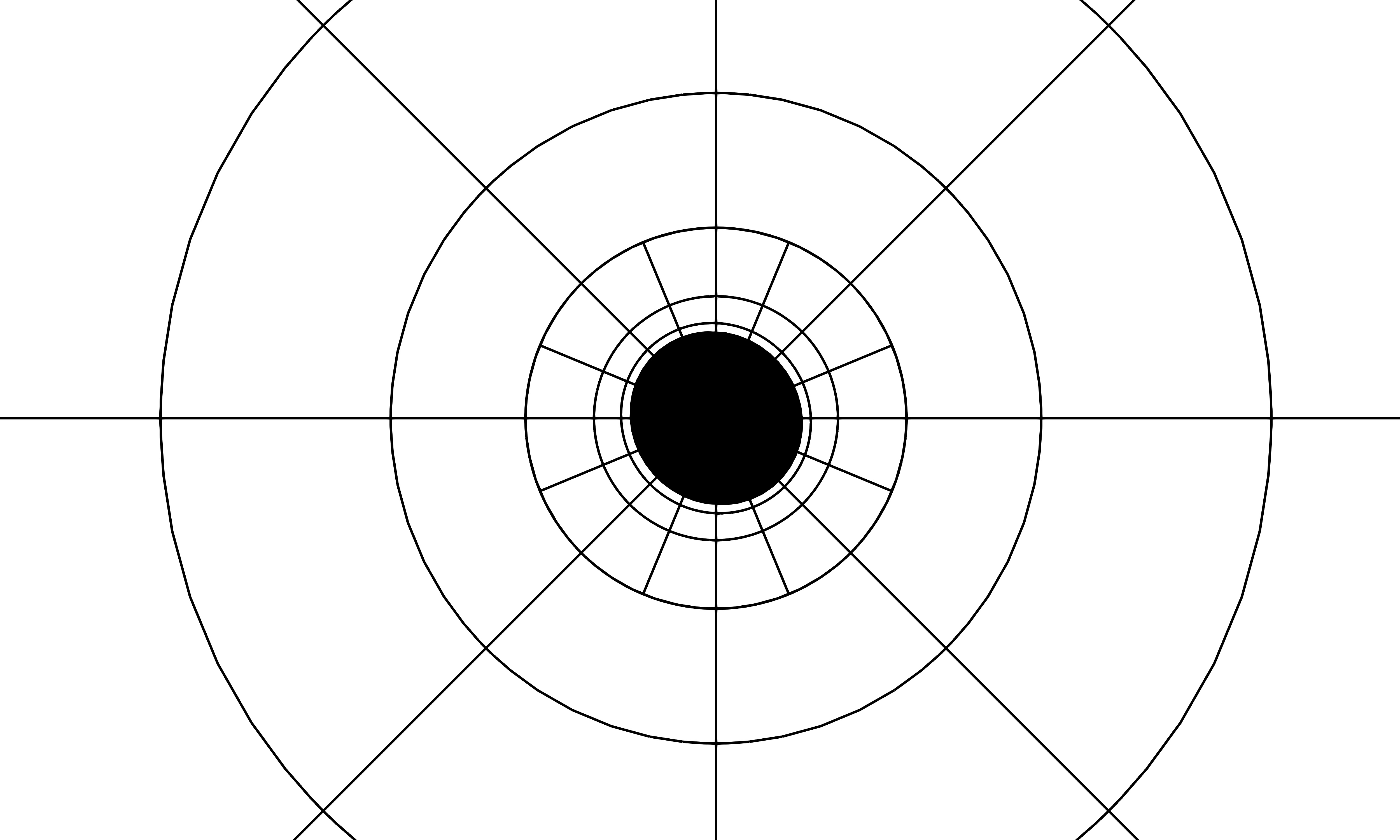}
\caption{ \emph{Spatial domain and apparent horizon.}
Domain decomposition for a black hole with (dimensionless) spin \(\boldsymbol{\chi} = 0.6 \, \hat{\boldsymbol{s}}\), where  \(\hat{\boldsymbol{s}} = (1/\sqrt{2}, 0, 1/\sqrt{2})\), in the \(xz\)-plane depicted above. The black lines correspond to discontinuous Galerkin element boundaries. The solid black region is delimited by the apparent horizon of the BH.
}
\label{fig: Domain}
\end{figure}

The full system is discretized following a discontinuous Galerkin scheme employing a numerical upwind flux.
Evolution in time is carried out by means of a fourth-order Adams-Bashforth time-stepper
with local adaptive time-stepping~\cite{throwe2020highorder}, and we apply a weak exponential filter on all evolved fields at each time step.
The spatial domain is illustrated in Fig.~\ref{fig: Domain} and consists of a series of concentric spherical shells. The outer boundary is located at \(R / M = 500\), while the inner boundary conforms to the shape of the apparent horizon. Each shell is further decomposed into elements isomorphic to unit cubes, endowed with a tensor product of Legendre polynomials.

The implementation of the scalar sector with a Klein-Gordon equation has been recently been described in detail in Ref.~\cite{Wittek:2023nyi}.
In particular, we impose constraint-preserving spherical-radiation boundary conditions~\cite{bayliss1980radiation} on the scalar field \(\Psi\).
The constraint damping parameters \(\{\gamma^{(\Psi)}_{1}, \gamma^{(\Psi)}_{2}\}\) are modulated by Gaussian profiles 
\begin{align} \label{eq: gaussian profile}
    \gamma_J(\boldsymbol{x}) \equiv C_J + A_J \exp\left(-\dfrac{\lvert\boldsymbol{x}-\boldsymbol{x}_{c, J}\rvert^2}{2 w_{J}^{2}}\right)~,
\end{align}
centered around the black hole \(\boldsymbol{x}_{c, J} = \boldsymbol{x}_\mathrm{BH}\). We set \(\gamma^{(\Psi)}_1 \equiv 0\) and specify \(\gamma^{(\Psi)}_{2}\) through the parameters
\(\{A = 6, \, w = 11M, \, C = 10^{-3}\}\).
For the metric sector, the full details of the numerical implementation will be presented in a specialized publication elsewhere~\cite{SXSinprep}, and we only broadly summarize the most relevant aspects here.
The evolution system for the metric variables in the Generalized Harmonic (GH) gauge is given in Ref.~\cite{Lindblom:2005qh} and is implemented in a dual-frame formulation~\cite{Scheel:2006gg}.
The constraint-damping parameters \(\{\gamma_0, \gamma_1, \gamma_2\}\)
follow a spatial distribution of the form~\eqref{eq: gaussian profile}. We choose \(\gamma_1 \equiv -1\) and 
\(\{A = 3, \, w/M = 11, \, C = 10^{-3}\}\)
for \(\gamma_0\) and \(\gamma_2\).
The gauge functions \(H_{a}\) in the GH system are evolved in Damped Harmonic (DH) gauge~\cite{Lindblom:2009tu, Choptuik:2009ww, Szilagyi:2009qz}.
A portion of the domain inside the apparent horizon is excised and
no boundary conditions are imposed at the excision sphere --we monitor, however, that at that boundary 
all characteristic fields have velocities flowing out of the computational domain, so that no boundary condition is required for the evolution to remain well-posed.
Finally, constraint-preserving Bjorhus boundary conditions~\cite{Lindblom:2005qh,Rinne:2007ui} are imposed at the outer boundary.

For the scalar driver part, the implementation is similar to that of the scalar field.
We monitor that the characteristic velocities are flowing out of the computational domain at the excision sphere and
impose
zero Dirichlet boundary conditions (i.e.\ \(\Sigma\rvert_\mathrm{out} \equiv 0\)) at the outer boundary of the spatial domain. 
Further, we allow for a Gaussian spatial dependence of the form~\eqref{eq: gaussian profile} for the parameters  \(\{\sigma, \tau \}\) introduced in the driver equation for \(\Sigma\). We choose 
\(\{w/M = 150, \, C = 1\}\)
for \(\sigma\), 
\(\{w/M = 150, \, C = 10^{-4}\}\)
for \(\tau\) and vary the amplitude \(A\) for both parameters.
For simplicity, when we quote values of \(\{\sigma, \tau\}\), we refer to their values at the center of the Gaussian profile.

%%%%%%%%%%%%%%%%%%%%%%%%%%%%%%%%%%%%%%%%%%%%%%%%%%%%%%%%%%%%
%%%%%%%%%%%%%%%%%%%%%%%%%%%%%%%%%%%%%%%%%%%%%%%%%%%%%%%%%%%%
%%%%%%%%%%%%%%%%%%%%%%%%%%%%%%%%%%%%%%%%%%%%%%%%%%%%%%%%%%%%

\subsection{Initial data}

We choose as initial data a \emph{hairless} Kerr black hole subject to an initial scalar perturbation on \(\Psi\). As remarked in Sec.~\ref{sec: Theory}, this solution is unstable and is expected to migrate during the evolution to a stable \emph{hairy} black hole.

For the metric variables, we choose Kerr-Schild initial data (with \(M = 1\)), in Cartesian coordinates \(\boldsymbol{x} = (x, y, z)\),
\begin{align}
    g_{ab} = \eta_{ab} + \mathcal{H} l_a l_b.
\end{align}
Here \(\eta_{ab} = \mathrm{diag}(-1, 1, 1, 1)\) is the Minkowski metric, and
the scalar function \(\mathcal{H}\) and one-form \(l_a\) (which satisfies \(l^c \partial_c l_a = l^c \nabla_c l_a = 0\)) are given by
\begin{align}
    \mathcal{H} &\equiv  \dfrac{M \rho^3}{\rho^4 +a^2 z^2}\,, \\
    l_a & \equiv \left(1, \dfrac{\rho x+ay}{\rho^2+a^2}, \dfrac{\rho y-ax}{\rho^2+a^2}, \dfrac{z}{\rho}\right),
\end{align}
where the spin direction is along the \(+z\) direction, \(\rho\) is implicitly defined through 
\(\rho^2(x^2+y^2) + (\rho^2+a^2)z^2 = \rho^2(\rho^2+a^2)\),
and \(a\) is the dimensionless spin parameter of the BH.
For the case of a BH with linear momentum, we obtain a boosted Kerr-Schild solution by applying the appropriate Lorentz boost to the coordinates \(x^a\) and  \(l_a\).

To induce the scalarization of the BH, we prescribe a scalar perturbation of the form~\cite{Scheel:2003vs}
\begin{align} \label{eq: scalar perturabtion profile}
    \Psi(r)\rvert_{t = 0} &= \Phi^{(\Psi)}_{i}(r)\rvert_{t = 0} \equiv 0 ~,\nonumber \\
    \Pi^{(\Psi)}(r)\rvert_{t = 0} &\equiv \mathcal{A} \, \mathrm{Re}\left[e^{-(r-r_0)^2/w^2}Y_{lm}\left(\theta, \phi\right)\right] ~,
\end{align}
where \(Y_{lm}(\theta, \phi)\) are spherical harmonics and \(\{\mathcal{A}, w, r_0\}\) are the amplitude, width and radius of the scalar profile.

For the quartic model~\eqref{eq: coupling function quartic model}, this initial data for \(\Psi\) implies that initially the scalar source vanishes, i.e.\ \(\mathcal{S}\rvert_{t = 0}  \equiv - \ell^2 f '(\Psi) \mathcal{G}_\mathrm{GB}= 0\).
Therefore, we prescribe zero initial conditions for the scalar driver, i.e.
\begin{align}
    \Sigma\rvert_{t = 0} = \Pi^{(\Sigma)}\rvert_{t = 0} \equiv  0~.
\end{align}

%%%%%%%%%%%%%%%%%%%%%%%%%%%%%%%%%%%%%%%%%%%%%%%%%%%%%%%%%%%%
%%%%%%%%%%%%%%%%%%%%%%%%%%%%%%%%%%%%%%%%%%%%%%%%%%%%%%%%%%%%
%%%%%%%%%%%%%%%%%%%%%%%%%%%%%%%%%%%%%%%%%%%%%%%%%%%%%%%%%%%%

\subsection{Comparison strategy and diagnostics}

For each of the examples in Sec.~\ref{sec: Results} we evolve three systems of equations with the same initial data. These are:
\begin{enumerate}
    \item The (unmodified) decoupling limit equations~\eqref{eq: decoupling limit equations}.
    We label these solutions \(\Psi^\text{(dec)}\).
    \item The ``fixed'' decoupling limit equations~\eqref{eq: fixed decoupling limit sGB system}
    with \(\Sigma\) evolved using Driver I [Eq.~\eqref{eq: advection squared driver}].
    We label these solutions \(\Psi^\text{(fix, I)}\).
    \item The ``fixed'' decoupling limit equations~\eqref{eq: fixed decoupling limit sGB system} with Driver II [Eq.~\eqref{eq: comving advection driver}].
    We label these solutions \(\Psi^\text{(fix, II)}\).
\end{enumerate}

We compare the scalar field volume configurations across the different systems (decoupling limit against the ``fixed'' versions of the decoupling limit) by computing a relative \emph{accuracy estimates} of the form
\begin{align}
    \mathcal{E}[Q] \equiv \dfrac{\lVert Q^\text{(dec)} - Q^\text{(fix)}\rVert}{\lVert Q^\text{(dec)}\rVert + \epsilon} ~,
\end{align}
where \(Q\) is the desired quantity to compare, \(\lVert \cdot \rVert\) is a suitable norm and \(\epsilon\) is a small number.

The accuracy estimate for the scalar field profile along a direction, for instance, in the \(x\) direction, is given by
\begin{align} \label{eq: relative error diagnostic radial}
    \mathcal{E}\left[\Psi(x)\right](t) \equiv \dfrac{\lvert \Psi^\mathrm{(dec)}(x) -\Psi^\mathrm{(fix)}(x) \rvert}{\lvert \Psi^\mathrm{(dec)}(x) \rvert + \epsilon}~.
\end{align}
When we compare the across the whole spatial domain we can define (see Refs.~\cite{Allwright:2018rut, Lara:2021piy} for similar measures)
\begin{align} \label{eq: relative error diagnostic}
    \mathcal{E}\left[\Psi\right](t) \equiv \dfrac{\lVert \Psi^\mathrm{(dec)} -\Psi^\mathrm{(fix)} \rVert_{2}}{\lVert \Psi^\mathrm{(dec)} \rVert_{2} + \epsilon}~,
\end{align}
\(\lVert\ \cdot ~\rVert_{2}\) is the \(L_{2}\)-norm with respect to the grid points \(\{x_i\}\),
\begin{align}
    \lVert y \rVert^{2}_{2} \equiv \dfrac{1}{N}\sum_{i =1}^{N} y(t, x_{i})^2~,
\end{align}
with \(N\) being the number of grid points.

We monitor the accuracy of the auxiliary field tracking the original value of the source term with (see Refs.~\cite{Lara:2021piy, Cayuso:2023aht} for similar measures)
\begin{align} \label{eq: relative tracking diagnostic}
    \mathcal{E}\left[\mathcal{S}\right](t) \equiv \dfrac{\lVert \Sigma- \mathcal{S}\rVert_{2}}{\lVert  \mathcal{S}\rVert_{2} + \epsilon}~.
\end{align}

Finally, we compute an accuracy estimate error for the scalar waveform, which is an example of an observable quantity.
We define the accuracy estimate for the root-mean-square (RMS) waveform \(\sqrt{\langle \Psi^2 \rangle}\) by
\begin{align} \label{eq: accuracy estimate for RMS wave}
    \mathcal{E}\left[ \sqrt{\langle \Psi^2 \rangle}\right](t; R) \equiv \dfrac{\left\lvert \sqrt{ \langle \Psi^2 \rangle^{\mathrm{(dec)}}} - \sqrt{\langle \Psi^2 \rangle^{\mathrm{(fix)}}} \right\rvert}{\left\lvert  \sqrt{\langle \Psi^2 \rangle^{\mathrm{(dec)}}} \right\rvert + \epsilon}~,
\end{align}
where \(\sqrt{\langle \Psi^2 \rangle}\)  is extracted at 6 equally-spaced spheres in the range \(R/M \in [50, 100]\) and for which the sphere average of the square of the scalar amplitude is computed as
\begin{align}
    \langle \Psi^2 \rangle \equiv \dfrac{1}{4 \pi R^2} \oint_{S_R} dS \, \Psi^2~.
\end{align}

%%%%%%%%%%%%%%%%%%%%%%%%%%%%%%%%%%%%%%%%%%%%%%%%%%%%%%%%%%%%
%%%%%%%%%%%%%%%%%%%%%%%%%%%%%%%%%%%%%%%%%%%%%%%%%%%%%%%%%%%%
%%%%%%%%%%%%%%%%%%%%%%%%%%%%%%%%%%%%%%%%%%%%%%%%%%%%%%%%%%%%
\subsubsection{Constraints}

We keep track of the first-order constraints for the scalar field to monitor the evolution. These are given by
\begin{align} \label{eq: scalar constraints}
    \mathcal{C}^{(\Psi)}_{i} &\equiv \partial_i \Psi- \Phi_{i}~, &
    \mathcal{C}^{(\Psi)}_{ij} &\equiv \partial_{i} \Phi_{j} - \partial_{j}\Phi_{i}~.
\end{align}
For the metric we keep track of 
the following constraints
\begin{align}
    \mathcal{C}_{a} &\equiv  H_{a} + \Gamma_{a} ~, &
    \mathcal{C}_{ia} &\simeq \partial_i C_{a}~, \nonumber\\
    \mathcal{C}_{iab} &\equiv \partial_{i}g_{ab} - \Phi_{iab} ~, &
    \mathcal{F}_{a} &\simeq n^{c} \partial_{c} \mathcal{C}_{a}~,
\end{align}
where \(\mathcal{C}_{ia}\) and \(\mathcal{F}_{a}\) are given in full in Eqs.~(43) and (44) of Ref.~\cite{Lindblom:2005qh}.
Here, \(\Gamma_{a} \equiv g_{ab} g^{cd}\Gamma^{b}_{cd}\) is a contraction of the 4-dimensional Christoffel symbol \(\Gamma^{a}_{bc}\).
Given the symmetric hyperbolic nature of the metric evolution equations, a symmetrizer can be constructed, and thus these constraints can be further condensed in a constraint energy \(\mathcal{E}_c\)
--given in Eq.~(53) of Ref.~\cite{Lindblom:2005qh}.

%%%%%%%%%%%%%%%%%%%%%%%%%%%%%%%%%%%%%%%%%%%%%%%%%%%%%%%%%%%%
%%%%%%%%%%%%%%%%%%%%%%%%%%%%%%%%%%%%%%%%%%%%%%%%%%%%%%%%%%%%
%%%%%%%%%%%%%%%%%%%%%%%%%%%%%%%%%%%%%%%%%%%%%%%%%%%%%%%%%%%%

\section{Results \label{sec: Results}}

\begin{figure}[]
    \centering
   \includegraphics[width=3.4 in]{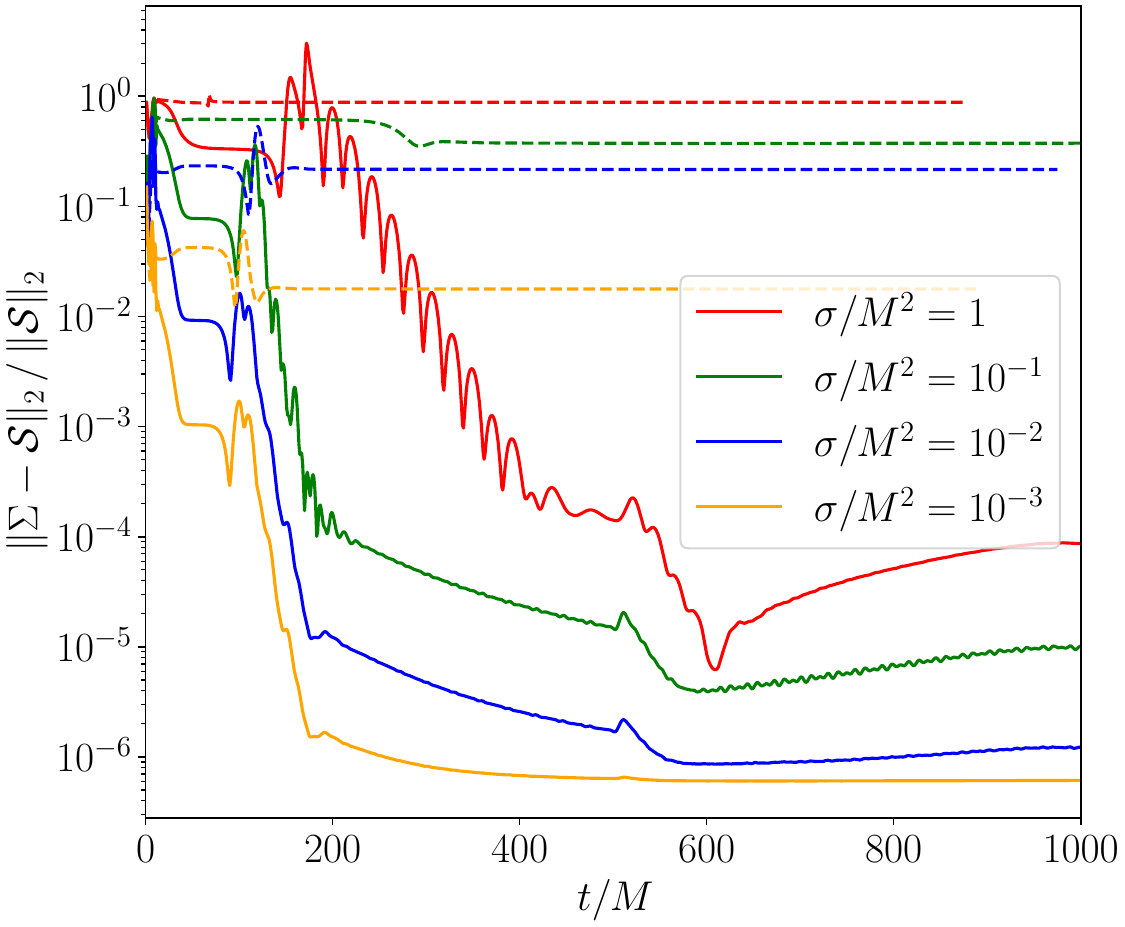}
   \caption{ \emph{\(\sigma\)-dependence for the accuracy estimate of the scalar source tracking (spinning).}
   Accuracy estimate for the scalar source (dashed lines) [Eq.~\eqref{eq: relative tracking diagnostic}] for Driver I [Eq.~\eqref{eq: advection squared driver}] as a function of time and for different values of sigma and where we fix \(\tau = \sigma/M\).
   Also in the plot, the same accuracy estimate (solid lines) for the Driver II [Eq.~\eqref{eq: comving advection driver}].
   }
   \label{fig: Relative tracking spinning}
   \end{figure}

\begin{figure}[]
 \centering
\includegraphics[width=3.4 in]{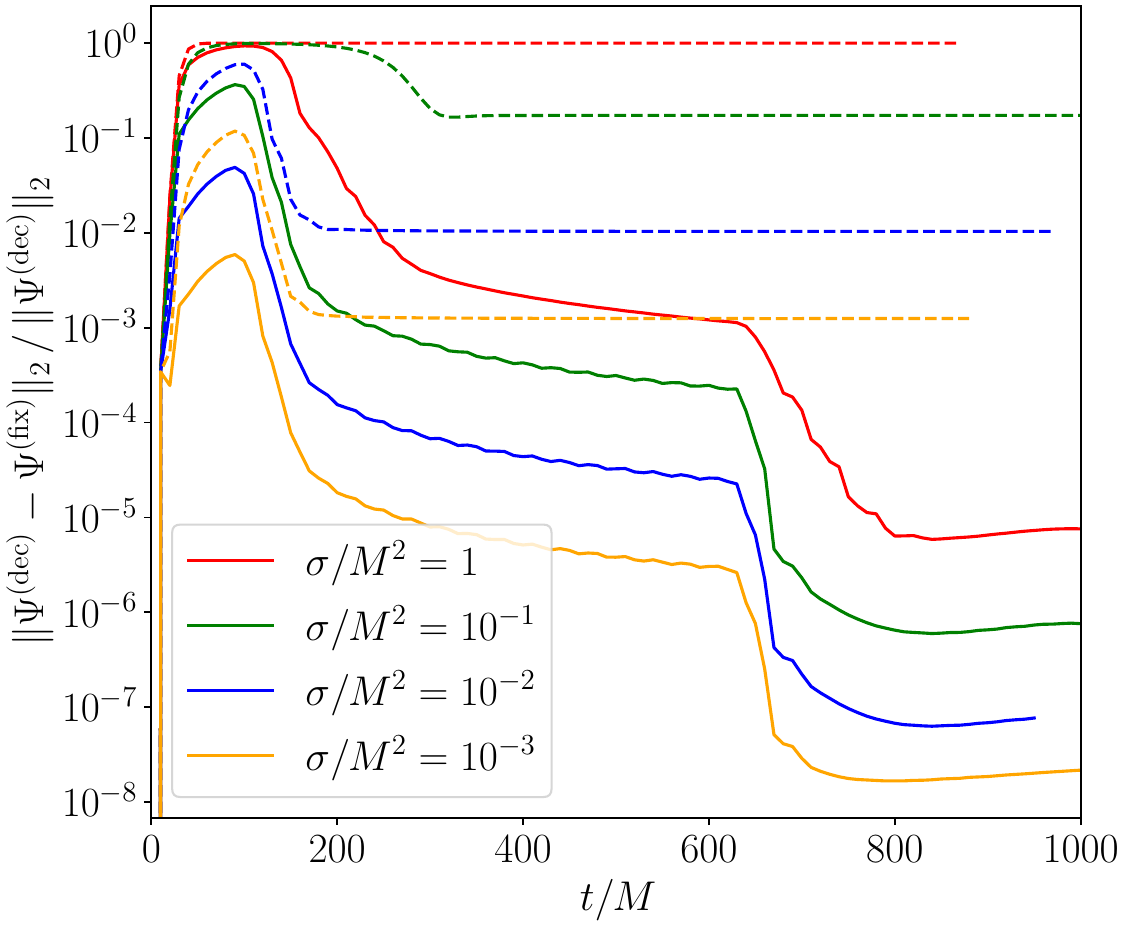}
\caption{ \emph{\(\sigma\)-dependence for the accuracy estimate in scalar field slices (spinning).}
Accuracy estimate for the scalar field 3-dimensional slices (dashed lines) [Eq.~\eqref{eq: relative error diagnostic}] for Driver I [Eq.~\eqref{eq: advection squared driver}] as a function of time and for different values of sigma and where we fix \(\tau = \sigma/M\). 
Also in the plot, the same accuracy estimate (solid lines) for the Driver II [Eq.~\eqref{eq: comving advection driver}].
}
\label{fig: Relative error spinning}
\end{figure}

\begin{figure}[]
 \centering
\includegraphics[width=3.4 in]{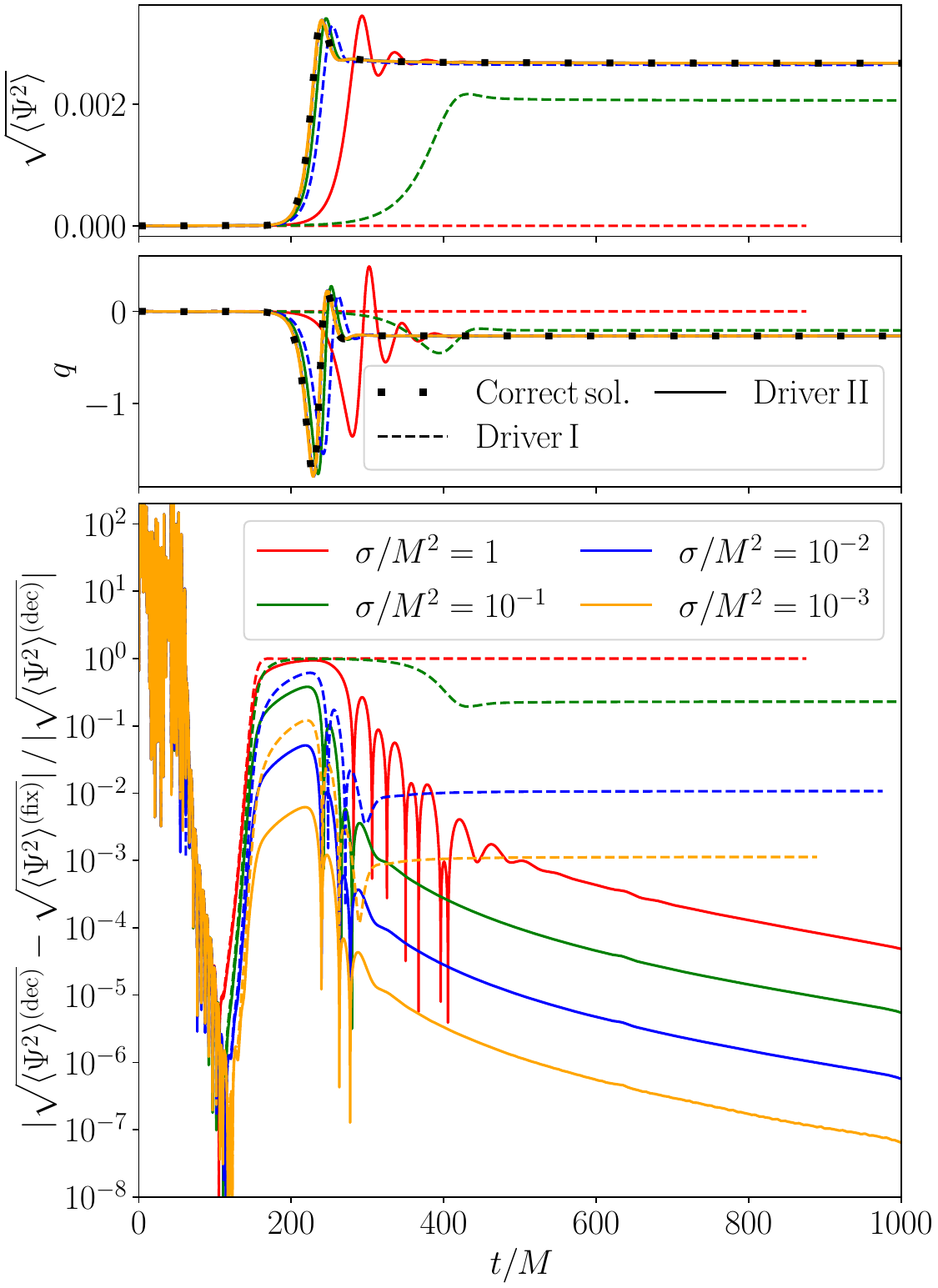}
\caption{ \emph{\(\sigma\)-dependence for the accuracy estimate of the scalar wave (spinning).}
Top panel: RMS scalar wave extracted at at \(R / M = 100\). We compare the signal between the correct solution (black dots) [evolved with Eqs.~\eqref{eq: decoupling limit equations}], the signal obtained for Driver I (dashed lines) [Eq.~\eqref{eq: advection squared driver}] and the signal obtained for Driver II (solid lines) [Eq.~\eqref{eq: comving advection driver}].
Middle panel: Corresponding scalar charge  (per unit mass squared) [Eq.~\eqref{eq: scalar charge definition}] extracted at the same finite radius and for which the correct solution asymptotes to \(q \simeq -0.267\).
Bottom panel:
The accuracy estimate for the RMS scalar wave [Eq.~\eqref{eq: accuracy estimate for RMS wave}] of the top panel shows that Driver II outperforms Driver I also in the dynamics.
}
\label{fig: Scalar wave relative error spinning}
\end{figure}

In the following, we illustrate with various examples the ability of our implementation of the \emph{fixing-the-equations} approach to reproduce the scalar dynamics of the decoupling limit of sGB theory.

\subsection{Spinning black holes \label{sec: Results spinning BHs}}

We start with an initially hairless Kerr BH with dimensionless spin parameter \(a = 0.6\) and with zero linear momentum.
We set the coupling parameters to 
\(\{\eta = 6, \, \zeta = -60\}\)
and induce
spontaneous scalarization by perturbing the BH with an initial scalar profile [Eq.~\eqref{eq: scalar perturabtion profile}] with parameters
\(\{\mathcal{A} = 10^{-5}, w/M = 5, r_0/M  = 30, (l, m) = (0,0)\}\).
In Fig.~\ref{fig: Scalarized BH example} we show the final stationary \emph{hairy} solution in the decoupling limit (the correct solution we are seeking to recover) against which we 
will compare the results obtained with different \emph{fixing-the-equations} schemes.
We then evolve the same set of initial data in the \emph{fixing-the-equations} approach for both Driver I and for our new Driver II to obtain stationary solutions.
Since the strongest dependence is on the \(\sigma\) parameter, for simplicity, we fix \(\tau  = \sigma / M\) and vary only \(\sigma\).
We choose values of \(\sigma / M^2 \leq 1\) such that the driver equations are in the underdamped regime \(\zeta_d < 1\) in the parallel made in Sec.~\ref{sec: theory advanced drivers and recovery of stationary solutions} with the damped harmonic oscillator.

The most straightforward comparison of the solutions is that of the scalar profile along a radial direction at late times. In the top panel of Fig.~\ref{fig: Scalarized BH example radial profile} we plot the scalar field along the \(x\)-axis (in the equatorial plane) for the exact solution (solid black) and for the different \emph{fixing-the-equations} approximations (colored lines).
In the bottom panel, we plot the accuracy estimate for the radial profile with respect to the exact solution.
We observe that as \(\sigma\) is decreased, the solutions obtained with both Driver I and II approach the desired solution.
For instance, for the largest choice of parameters, \(\sigma / M^2 = 1\), the solution with Driver I (dashed red) completely fails to reproduce the scalarized solution, giving a vanishing scalar profile everywhere.
As sigma is decreased to \(\sigma / M^2 = 10^{-1}\) (dashed green), the solution obtained with this driver is qualitatively similar in shape to the exact solution, but underestimated in magnitude~by about \(20\%\).
For smaller values of \(\sigma\), the agreement with the exact solution is close enough that it is no longer apparent in the top panel, and they yield an accuracy estimate of \(\lesssim 1 \%\).
This is to be expected since \(\sqrt{\sigma}\) is a timescale (lengthscale) that controls how closely the driver variable \(\Sigma\) is tracking the intended source term \(\mathcal{S}\).
Nevertheless, it is apparent that Driver II (solid lines) reproduces the correct solution far more accurately than Driver I.
Even for the largest parameters \(\sigma / M^2 = 1\), the agreement with the stationary solution is better than the best solution obtained with Driver I.

Having focused on late times in the previous paragraph, 
we now turn to analyzing the tracking in time throughout all of the scalarization event.
In our implementation of the \emph{fixing-the-equations} approach, \(\Sigma\) acts as the source term of the scalar field  \(\Psi\). To correctly capture the physics of the system, it needs to accurately track the original value of the source term \(\mathcal{S} \equiv - \ell^2 f'(\Psi)\mathcal{G}\).
In order to evaluate how well \(\Sigma\) is tracking \(\mathcal{S}\), in Fig.~\ref{fig: Relative tracking spinning} we compute the accuracy estimate for the scalar source term [Eq.~\eqref{eq: relative tracking diagnostic}] as a function of time.
The accuracy estimate is largest at the beginning of the simulation, when the scalar field is almost vanishing everywhere, and therefore one should ignore this part. A second peak at \(t / M \simeq 100 \) 
corresponds to the time when the tachyonic instability in the scalar is active and the scalar hair is growing at the fastest rate. At late times, as the scalar settles into a stationary situation, the estimate tends to an (almost) constant value.
As before, the accuracy estimate decreases as \(\sigma\) is decreased.
Whereas Driver I is limited at late times to approach \(\mathcal{S}\) exactly by non-vanishing spatial derivative terms, our new Driver II was designed to avoid this issue --recall the discussion in Sec~\ref{sec: theory advanced drivers and recovery of stationary solutions}. 
Therefore,
the improved behavior of our new Driver II (in solid lines) is apparent throughout the entire evolution: the accuracy estimate of \(\mathcal{S}\) is generally smaller, decreases faster and achieves much lower values at late times.
We have observed similar behaviour when we decrease the value of \(\tau\) with respect to \(\sigma\).

Since \(\Psi\) and \(\Sigma\) are related through a differential equation, the tracking the source term is an indirect comparison of the dynamics of the scalar field.
A more direct comparison however is to compare the full 3-dimensional scalar configuration of the \emph{fixing-the-equations} approximations against that of the correct solution.
In Fig.~\ref{fig: Relative error spinning}, we plot the accuracy estimate in time of the spatial scalar profiles with respect to the reference,
which confirms that the tracking behavior of the source term \(\Sigma \) observed before directly translates into a good tracking of the dynamics of \(\Psi\).

Finally, we turn to comparing an observable quantity: the scalar waveform.
In the top panel of Fig.~\ref{fig: Scalar wave relative error spinning}, we plot the RMS scalar waveform extracted at \(R / M = 100\) as well as the waveform in the exact solution.
For completeness, we also show in the middle panel the corresponding finite-radius approximation of the scalar charge (per unit mass squared) [c.f.~Eq.~\eqref{eq: scalar charge definition}], which reaches an asymptotic value of \(q \simeq -0.267\). 
In the bottom panel, the accuracy estimate for the scalar waveform against the correct solution is shown. 
At early times, the seemingly large accuracy estimate is due to the scalar field being almost zero at the location of the extraction sphere --and as before, this part of the estimate should be disregarded.
The scalar field acquires non-zero values (and the estimate quickly decreases) as the outgoing initial perturbation and the scalarization front reach the extraction sphere. 
In line with the previous discussion, the scalar wave is better tracked with the new Driver II (solid lines).
For instance,
the accuracy estimate for the best case of Driver II (orange solid lines) reaches \(< 10^{-2} \) at the peak, and is as low as \(\lesssim 10^{-7}\)  towards the end the evolution.
In comparison, the accuracy in the estimate for the corresponding best case for Driver I (orange dashed lines) saturates at \(\sim 10^{-3}\) after a peak of \(\sim 10 \% \) in accuracy.

%%%%%%%%%%%%%%%%%%%%%%%%%%%%%%%%%%%%%%%%%%%%%%%%%%%%%%%%%%%%
%%%%%%%%%%%%%%%%%%%%%%%%%%%%%%%%%%%%%%%%%%%%%%%%%%%%%%%%%%%%
%%%%%%%%%%%%%%%%%%%%%%%%%%%%%%%%%%%%%%%%%%%%%%%%%%%%%%%%%%%%

\subsection{Boosted black holes}

\begin{figure}[]
    \centering
   \includegraphics[width=3.4 in]{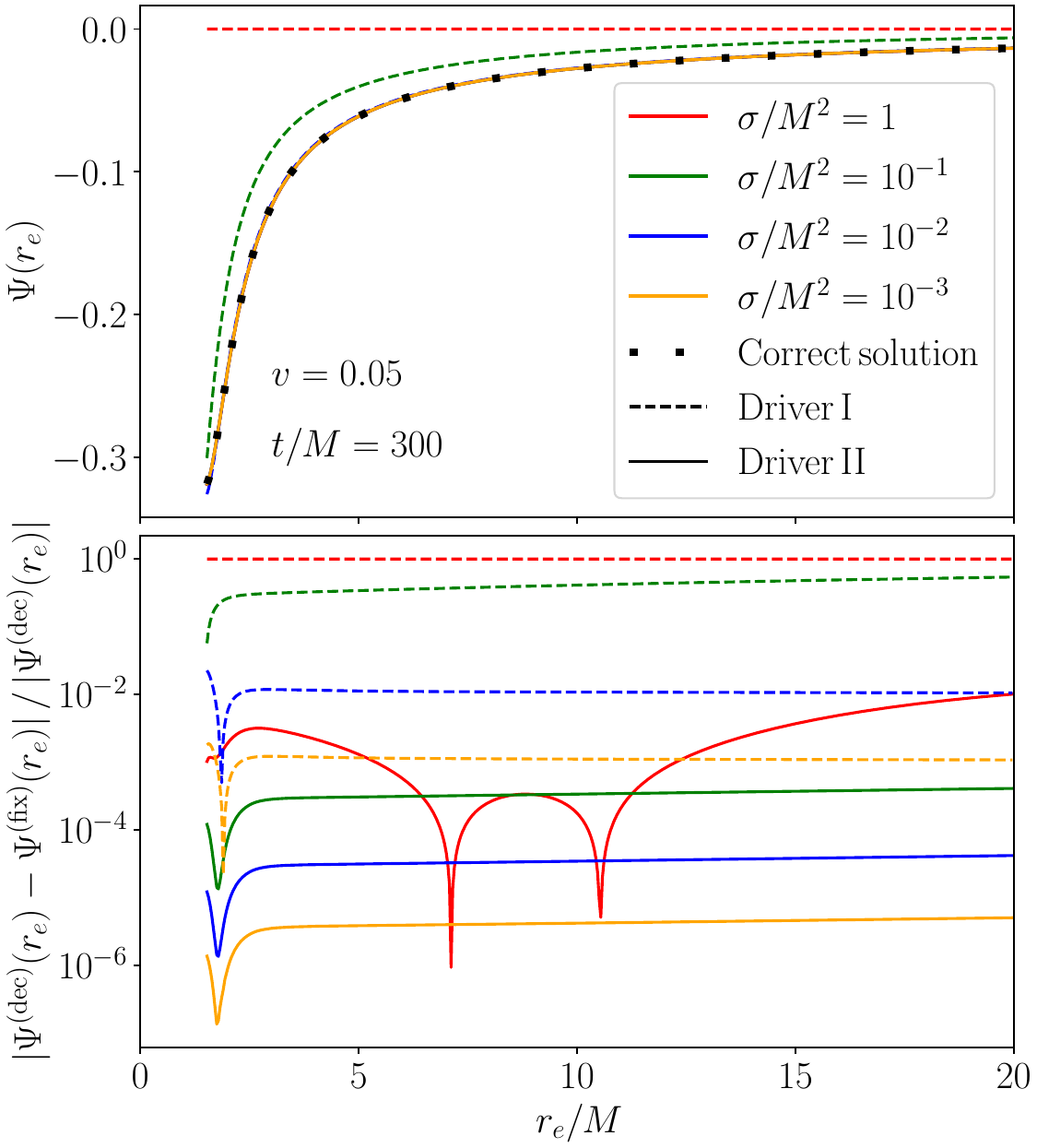}
   \caption{ \emph{Boosted scalarized spinning BH (equatorial profile).}
   In this example the BH is spinning \(\boldsymbol{\chi} = 0.6 \, \hat{\boldsymbol{s}}\), along the direction  \(\hat{\boldsymbol{s}} = (1/\sqrt{2}, 0, 1/\sqrt{2})\) and is also boosted with velocity \(\boldsymbol{v} = (0.05, 0, 0)\).
   Top: Scalar profile (solid black) for the BH of Fig.~\ref{fig: Scalarized BH example} along the equatorial direction \(r_e\) (at \(-45^{\circ}\) in the \(xz\)-plane) at \(t / M  = 300\).
   We show the scalar profile (dashed lines) corresponding to Driver I [Eq.~\eqref{eq: advection squared driver}] as well as the scalar profile (solid lines) corresponding to our \emph{new} Driver II [Eq.~\eqref{eq: comving advection driver}]. We fix \(\tau = \sigma / M\). 
   The scalar profiles are plotted starting from the excision sphere, slightly inside the apparent horizon of the BH.
   Bottom: Relative error in the scalar profile with respect to the correct solution given by the decoupling limit [Eqs.~\eqref{eq: decoupling limit equations}].
   }
   \label{fig: Moving BH example radial profile}
\end{figure}

\begin{figure}[]
 \centering
\includegraphics[width=3.4 in]{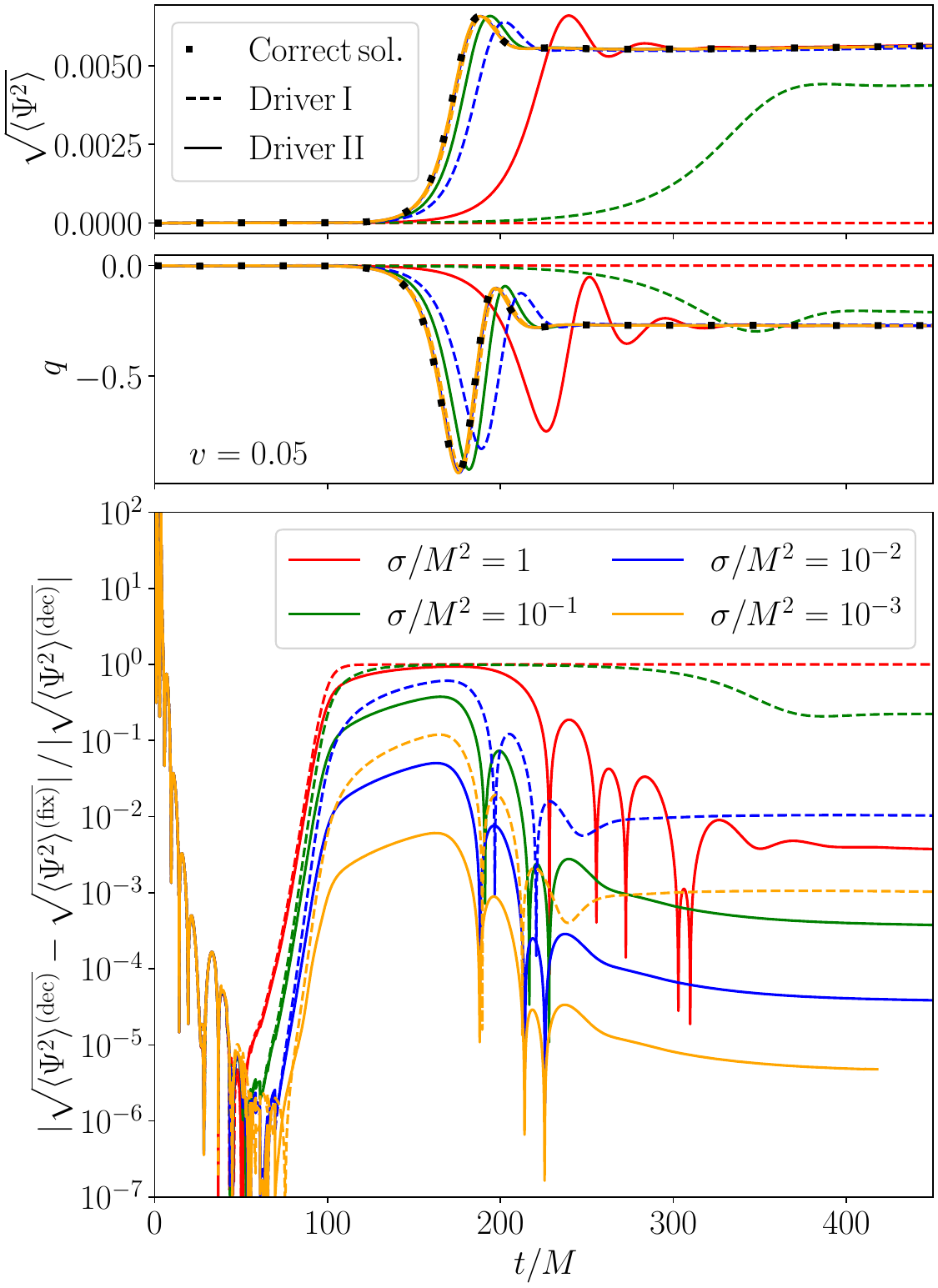}
\caption{ \emph{\(\sigma\)-dependence for the accuracy estimate of the scalar wave (boosted BH).}
Top panel: RMS scalar wave extracted at at \(R / M = 50\). We compare the signal between the correct solution (black dots) [evolved with Eqs.~\eqref{eq: decoupling limit equations}], the signal obtained for Driver I (dashed lines) [Eq.~\eqref{eq: advection squared driver}] and the signal obtained for Driver II (solid lines) [Eq.~\eqref{eq: comving advection driver}].
Middle panel: Corresponding scalar charge  (per unit mass squared) [Eq.~\eqref{eq: scalar charge definition}] extracted at the same finite radius.
and for which the correct solution asymptotes to \(q \simeq -0.273\).
Bottom panel:
The accuracy estimate for the RMS scalar wave [Eq.~\eqref{eq: accuracy estimate for RMS wave}] of the top panel shows that Driver II outperforms Driver I also in the dynamics.
}
\label{fig: Scalar wave relative error moving BH}
\end{figure}

We now study the case of boosted BH as a 
test of the applicability of our drivers to the early inspiral phase in BBH system.
In addition, this configuration will also test terms proportional to the BH velocity in the convective derivative on which our new Driver II is based.
We choose an initial boost velocity \(\boldsymbol{v} = (0.05, 0, 0)\) and spin unaligned with the direction of motion given by \(\boldsymbol{\chi} = 0.6 \, \hat{\boldsymbol{s}}\), along the direction  \(\hat{\boldsymbol{s}} = (1/\sqrt{2}, 0, 1/\sqrt{2})\) --illustrated in Fig.~\ref{fig: Domain}.
We use the same set of coupling parameters and initial scalar perturbation parameters as in the previous example.

In Fig.~\ref{fig: Moving BH example radial profile} we show the scalar profiles along the equator (which in this case is not aligned with the \(x\)-axis), and in Fig.~\ref{fig: Scalar wave relative error moving BH} we plot the scalar waveform and corresponding accuracy estimate for this case.
Qualitatively similar behaviour is seen as in the previous example.
Namely, that for both drivers the accuracy improves as \(\sigma/M^2\) is decreased and that for a given value of \(\sigma/M^2\) our new Driver II improves the accuracy of the scalar equatorial profile and scalar wave by several orders of magnitude compared to the same case evolved with Driver I.

There are two noticeable differences with respect to the vanishing BH velocity case.
First, the accuracy estimates corresponding to Driver II for the equatorial profile and for the scalar wave are larger.
Nevertheless, they remain clearly well below the estimate for Driver I.
Second, the length of the simulations is shorter than in the previous case.
In that regard, our simulations currently are limited  for larger speeds and longer later times due to an imperfect alignment of the excision sphere with respect to the moving apparent horizon of the BH.
Further improvements in automatic control systems are underway to dynamically adjust the location and shape of the excision sphere.
This will allow to control this source of error and to extend these simulations to \(v \gtrsim 0.05\) and \(t / M \sim \mathcal{O}(10^3)\).

%%%%%%%%%%%%%%%%%%%%%%%%%%%%%%%%%%%%%%%%%%%%%%%%%%%%%%%%%%%%
%%%%%%%%%%%%%%%%%%%%%%%%%%%%%%%%%%%%%%%%%%%%%%%%%%%%%%%%%%%%
%%%%%%%%%%%%%%%%%%%%%%%%%%%%%%%%%%%%%%%%%%%%%%%%%%%%%%%%%%%%

\section{Conclusion \label{sec: Conclusion}}

In this paper, we have reported on recent progress in implementing the \emph{fixing-the-equations} approach~\cite{Cayuso:2017iqc} in the numerical relativity code \texttt{SpECTRE}~\cite{deppe_2024_10619885}
with the aim of obtaining
quantitative predictions from BBH coalescence in scalar Gauss-Bonnet theory.
In this scheme, the equations of motion are modified by introducing additional variables
that replace the correction terms to the GR principal part that can lead to unstable solutions.
The evolution of these new variables is controlled by \emph{ad hoc} evolution equations that drive their values to those of the original terms within a certain timescale, effectively damping the high frequency modes that lead to instabilities in the original equations.
In its current stage, our code is able to evolve the scalar, metric and the scalar part of the driver systems in full 3-dimensional space using a discontinuous Galerkin scheme.

Since the driver equations in the \emph{fixing-the-equations} approach are not uniquely prescribed, another focus of this paper has been to identify effective driver equations that are able to accurately describe the stationary solutions of the theory.
Restricting to the decoupling limit of the theory (where we have neglected the back reaction of the scalar field on the metric), we were able to evolve both the original theory and the corresponding ``fixed'' theory, and thus evaluate the accuracy of the stationary solutions obtained with different driver equations.
By studying different examples of isolated BH that undergo spontaneous scalarization we have confirmed that, both in the stationary case as well as during the scalar dynamics of the scalarization event, one can recover the solutions of the original theory with increasing accuracy by tightening the timescales entering the driver equations.

Moreover, we have presented a new driver equation with an improved stationary limit and have shown that it performs remarkably well in recovering the solutions of the original theory in full 3-dimensional space without symmetries, in comparison with similar drivers in the literature~\cite{Franchini:2022ukz, Cayuso:2023aht}.
While our tests are encouraging for low BH speeds, further testing at higher speeds would be ideal to confirm that the improved behaviour persists in more challenging situations.
We leave this for future work as this requires improved control systems to adjust the location of the excision sphere inside the BH.

The results presented here will inform the next stages of our program to simulate BBH systems in the \emph{fixing-the-equations} approach.
From our experience, we have observed that there is a trade-off between increasing accuracy and computational resources. 
As we have shown here, by decreasing the timescales in the driver equations, one can obtain more accurate solutions.
This comes at the cost of a smaller (minimum) timestep required for the evolution --mainly because the driver equation becomes increasingly stiff.
By allowing for greater accuracy with larger values of the ``fixing'' timescales, we  anticipate that the new scalar driver equation presented here (and their required tensorial generalizations needed to handle the back-reaction of the scalar field) will have a beneficial effect in the computational resources needed to evolve such systems, at least during the inspiral stage.

%%%%%%%%%%%%%%%%%%%%%%%%%%%%%%%%%%%%%%%%%%%%%%%%%%%%%%%%%%%%
%%%%%%%%%%%%%%%%%%%%%%%%%%%%%%%%%%%%%%%%%%%%%%%%%%%%%%%%%%%%
%%%%%%%%%%%%%%%%%%%%%%%%%%%%%%%%%%%%%%%%%%%%%%%%%%%%%%%%%%%%

%%%%%%%%%%%%%%%%%%%%%%%%%%%%%%%%%%%%%%%%%%%%%%%%%%%%%%%%%%%%
%%%%%%%%%%%%%%%%%%%%%%%%%%%%%%%%%%%%%%%%%%%%%%%%%%%%%%%%%%%%
%%%%%%%%%%%%%%%%%%%%%%%%%%%%%%%%%%%%%%%%%%%%%%%%%%%%%%%%%%%%

\begin{acknowledgments}
  We would like to thank Enrico Barausse, Miguel Bezares, Ramiro
  Cayuso, Alexandru Dima, Nicola Franchini, Aaron Held, Aron D. Kovacs,
  Luis Lehner and Peter J. Nee for useful discussions about the
  fixing-the-equations approach.
  Computations were performed on the Urania and Raven HPC systems at the Max Planck Computing and Data Facility.
  This work was supported in part by the Sherman Fairchild Foundation, and by NSF Grants PHY-2309211, PHY-2309231, and OAC-2209656 at Caltech.
  Alexander Carpenter and Geoffrey Lovelace acknowledge support from NSF award PHY-2208014, the Dan Black Family Trust and Nicholas and Lee Begovich.
\end{acknowledgments}

%%%%%%%%%%%%%%%%%%%%%%%%%%%%%%%%%%%%%%%%%%%%%%%%%%%%%%%%%%%%
%%%%%%%%%%%%%%%%%%%%%%%%%%%%%%%%%%%%%%%%%%%%%%%%%%%%%%%%%%%%
%%%%%%%%%%%%%%%%%%%%%%%%%%%%%%%%%%%%%%%%%%%%%%%%%%%%%%%%%%%%

\FloatBarrier

\appendix

\section{ Comparison with a wavelike driver \label{sec: Appendix comparison with box driver}}

In this section we compare our accuracy estimates with against a wavelike drive equation of the form of Eq.~\eqref{eq: box driver} for the case of Sec.~\ref{sec: Results spinning BHs}.
Similar conclusions as in the main text hold.

In first-order form, the wavelike driver scalar system is given by
\begin{align} \label{eq: box driver first order}
    \left(\partial_t - \beta^i \partial_i\right) &\Sigma = -\alpha \Pi^{(\Sigma)} \nonumber \\
    \left(\partial_t - \beta^j \partial_j\right) &\Phi^{(\Sigma)}_{j} = -\alpha \partial_i \Pi^{(\Sigma)} - \Pi^{(\Sigma)}\partial_i \alpha + \Phi_j \partial_i \beta^j \nonumber \\
    & + \gamma^{(\Sigma)}_{2} \alpha \left(\partial_i \Psi - \Phi_i\right) \nonumber\\
    \sigma \left(\partial_t - \beta^i \partial_i \right)&\Pi^{(\Sigma)} = - \alpha^2\tau   \Pi^{(\Sigma)} + \alpha \left(\Sigma - \mathcal{S}\right) - \alpha \gamma^{ij}\partial_i\Phi^{(\Sigma)}_j \nonumber \\
    & + \alpha {^{(3)}\Gamma^{i}}\Phi^{(\Sigma)}_i 
    + \alpha K \Pi^{(\Sigma)} 
    - \gamma^{ij} \Phi^{(\Sigma)}_{i} \partial_{j} \alpha~, 
\end{align}
where \(^{(3)}\Gamma^{i} \equiv \gamma^{jk}{^{(3)}\Gamma}^{i}_{jk}\)  is a contraction of the spatial Christoffel symbol, \(K \equiv \gamma^{ij}K_{ij}\) is the trace of the extrinsic curvature \(K_{ij}\), and \(\gamma^{(\Sigma)}_{1, 2}\) are parameters which are taken to have the same spatial dependence and value as for the system of the sGB scalar \(\Psi\). Given the similarity of this driver equation to the Klein-Gordon equation, we implement constraint-preserving spherical radiation boundary conditions at the outer edge of the domain for this case instead of Dirichlet boundary conditions.

\begin{figure}[]
 \centering
\includegraphics[width=3.4 in]{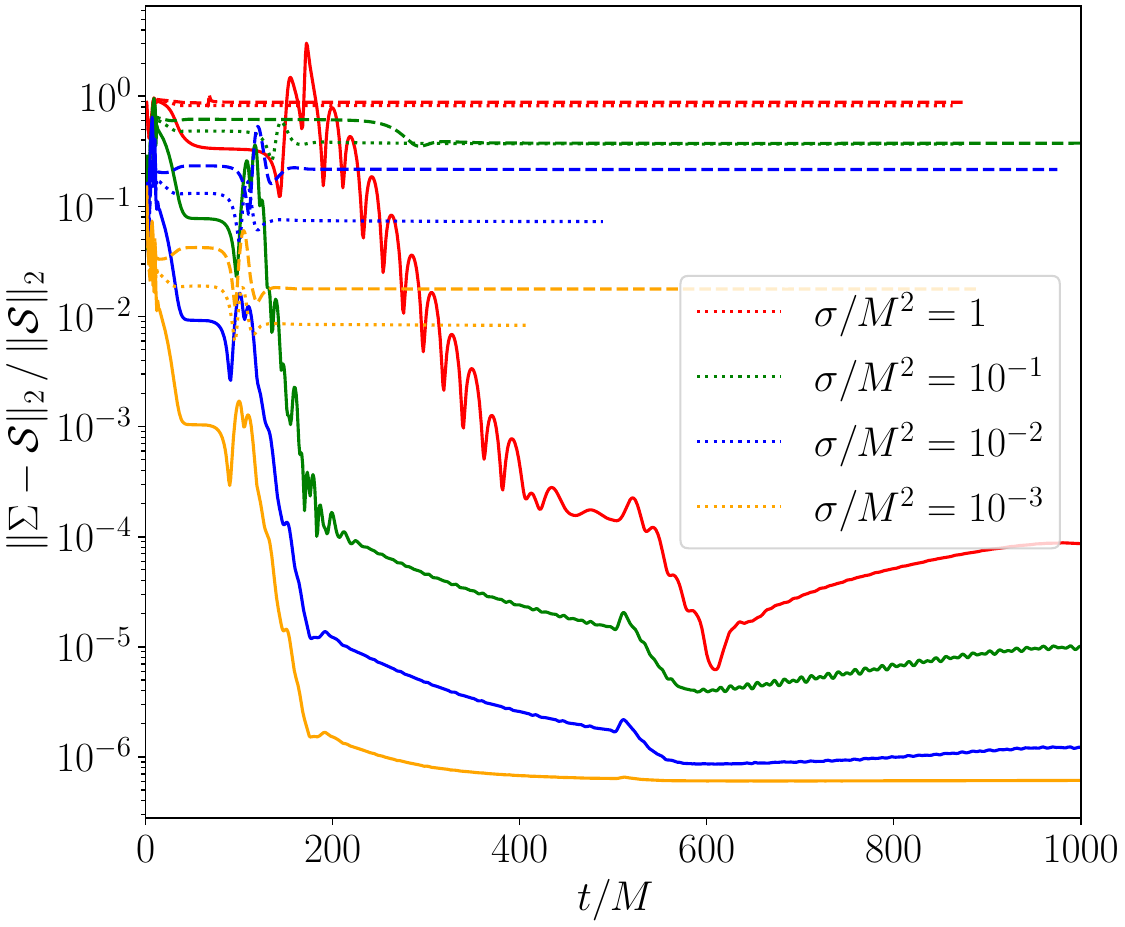}
\caption{ \emph{Comparison of the \(\sigma\)-dependence for the relative error the scalar source tracking with a wavelike driver (spinning).}
Same as Fig.~\ref{fig: Relative tracking spinning} in the main text, but we include the accuracy estimate (dotted lines) for the wavelike driver [Eq.~\eqref{eq: box driver}].
}
\label{fig: Relative tracking comparison with box driver}
\end{figure}

\begin{figure}[]
 \centering
\includegraphics[width=3.4 in]{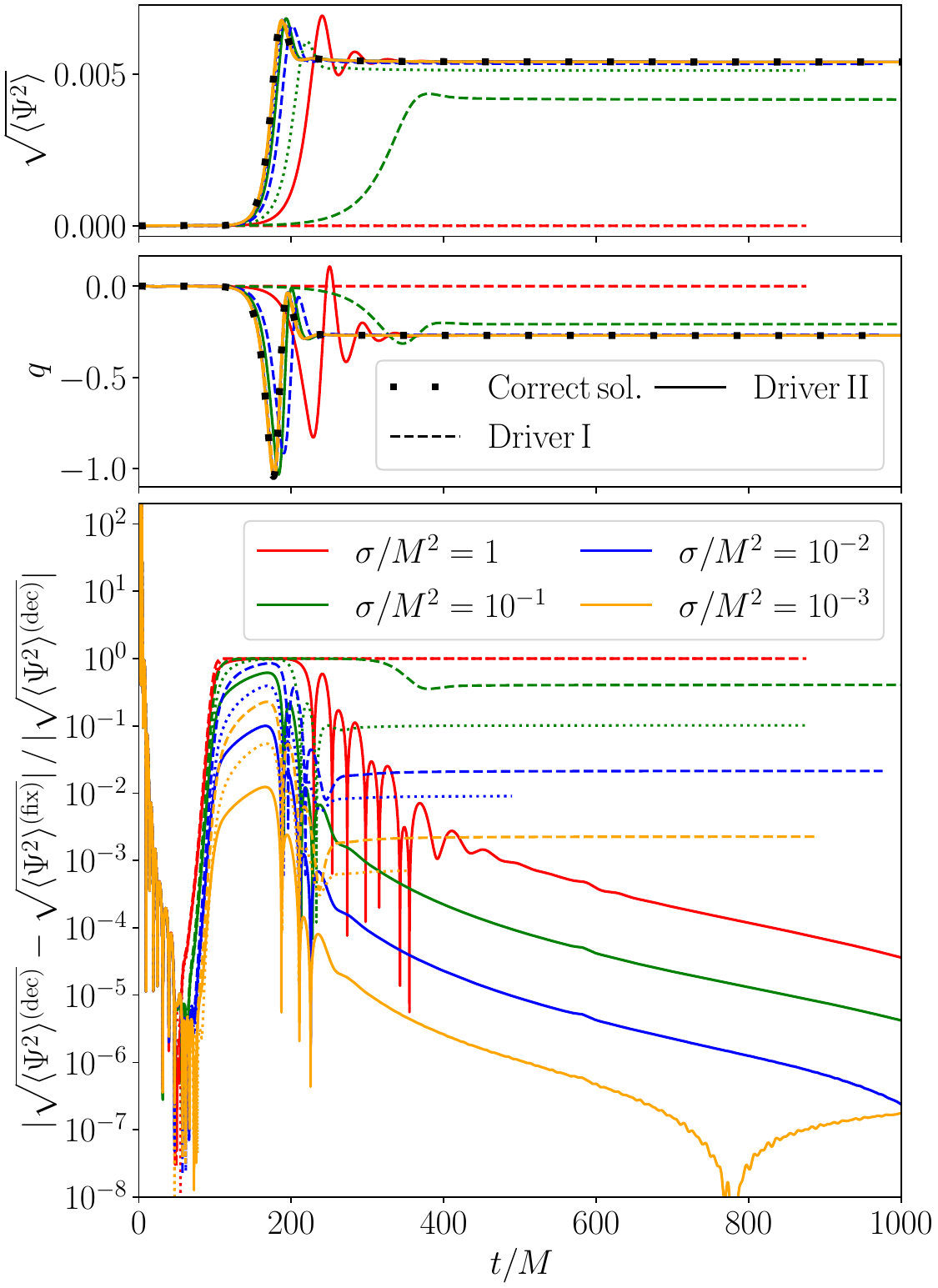}
\caption{ \emph{Comparison of the \(\sigma\)-dependence for the relative error of the scalar wave with a wavelike driver (spinning).}
Same as Fig.~\ref{fig: Scalar wave relative error spinning} in the main text, but extract at \(R/M = 50\) and include the accuracy estimate (dotted lines) for the wavelike driver [Eq.~\eqref{eq: box driver}].
}
\label{fig: Scalar wave relative error comparison with box driver}
\end{figure}

In Fig.~\ref{fig: Relative tracking comparison with box driver}, we show the relative tracking of the source for all driver equations explored in this work. When using the wavelike driver [Eq.~\eqref{eq: box driver}], the solution (in dotted lines) is subject to an asymptotic error with respect to the stationary solution of the decoupling limit. While it improves with respect to Driver I [Eq.~\eqref{eq: advection squared driver}], it is still outperformed by Driver II [Eq.~\eqref{eq: comving advection driver}].
In Fig.~\ref{fig: Scalar wave relative error comparison with box driver}, we show the corresponding accuracy estimate for the scalar wave. 

%%%%%%%%%%%%%%%%%%%%%%%%%%%%%%%%%%%%%%%%%%%%%%%%%%%%%%%%%%%%
%%%%%%%%%%%%%%%%%%%%%%%%%%%%%%%%%%%%%%%%%%%%%%%%%%%%%%%%%%%%
%%%%%%%%%%%%%%%%%%%%%%%%%%%%%%%%%%%%%%%%%%%%%%%%%%%%%%%%%%%%

\section{Convergence}\label{sec: Appendix Convergence}

In this Appendix, we provide more details on the convergence of the simulations presented in the main text.

\begin{figure}[]
 \centering
\includegraphics[width=3.4 in]{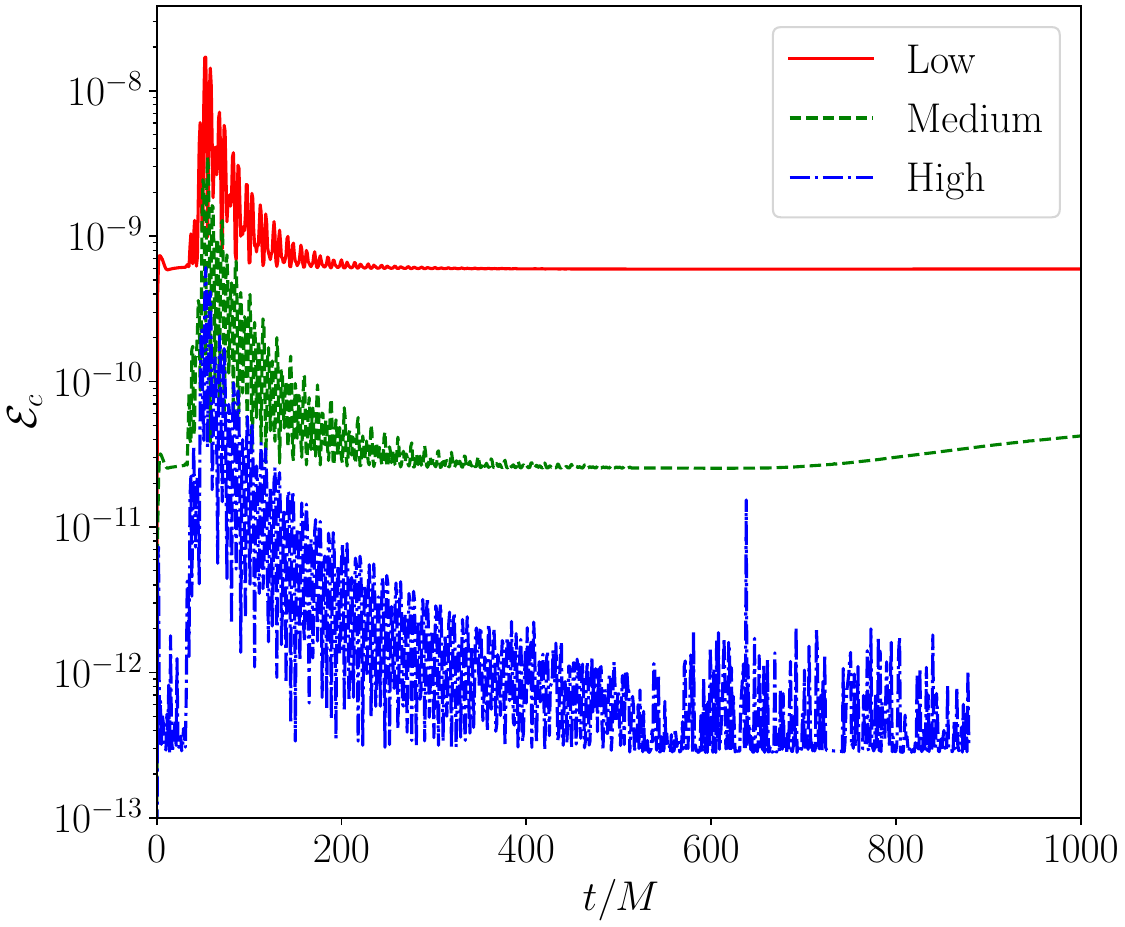}
\caption{ \emph{Convergence test for the constraint energy \(\mathcal{E}_{c}\).}
The constraint energy [Eq.~(53) of Ref.~\cite{Lindblom:2005qh}], which combines the \(\mathcal{C}_{a}\), \(\mathcal{F}_{a}\), \(\mathcal{C}_{ab}\), \(\mathcal{C}_{ibc}\), for different \(p\)-refinement resolutions. Here \(p_\mathrm{mid} = p_\mathrm{low} + \Delta p \) and \(p_\mathrm{high} = p_\mathrm{mid} + \Delta p = p_\mathrm{low} + 2 \Delta p\), for each DG element  and direction, with constant increase \(\Delta p = 1\).
}
\label{fig: Constraint energy convergence adv comov}
\end{figure}

\begin{figure}[]
 \centering
\includegraphics[width=3.4 in]{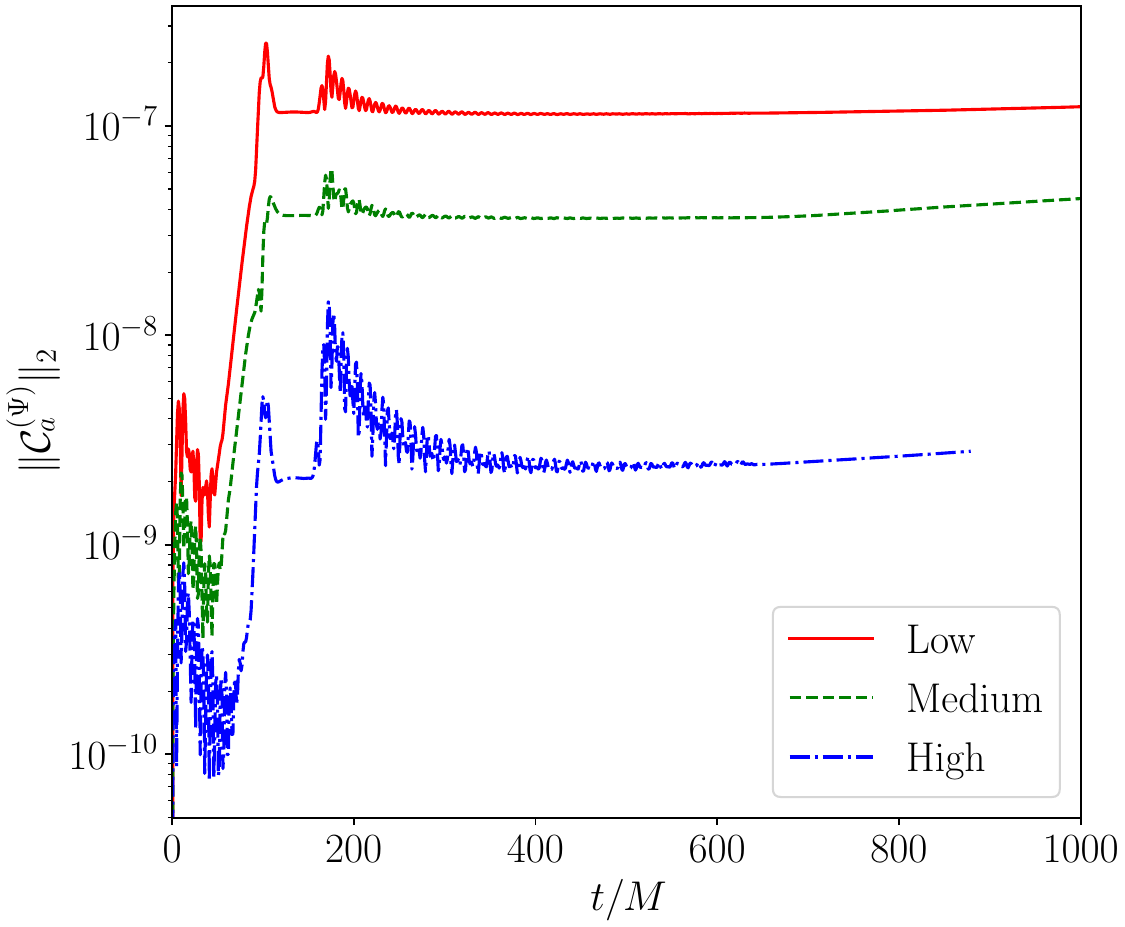}
\caption{ \emph{Convergence test for the one-index constraint \(\mathcal{C}^{(\Psi)}_a\).}
The \(L_2\)-norm , in the components and over the spatial gridpoints, of the \(\mathcal{C}^{(\Psi)}_a\) is plotted for different resolutions. Here \(p_\mathrm{mid} = p_\mathrm{low} + \Delta p \) and \(p_\mathrm{high} = p_\mathrm{mid} + \Delta p = p_\mathrm{low} + 2 \Delta p\), for each DG element  and direction, with constant increase \(\Delta p = 1\).
}
\label{fig: One index constraint convergence adv comov}
\end{figure}

Since we are using pseudo-spectral methods to evolve the equations of motion, we expect exponential convergence with increasing number \(p\) of basis functions used (\(p\)-refinement).
In Figs.~\ref{fig: Constraint energy convergence adv comov} and \ref{fig: One index constraint convergence adv comov}, we show the convergence of the constraint energy and one-index constraint for \(\Psi\).

%%%%%%%%%%%%%%%%%%%%%%%%%%%%%%%%%%%%%%%%%%%%%%%%%%%%%%%%%%%%
%%%%%%%%%%%%%%%%%%%%%%%%%%%%%%%%%%%%%%%%%%%%%%%%%%%%%%%%%%%%
%%%%%%%%%%%%%%%%%%%%%%%%%%%%%%%%%%%%%%%%%%%%%%%%%%%%%%%%%%%%

\FloatBarrier

\bibliography{bibliography}

\begin{thebibliography}{100}%
\makeatletter
\providecommand \@ifxundefined [1]{%
 \@ifx{#1\undefined}
}%
\providecommand \@ifnum [1]{%
 \ifnum #1\expandafter \@firstoftwo
 \else \expandafter \@secondoftwo
 \fi
}%
\providecommand \@ifx [1]{%
 \ifx #1\expandafter \@firstoftwo
 \else \expandafter \@secondoftwo
 \fi
}%
\providecommand \natexlab [1]{#1}%
\providecommand \enquote  [1]{``#1''}%
\providecommand \bibnamefont  [1]{#1}%
\providecommand \bibfnamefont [1]{#1}%
\providecommand \citenamefont [1]{#1}%
\providecommand \href@noop [0]{\@secondoftwo}%
\providecommand \href [0]{\begingroup \@sanitize@url \@href}%
\providecommand \@href[1]{\@@startlink{#1}\@@href}%
\providecommand \@@href[1]{\endgroup#1\@@endlink}%
\providecommand \@sanitize@url [0]{\catcode `\\12\catcode `\$12\catcode
  `\&12\catcode `\#12\catcode `\^12\catcode `\_12\catcode `\%12\relax}%
\providecommand \@@startlink[1]{}%
\providecommand \@@endlink[0]{}%
\providecommand \url  [0]{\begingroup\@sanitize@url \@url }%
\providecommand \@url [1]{\endgroup\@href {#1}{\urlprefix }}%
\providecommand \urlprefix  [0]{URL }%
\providecommand \Eprint [0]{\href }%
\providecommand \doibase [0]{https://doi.org/}%
\providecommand \selectlanguage [0]{\@gobble}%
\providecommand \bibinfo  [0]{\@secondoftwo}%
\providecommand \bibfield  [0]{\@secondoftwo}%
\providecommand \translation [1]{[#1]}%
\providecommand \BibitemOpen [0]{}%
\providecommand \bibitemStop [0]{}%
\providecommand \bibitemNoStop [0]{.\EOS\space}%
\providecommand \EOS [0]{\spacefactor3000\relax}%
\providecommand \BibitemShut  [1]{\csname bibitem#1\endcsname}%
\let\auto@bib@innerbib\@empty
%</preamble>
\bibitem [{\citenamefont {Aasi}\ \emph {et~al.}(2015)\citenamefont {Aasi} \emph
  {et~al.}}]{LIGOScientific:2014pky}%
  \BibitemOpen
  \bibfield  {author} {\bibinfo {author} {\bibfnamefont {J.}~\bibnamefont
  {Aasi}} \emph {et~al.} (\bibinfo {collaboration} {LIGO Scientific}),\
  }\bibfield  {title} {\bibinfo {title} {{Advanced LIGO}},\ }\href
  {https://doi.org/10.1088/0264-9381/32/7/074001} {\bibfield  {journal}
  {\bibinfo  {journal} {Class. Quant. Grav.}\ }\textbf {\bibinfo {volume}
  {32}},\ \bibinfo {pages} {074001} (\bibinfo {year} {2015})},\ \Eprint
  {https://arxiv.org/abs/1411.4547} {arXiv:1411.4547 [gr-qc]} \BibitemShut
  {NoStop}%
\bibitem [{\citenamefont {Acernese}\ \emph {et~al.}(2015)\citenamefont
  {Acernese} \emph {et~al.}}]{VIRGO:2014yos}%
  \BibitemOpen
  \bibfield  {author} {\bibinfo {author} {\bibfnamefont {F.}~\bibnamefont
  {Acernese}} \emph {et~al.} (\bibinfo {collaboration} {VIRGO}),\ }\bibfield
  {title} {\bibinfo {title} {{Advanced Virgo: a second-generation
  interferometric gravitational wave detector}},\ }\href
  {https://doi.org/10.1088/0264-9381/32/2/024001} {\bibfield  {journal}
  {\bibinfo  {journal} {Class. Quant. Grav.}\ }\textbf {\bibinfo {volume}
  {32}},\ \bibinfo {pages} {024001} (\bibinfo {year} {2015})},\ \Eprint
  {https://arxiv.org/abs/1408.3978} {arXiv:1408.3978 [gr-qc]} \BibitemShut
  {NoStop}%
\bibitem [{\citenamefont {Akutsu}\ \emph {et~al.}(2021)\citenamefont {Akutsu}
  \emph {et~al.}}]{KAGRA:2020tym}%
  \BibitemOpen
  \bibfield  {author} {\bibinfo {author} {\bibfnamefont {T.}~\bibnamefont
  {Akutsu}} \emph {et~al.} (\bibinfo {collaboration} {KAGRA}),\ }\bibfield
  {title} {\bibinfo {title} {{Overview of KAGRA: Detector design and
  construction history}},\ }\href {https://doi.org/10.1093/ptep/ptaa125}
  {\bibfield  {journal} {\bibinfo  {journal} {PTEP}\ }\textbf {\bibinfo
  {volume} {2021}},\ \bibinfo {pages} {05A101} (\bibinfo {year} {2021})},\
  \Eprint {https://arxiv.org/abs/2005.05574} {arXiv:2005.05574
  [physics.ins-det]} \BibitemShut {NoStop}%
\bibitem [{\citenamefont {Amaro-Seoane}\ \emph {et~al.}(2017)\citenamefont
  {Amaro-Seoane} \emph {et~al.}}]{LISA:2017pwj}%
  \BibitemOpen
  \bibfield  {author} {\bibinfo {author} {\bibfnamefont {P.}~\bibnamefont
  {Amaro-Seoane}} \emph {et~al.} (\bibinfo {collaboration} {LISA}),\ }\bibfield
   {title} {\bibinfo {title} {{Laser Interferometer Space Antenna}},\
  }\href@noop {} {\  (\bibinfo {year} {2017})},\ \Eprint
  {https://arxiv.org/abs/1702.00786} {arXiv:1702.00786 [astro-ph.IM]}
  \BibitemShut {NoStop}%
\bibitem [{\citenamefont {Punturo}\ \emph {et~al.}(2010)\citenamefont {Punturo}
  \emph {et~al.}}]{Punturo:2010zz}%
  \BibitemOpen
  \bibfield  {author} {\bibinfo {author} {\bibfnamefont {M.}~\bibnamefont
  {Punturo}} \emph {et~al.},\ }\bibfield  {title} {\bibinfo {title} {{The
  Einstein Telescope: A third-generation gravitational wave observatory}},\
  }\href {https://doi.org/10.1088/0264-9381/27/19/194002} {\bibfield  {journal}
  {\bibinfo  {journal} {Class. Quant. Grav.}\ }\textbf {\bibinfo {volume}
  {27}},\ \bibinfo {pages} {194002} (\bibinfo {year} {2010})}\BibitemShut
  {NoStop}%
\bibitem [{\citenamefont {Reitze}\ \emph {et~al.}(2019)\citenamefont {Reitze}
  \emph {et~al.}}]{Reitze:2019iox}%
  \BibitemOpen
  \bibfield  {author} {\bibinfo {author} {\bibfnamefont {D.}~\bibnamefont
  {Reitze}} \emph {et~al.},\ }\bibfield  {title} {\bibinfo {title} {{Cosmic
  Explorer: The U.S. Contribution to Gravitational-Wave Astronomy beyond
  LIGO}},\ }\href@noop {} {\bibfield  {journal} {\bibinfo  {journal} {Bull. Am.
  Astron. Soc.}\ }\textbf {\bibinfo {volume} {51}},\ \bibinfo {pages} {035}
  (\bibinfo {year} {2019})},\ \Eprint {https://arxiv.org/abs/1907.04833}
  {arXiv:1907.04833 [astro-ph.IM]} \BibitemShut {NoStop}%
\bibitem [{\citenamefont {Will}(2018)}]{Will:2018bme}%
  \BibitemOpen
  \bibfield  {author} {\bibinfo {author} {\bibfnamefont {C.~M.}\ \bibnamefont
  {Will}},\ }\href@noop {} {\emph {\bibinfo {title} {{Theory and Experiment in
  Gravitational Physics}}}}\ (\bibinfo  {publisher} {Cambridge University
  Press},\ \bibinfo {year} {2018})\BibitemShut {NoStop}%
\bibitem [{\citenamefont {Abbott}\ \emph {et~al.}(2019)\citenamefont {Abbott}
  \emph {et~al.}}]{LIGOScientific:2019fpa}%
  \BibitemOpen
  \bibfield  {author} {\bibinfo {author} {\bibfnamefont {B.~P.}\ \bibnamefont
  {Abbott}} \emph {et~al.} (\bibinfo {collaboration} {LIGO Scientific,
  Virgo}),\ }\bibfield  {title} {\bibinfo {title} {{Tests of General Relativity
  with the Binary Black Hole Signals from the LIGO-Virgo Catalog GWTC-1}},\
  }\href {https://doi.org/10.1103/PhysRevD.100.104036} {\bibfield  {journal}
  {\bibinfo  {journal} {Phys. Rev.}\ }\textbf {\bibinfo {volume} {D100}},\
  \bibinfo {pages} {104036} (\bibinfo {year} {2019})},\ \Eprint
  {https://arxiv.org/abs/1903.04467} {arXiv:1903.04467 [gr-qc]} \BibitemShut
  {NoStop}%
%%CITATION = ARXIV:1903.04467;%%
\bibitem [{\citenamefont {Abbott}\ \emph
  {et~al.}(2021{\natexlab{a}})\citenamefont {Abbott} \emph
  {et~al.}}]{LIGOScientific:2020tif}%
  \BibitemOpen
  \bibfield  {author} {\bibinfo {author} {\bibfnamefont {R.}~\bibnamefont
  {Abbott}} \emph {et~al.} (\bibinfo {collaboration} {LIGO Scientific,
  Virgo}),\ }\bibfield  {title} {\bibinfo {title} {{Tests of general relativity
  with binary black holes from the second LIGO-Virgo gravitational-wave
  transient catalog}},\ }\href {https://doi.org/10.1103/PhysRevD.103.122002}
  {\bibfield  {journal} {\bibinfo  {journal} {Phys. Rev. D}\ }\textbf {\bibinfo
  {volume} {103}},\ \bibinfo {pages} {122002} (\bibinfo {year}
  {2021}{\natexlab{a}})},\ \Eprint {https://arxiv.org/abs/2010.14529}
  {arXiv:2010.14529 [gr-qc]} \BibitemShut {NoStop}%
\bibitem [{\citenamefont {Abbott}\ \emph
  {et~al.}(2021{\natexlab{b}})\citenamefont {Abbott} \emph
  {et~al.}}]{LIGOScientific:2021sio}%
  \BibitemOpen
  \bibfield  {author} {\bibinfo {author} {\bibfnamefont {R.}~\bibnamefont
  {Abbott}} \emph {et~al.} (\bibinfo {collaboration} {LIGO Scientific, VIRGO,
  KAGRA}),\ }\bibfield  {title} {\bibinfo {title} {{Tests of General Relativity
  with GWTC-3}},\ }\href@noop {} {\  (\bibinfo {year} {2021}{\natexlab{b}})},\
  \Eprint {https://arxiv.org/abs/2112.06861} {arXiv:2112.06861 [gr-qc]}
  \BibitemShut {NoStop}%
\bibitem [{\citenamefont {Kocherlakota}\ \emph {et~al.}(2021)\citenamefont
  {Kocherlakota} \emph {et~al.}}]{EventHorizonTelescope:2021dqv}%
  \BibitemOpen
  \bibfield  {author} {\bibinfo {author} {\bibfnamefont {P.}~\bibnamefont
  {Kocherlakota}} \emph {et~al.} (\bibinfo {collaboration} {Event Horizon
  Telescope}),\ }\bibfield  {title} {\bibinfo {title} {{Constraints on
  black-hole charges with the 2017 EHT observations of M87*}},\ }\href
  {https://doi.org/10.1103/PhysRevD.103.104047} {\bibfield  {journal} {\bibinfo
   {journal} {Phys. Rev. D}\ }\textbf {\bibinfo {volume} {103}},\ \bibinfo
  {pages} {104047} (\bibinfo {year} {2021})},\ \Eprint
  {https://arxiv.org/abs/2105.09343} {arXiv:2105.09343 [gr-qc]} \BibitemShut
  {NoStop}%
\bibitem [{\citenamefont {Akiyama}\ \emph {et~al.}(2022)\citenamefont {Akiyama}
  \emph {et~al.}}]{EventHorizonTelescope:2022xqj}%
  \BibitemOpen
  \bibfield  {author} {\bibinfo {author} {\bibfnamefont {K.}~\bibnamefont
  {Akiyama}} \emph {et~al.} (\bibinfo {collaboration} {Event Horizon
  Telescope}),\ }\bibfield  {title} {\bibinfo {title} {{First Sagittarius A*
  Event Horizon Telescope Results. VI. Testing the Black Hole Metric}},\ }\href
  {https://doi.org/10.3847/2041-8213/ac6756} {\bibfield  {journal} {\bibinfo
  {journal} {Astrophys. J. Lett.}\ }\textbf {\bibinfo {volume} {930}},\
  \bibinfo {pages} {L17} (\bibinfo {year} {2022})},\ \Eprint
  {https://arxiv.org/abs/2311.09484} {arXiv:2311.09484 [astro-ph.HE]}
  \BibitemShut {NoStop}%
\bibitem [{\citenamefont {Crisostomi}\ and\ \citenamefont
  {Koyama}(2018)}]{Crisostomi:2017pjs}%
  \BibitemOpen
  \bibfield  {author} {\bibinfo {author} {\bibfnamefont {M.}~\bibnamefont
  {Crisostomi}}\ and\ \bibinfo {author} {\bibfnamefont {K.}~\bibnamefont
  {Koyama}},\ }\bibfield  {title} {\bibinfo {title} {{Self-accelerating
  universe in scalar-tensor theories after GW170817}},\ }\href
  {https://doi.org/10.1103/PhysRevD.97.084004} {\bibfield  {journal} {\bibinfo
  {journal} {Phys. Rev.}\ }\textbf {\bibinfo {volume} {D97}},\ \bibinfo {pages}
  {084004} (\bibinfo {year} {2018})},\ \Eprint
  {https://arxiv.org/abs/1712.06556} {arXiv:1712.06556 [astro-ph.CO]}
  \BibitemShut {NoStop}%
%%CITATION = ARXIV:1712.06556;%%
\bibitem [{\citenamefont {Langlois}\ and\ \citenamefont
  {Noui}(2016)}]{Langlois:2015cwa}%
  \BibitemOpen
  \bibfield  {author} {\bibinfo {author} {\bibfnamefont {D.}~\bibnamefont
  {Langlois}}\ and\ \bibinfo {author} {\bibfnamefont {K.}~\bibnamefont
  {Noui}},\ }\bibfield  {title} {\bibinfo {title} {{Degenerate higher
  derivative theories beyond Horndeski: evading the Ostrogradski
  instability}},\ }\href {https://doi.org/10.1088/1475-7516/2016/02/034}
  {\bibfield  {journal} {\bibinfo  {journal} {JCAP}\ }\textbf {\bibinfo
  {volume} {02}},\ \bibinfo {pages} {034}},\ \Eprint
  {https://arxiv.org/abs/1510.06930} {arXiv:1510.06930 [gr-qc]} \BibitemShut
  {NoStop}%
\bibitem [{\citenamefont {Crisostomi}\ \emph {et~al.}(2016)\citenamefont
  {Crisostomi}, \citenamefont {Koyama},\ and\ \citenamefont
  {Tasinato}}]{Crisostomi:2016czh}%
  \BibitemOpen
  \bibfield  {author} {\bibinfo {author} {\bibfnamefont {M.}~\bibnamefont
  {Crisostomi}}, \bibinfo {author} {\bibfnamefont {K.}~\bibnamefont {Koyama}},\
  and\ \bibinfo {author} {\bibfnamefont {G.}~\bibnamefont {Tasinato}},\
  }\bibfield  {title} {\bibinfo {title} {{Extended Scalar-Tensor Theories of
  Gravity}},\ }\href {https://doi.org/10.1088/1475-7516/2016/04/044} {\bibfield
   {journal} {\bibinfo  {journal} {JCAP}\ }\textbf {\bibinfo {volume} {1604}},\
  \bibinfo {pages} {044}},\ \Eprint {https://arxiv.org/abs/1602.03119}
  {arXiv:1602.03119 [hep-th]} \BibitemShut {NoStop}%
%%CITATION = ARXIV:1602.03119;%%
\bibitem [{\citenamefont {Ben~Achour}\ \emph {et~al.}(2016)\citenamefont
  {Ben~Achour}, \citenamefont {Crisostomi}, \citenamefont {Koyama},
  \citenamefont {Langlois}, \citenamefont {Noui},\ and\ \citenamefont
  {Tasinato}}]{BenAchour:2016fzp}%
  \BibitemOpen
  \bibfield  {author} {\bibinfo {author} {\bibfnamefont {J.}~\bibnamefont
  {Ben~Achour}}, \bibinfo {author} {\bibfnamefont {M.}~\bibnamefont
  {Crisostomi}}, \bibinfo {author} {\bibfnamefont {K.}~\bibnamefont {Koyama}},
  \bibinfo {author} {\bibfnamefont {D.}~\bibnamefont {Langlois}}, \bibinfo
  {author} {\bibfnamefont {K.}~\bibnamefont {Noui}},\ and\ \bibinfo {author}
  {\bibfnamefont {G.}~\bibnamefont {Tasinato}},\ }\bibfield  {title} {\bibinfo
  {title} {{Degenerate higher order scalar-tensor theories beyond Horndeski up
  to cubic order}},\ }\href {https://doi.org/10.1007/JHEP12(2016)100}
  {\bibfield  {journal} {\bibinfo  {journal} {JHEP}\ }\textbf {\bibinfo
  {volume} {12}},\ \bibinfo {pages} {100}},\ \Eprint
  {https://arxiv.org/abs/1608.08135} {arXiv:1608.08135 [hep-th]} \BibitemShut
  {NoStop}%
\bibitem [{\citenamefont {Horndeski}(1974)}]{Horndeski:1974wa}%
  \BibitemOpen
  \bibfield  {author} {\bibinfo {author} {\bibfnamefont {G.~W.}\ \bibnamefont
  {Horndeski}},\ }\bibfield  {title} {\bibinfo {title} {{Second-order
  scalar-tensor field equations in a four-dimensional space}},\ }\href
  {https://doi.org/10.1007/BF01807638} {\bibfield  {journal} {\bibinfo
  {journal} {Int. J. Theor. Phys.}\ }\textbf {\bibinfo {volume} {10}},\
  \bibinfo {pages} {363} (\bibinfo {year} {1974})}\BibitemShut {NoStop}%
\bibitem [{\citenamefont {Endlich}\ \emph {et~al.}(2017)\citenamefont
  {Endlich}, \citenamefont {Gorbenko}, \citenamefont {Huang},\ and\
  \citenamefont {Senatore}}]{Endlich:2017tqa}%
  \BibitemOpen
  \bibfield  {author} {\bibinfo {author} {\bibfnamefont {S.}~\bibnamefont
  {Endlich}}, \bibinfo {author} {\bibfnamefont {V.}~\bibnamefont {Gorbenko}},
  \bibinfo {author} {\bibfnamefont {J.}~\bibnamefont {Huang}},\ and\ \bibinfo
  {author} {\bibfnamefont {L.}~\bibnamefont {Senatore}},\ }\bibfield  {title}
  {\bibinfo {title} {{An effective formalism for testing extensions to General
  Relativity with gravitational waves}},\ }\href
  {https://doi.org/10.1007/JHEP09(2017)122} {\bibfield  {journal} {\bibinfo
  {journal} {JHEP}\ }\textbf {\bibinfo {volume} {09}},\ \bibinfo {pages}
  {122}},\ \Eprint {https://arxiv.org/abs/1704.01590} {arXiv:1704.01590
  [gr-qc]} \BibitemShut {NoStop}%
\bibitem [{\citenamefont {Weinberg}(2008)}]{Weinberg:2008hq}%
  \BibitemOpen
  \bibfield  {author} {\bibinfo {author} {\bibfnamefont {S.}~\bibnamefont
  {Weinberg}},\ }\bibfield  {title} {\bibinfo {title} {{Effective Field Theory
  for Inflation}},\ }\href {https://doi.org/10.1103/PhysRevD.77.123541}
  {\bibfield  {journal} {\bibinfo  {journal} {Phys. Rev. D}\ }\textbf {\bibinfo
  {volume} {77}},\ \bibinfo {pages} {123541} (\bibinfo {year} {2008})},\
  \Eprint {https://arxiv.org/abs/0804.4291} {arXiv:0804.4291 [hep-th]}
  \BibitemShut {NoStop}%
\bibitem [{\citenamefont {Abbott}\ \emph
  {et~al.}(2017{\natexlab{a}})\citenamefont {Abbott} \emph
  {et~al.}}]{Monitor:2017mdv}%
  \BibitemOpen
  \bibfield  {author} {\bibinfo {author} {\bibfnamefont {B.}~\bibnamefont
  {Abbott}} \emph {et~al.} (\bibinfo {collaboration} {LIGO Scientific, Virgo,
  Fermi-GBM, INTEGRAL}),\ }\bibfield  {title} {\bibinfo {title} {{Gravitational
  Waves and Gamma-rays from a Binary Neutron Star Merger: GW170817 and GRB
  170817A}},\ }\href {https://doi.org/10.3847/2041-8213/aa920c} {\bibfield
  {journal} {\bibinfo  {journal} {Astrophys. J. Lett.}\ }\textbf {\bibinfo
  {volume} {848}},\ \bibinfo {pages} {L13} (\bibinfo {year}
  {2017}{\natexlab{a}})},\ \Eprint {https://arxiv.org/abs/1710.05834}
  {arXiv:1710.05834 [astro-ph.HE]} \BibitemShut {NoStop}%
\bibitem [{\citenamefont {Abbott}\ \emph
  {et~al.}(2017{\natexlab{b}})\citenamefont {Abbott} \emph
  {et~al.}}]{TheLIGOScientific:2017qsa}%
  \BibitemOpen
  \bibfield  {author} {\bibinfo {author} {\bibfnamefont {B.}~\bibnamefont
  {Abbott}} \emph {et~al.} (\bibinfo {collaboration} {LIGO Scientific,
  Virgo}),\ }\bibfield  {title} {\bibinfo {title} {{GW170817: Observation of
  Gravitational Waves from a Binary Neutron Star Inspiral}},\ }\href
  {https://doi.org/10.1103/PhysRevLett.119.161101} {\bibfield  {journal}
  {\bibinfo  {journal} {Phys. Rev. Lett.}\ }\textbf {\bibinfo {volume} {119}},\
  \bibinfo {pages} {161101} (\bibinfo {year} {2017}{\natexlab{b}})},\ \Eprint
  {https://arxiv.org/abs/1710.05832} {arXiv:1710.05832 [gr-qc]} \BibitemShut
  {NoStop}%
\bibitem [{\citenamefont {Creminelli}\ and\ \citenamefont
  {Vernizzi}(2017)}]{Creminelli:2017sry}%
  \BibitemOpen
  \bibfield  {author} {\bibinfo {author} {\bibfnamefont {P.}~\bibnamefont
  {Creminelli}}\ and\ \bibinfo {author} {\bibfnamefont {F.}~\bibnamefont
  {Vernizzi}},\ }\bibfield  {title} {\bibinfo {title} {{Dark Energy after
  GW170817 and GRB170817A}},\ }\href
  {https://doi.org/10.1103/PhysRevLett.119.251302} {\bibfield  {journal}
  {\bibinfo  {journal} {Phys. Rev. Lett.}\ }\textbf {\bibinfo {volume} {119}},\
  \bibinfo {pages} {251302} (\bibinfo {year} {2017})},\ \Eprint
  {https://arxiv.org/abs/1710.05877} {arXiv:1710.05877 [astro-ph.CO]}
  \BibitemShut {NoStop}%
\bibitem [{\citenamefont {Ezquiaga}\ and\ \citenamefont
  {Zumalac\'arregui}(2017)}]{Ezquiaga:2017ekz}%
  \BibitemOpen
  \bibfield  {author} {\bibinfo {author} {\bibfnamefont {J.~M.}\ \bibnamefont
  {Ezquiaga}}\ and\ \bibinfo {author} {\bibfnamefont {M.}~\bibnamefont
  {Zumalac\'arregui}},\ }\bibfield  {title} {\bibinfo {title} {{Dark Energy
  After GW170817: Dead Ends and the Road Ahead}},\ }\href
  {https://doi.org/10.1103/PhysRevLett.119.251304} {\bibfield  {journal}
  {\bibinfo  {journal} {Phys. Rev. Lett.}\ }\textbf {\bibinfo {volume} {119}},\
  \bibinfo {pages} {251304} (\bibinfo {year} {2017})},\ \Eprint
  {https://arxiv.org/abs/1710.05901} {arXiv:1710.05901 [astro-ph.CO]}
  \BibitemShut {NoStop}%
\bibitem [{\citenamefont {Baker}\ \emph {et~al.}(2017)\citenamefont {Baker},
  \citenamefont {Bellini}, \citenamefont {Ferreira}, \citenamefont {Lagos},
  \citenamefont {Noller},\ and\ \citenamefont {Sawicki}}]{Baker:2017hug}%
  \BibitemOpen
  \bibfield  {author} {\bibinfo {author} {\bibfnamefont {T.}~\bibnamefont
  {Baker}}, \bibinfo {author} {\bibfnamefont {E.}~\bibnamefont {Bellini}},
  \bibinfo {author} {\bibfnamefont {P.~G.}\ \bibnamefont {Ferreira}}, \bibinfo
  {author} {\bibfnamefont {M.}~\bibnamefont {Lagos}}, \bibinfo {author}
  {\bibfnamefont {J.}~\bibnamefont {Noller}},\ and\ \bibinfo {author}
  {\bibfnamefont {I.}~\bibnamefont {Sawicki}},\ }\bibfield  {title} {\bibinfo
  {title} {{Strong constraints on cosmological gravity from GW170817 and GRB
  170817A}},\ }\href {https://doi.org/10.1103/PhysRevLett.119.251301}
  {\bibfield  {journal} {\bibinfo  {journal} {Phys. Rev. Lett.}\ }\textbf
  {\bibinfo {volume} {119}},\ \bibinfo {pages} {251301} (\bibinfo {year}
  {2017})},\ \Eprint {https://arxiv.org/abs/1710.06394} {arXiv:1710.06394
  [astro-ph.CO]} \BibitemShut {NoStop}%
\bibitem [{\citenamefont {Sakstein}\ and\ \citenamefont
  {Jain}(2017)}]{Sakstein:2017xjx}%
  \BibitemOpen
  \bibfield  {author} {\bibinfo {author} {\bibfnamefont {J.}~\bibnamefont
  {Sakstein}}\ and\ \bibinfo {author} {\bibfnamefont {B.}~\bibnamefont
  {Jain}},\ }\bibfield  {title} {\bibinfo {title} {{Implications of the Neutron
  Star Merger GW170817 for Cosmological Scalar-Tensor Theories}},\ }\href
  {https://doi.org/10.1103/PhysRevLett.119.251303} {\bibfield  {journal}
  {\bibinfo  {journal} {Phys. Rev. Lett.}\ }\textbf {\bibinfo {volume} {119}},\
  \bibinfo {pages} {251303} (\bibinfo {year} {2017})},\ \Eprint
  {https://arxiv.org/abs/1710.05893} {arXiv:1710.05893 [astro-ph.CO]}
  \BibitemShut {NoStop}%
\bibitem [{\citenamefont {Varma}\ \emph {et~al.}(2019)\citenamefont {Varma},
  \citenamefont {Field}, \citenamefont {Scheel}, \citenamefont {Blackman},
  \citenamefont {Kidder},\ and\ \citenamefont {Pfeiffer}}]{Varma:2018mmi}%
  \BibitemOpen
  \bibfield  {author} {\bibinfo {author} {\bibfnamefont {V.}~\bibnamefont
  {Varma}}, \bibinfo {author} {\bibfnamefont {S.~E.}\ \bibnamefont {Field}},
  \bibinfo {author} {\bibfnamefont {M.~A.}\ \bibnamefont {Scheel}}, \bibinfo
  {author} {\bibfnamefont {J.}~\bibnamefont {Blackman}}, \bibinfo {author}
  {\bibfnamefont {L.~E.}\ \bibnamefont {Kidder}},\ and\ \bibinfo {author}
  {\bibfnamefont {H.~P.}\ \bibnamefont {Pfeiffer}},\ }\bibfield  {title}
  {\bibinfo {title} {{Surrogate model of hybridized numerical relativity binary
  black hole waveforms}},\ }\href {https://doi.org/10.1103/PhysRevD.99.064045}
  {\bibfield  {journal} {\bibinfo  {journal} {Phys. Rev. D}\ }\textbf {\bibinfo
  {volume} {99}},\ \bibinfo {pages} {064045} (\bibinfo {year} {2019})},\
  \Eprint {https://arxiv.org/abs/1812.07865} {arXiv:1812.07865 [gr-qc]}
  \BibitemShut {NoStop}%
\bibitem [{\citenamefont {Ossokine}\ \emph {et~al.}(2020)\citenamefont
  {Ossokine} \emph {et~al.}}]{Ossokine:2020kjp}%
  \BibitemOpen
  \bibfield  {author} {\bibinfo {author} {\bibfnamefont {S.}~\bibnamefont
  {Ossokine}} \emph {et~al.},\ }\bibfield  {title} {\bibinfo {title}
  {{Multipolar Effective-One-Body Waveforms for Precessing Binary Black Holes:
  Construction and Validation}},\ }\href
  {https://doi.org/10.1103/PhysRevD.102.044055} {\bibfield  {journal} {\bibinfo
   {journal} {Phys. Rev. D}\ }\textbf {\bibinfo {volume} {102}},\ \bibinfo
  {pages} {044055} (\bibinfo {year} {2020})},\ \Eprint
  {https://arxiv.org/abs/2004.09442} {arXiv:2004.09442 [gr-qc]} \BibitemShut
  {NoStop}%
\bibitem [{\citenamefont {Pratten}\ \emph {et~al.}(2021)\citenamefont {Pratten}
  \emph {et~al.}}]{Pratten:2020ceb}%
  \BibitemOpen
  \bibfield  {author} {\bibinfo {author} {\bibfnamefont {G.}~\bibnamefont
  {Pratten}} \emph {et~al.},\ }\bibfield  {title} {\bibinfo {title}
  {{Computationally efficient models for the dominant and subdominant harmonic
  modes of precessing binary black holes}},\ }\href
  {https://doi.org/10.1103/PhysRevD.103.104056} {\bibfield  {journal} {\bibinfo
   {journal} {Phys. Rev. D}\ }\textbf {\bibinfo {volume} {103}},\ \bibinfo
  {pages} {104056} (\bibinfo {year} {2021})},\ \Eprint
  {https://arxiv.org/abs/2004.06503} {arXiv:2004.06503 [gr-qc]} \BibitemShut
  {NoStop}%
\bibitem [{\citenamefont {Gamba}\ \emph {et~al.}(2022)\citenamefont {Gamba},
  \citenamefont {Ak\c{c}ay}, \citenamefont {Bernuzzi},\ and\ \citenamefont
  {Williams}}]{Gamba:2021ydi}%
  \BibitemOpen
  \bibfield  {author} {\bibinfo {author} {\bibfnamefont {R.}~\bibnamefont
  {Gamba}}, \bibinfo {author} {\bibfnamefont {S.}~\bibnamefont {Ak\c{c}ay}},
  \bibinfo {author} {\bibfnamefont {S.}~\bibnamefont {Bernuzzi}},\ and\
  \bibinfo {author} {\bibfnamefont {J.}~\bibnamefont {Williams}},\ }\bibfield
  {title} {\bibinfo {title} {{Effective-one-body waveforms for precessing
  coalescing compact binaries with post-Newtonian twist}},\ }\href
  {https://doi.org/10.1103/PhysRevD.106.024020} {\bibfield  {journal} {\bibinfo
   {journal} {Phys. Rev. D}\ }\textbf {\bibinfo {volume} {106}},\ \bibinfo
  {pages} {024020} (\bibinfo {year} {2022})},\ \Eprint
  {https://arxiv.org/abs/2111.03675} {arXiv:2111.03675 [gr-qc]} \BibitemShut
  {NoStop}%
\bibitem [{\citenamefont {Baumgarte}\ and\ \citenamefont
  {Shapiro}(2010)}]{Baumgarte:2010ndz}%
  \BibitemOpen
  \bibfield  {author} {\bibinfo {author} {\bibfnamefont {T.~W.}\ \bibnamefont
  {Baumgarte}}\ and\ \bibinfo {author} {\bibfnamefont {S.~L.}\ \bibnamefont
  {Shapiro}},\ }\href {https://doi.org/10.1017/CBO9781139193344} {\emph
  {\bibinfo {title} {{Numerical Relativity: Solving Einstein's Equations on the
  Computer}}}}\ (\bibinfo  {publisher} {Cambridge University Press},\ \bibinfo
  {year} {2010})\BibitemShut {NoStop}%
\bibitem [{\citenamefont {Hadamard}(1902)}]{Hadamard10030321135}%
  \BibitemOpen
  \bibfield  {author} {\bibinfo {author} {\bibfnamefont {J.}~\bibnamefont
  {Hadamard}},\ }\bibfield  {title} {\bibinfo {title} {Sur les problemes aux
  derivees partielles et leur signification physique},\ }\href
  {https://ci.nii.ac.jp/naid/10030321135/en/} {\bibfield  {journal} {\bibinfo
  {journal} {Princeton university bulletin}\ ,\ \bibinfo {pages} {49}}
  (\bibinfo {year} {1902})}\BibitemShut {NoStop}%
\bibitem [{\citenamefont {East}\ and\ \citenamefont
  {Ripley}(2021{\natexlab{a}})}]{East:2021bqk}%
  \BibitemOpen
  \bibfield  {author} {\bibinfo {author} {\bibfnamefont {W.~E.}\ \bibnamefont
  {East}}\ and\ \bibinfo {author} {\bibfnamefont {J.~L.}\ \bibnamefont
  {Ripley}},\ }\bibfield  {title} {\bibinfo {title} {{Dynamics of Spontaneous
  Black Hole Scalarization and Mergers in Einstein-Scalar-Gauss-Bonnet
  Gravity}},\ }\href {https://doi.org/10.1103/PhysRevLett.127.101102}
  {\bibfield  {journal} {\bibinfo  {journal} {Phys. Rev. Lett.}\ }\textbf
  {\bibinfo {volume} {127}},\ \bibinfo {pages} {101102} (\bibinfo {year}
  {2021}{\natexlab{a}})},\ \Eprint {https://arxiv.org/abs/2105.08571}
  {arXiv:2105.08571 [gr-qc]} \BibitemShut {NoStop}%
\bibitem [{\citenamefont {Bernard}\ \emph {et~al.}(2019)\citenamefont
  {Bernard}, \citenamefont {Lehner},\ and\ \citenamefont
  {Luna}}]{Bernard:2019fjb}%
  \BibitemOpen
  \bibfield  {author} {\bibinfo {author} {\bibfnamefont {L.}~\bibnamefont
  {Bernard}}, \bibinfo {author} {\bibfnamefont {L.}~\bibnamefont {Lehner}},\
  and\ \bibinfo {author} {\bibfnamefont {R.}~\bibnamefont {Luna}},\ }\bibfield
  {title} {\bibinfo {title} {{Challenges to global solutions in Horndeski's
  theory}},\ }\href {https://doi.org/10.1103/PhysRevD.100.024011} {\bibfield
  {journal} {\bibinfo  {journal} {Phys. Rev. D}\ }\textbf {\bibinfo {volume}
  {100}},\ \bibinfo {pages} {024011} (\bibinfo {year} {2019})},\ \Eprint
  {https://arxiv.org/abs/1904.12866} {arXiv:1904.12866 [gr-qc]} \BibitemShut
  {NoStop}%
\bibitem [{\citenamefont {Ripley}\ and\ \citenamefont
  {Pretorius}(2019)}]{Ripley:2019irj}%
  \BibitemOpen
  \bibfield  {author} {\bibinfo {author} {\bibfnamefont {J.~L.}\ \bibnamefont
  {Ripley}}\ and\ \bibinfo {author} {\bibfnamefont {F.}~\bibnamefont
  {Pretorius}},\ }\bibfield  {title} {\bibinfo {title} {{Gravitational collapse
  in Einstein dilaton-Gauss\textendash{}Bonnet gravity}},\ }\href
  {https://doi.org/10.1088/1361-6382/ab2416} {\bibfield  {journal} {\bibinfo
  {journal} {Class. Quant. Grav.}\ }\textbf {\bibinfo {volume} {36}},\ \bibinfo
  {pages} {134001} (\bibinfo {year} {2019})},\ \Eprint
  {https://arxiv.org/abs/1903.07543} {arXiv:1903.07543 [gr-qc]} \BibitemShut
  {NoStop}%
\bibitem [{\citenamefont {Ripley}\ and\ \citenamefont
  {Pretorius}(2020{\natexlab{a}})}]{Ripley:2019aqj}%
  \BibitemOpen
  \bibfield  {author} {\bibinfo {author} {\bibfnamefont {J.~L.}\ \bibnamefont
  {Ripley}}\ and\ \bibinfo {author} {\bibfnamefont {F.}~\bibnamefont
  {Pretorius}},\ }\bibfield  {title} {\bibinfo {title} {{Scalarized Black Hole
  dynamics in Einstein dilaton Gauss-Bonnet Gravity}},\ }\href
  {https://doi.org/10.1103/PhysRevD.101.044015} {\bibfield  {journal} {\bibinfo
   {journal} {Phys. Rev. D}\ }\textbf {\bibinfo {volume} {101}},\ \bibinfo
  {pages} {044015} (\bibinfo {year} {2020}{\natexlab{a}})},\ \Eprint
  {https://arxiv.org/abs/1911.11027} {arXiv:1911.11027 [gr-qc]} \BibitemShut
  {NoStop}%
\bibitem [{\citenamefont {Ripley}\ and\ \citenamefont
  {Pretorius}(2020{\natexlab{b}})}]{Ripley:2020vpk}%
  \BibitemOpen
  \bibfield  {author} {\bibinfo {author} {\bibfnamefont {J.~L.}\ \bibnamefont
  {Ripley}}\ and\ \bibinfo {author} {\bibfnamefont {F.}~\bibnamefont
  {Pretorius}},\ }\bibfield  {title} {\bibinfo {title} {{Dynamics of a $\mathbb
  Z_2$ symmetric EdGB gravity in spherical symmetry}},\ }\href
  {https://doi.org/10.1088/1361-6382/ab9bbb} {\bibfield  {journal} {\bibinfo
  {journal} {Class. Quant. Grav.}\ }\textbf {\bibinfo {volume} {37}},\ \bibinfo
  {pages} {155003} (\bibinfo {year} {2020}{\natexlab{b}})},\ \Eprint
  {https://arxiv.org/abs/2005.05417} {arXiv:2005.05417 [gr-qc]} \BibitemShut
  {NoStop}%
\bibitem [{\citenamefont {R}\ \emph {et~al.}(2023)\citenamefont {R},
  \citenamefont {Ripley},\ and\ \citenamefont {Yunes}}]{R:2022hlf}%
  \BibitemOpen
  \bibfield  {author} {\bibinfo {author} {\bibfnamefont {A.~H.~K.}\
  \bibnamefont {R}}, \bibinfo {author} {\bibfnamefont {J.~L.}\ \bibnamefont
  {Ripley}},\ and\ \bibinfo {author} {\bibfnamefont {N.}~\bibnamefont
  {Yunes}},\ }\bibfield  {title} {\bibinfo {title} {{Where and why does
  Einstein-scalar-Gauss-Bonnet theory break down?}},\ }\href
  {https://doi.org/10.1103/PhysRevD.107.044044} {\bibfield  {journal} {\bibinfo
   {journal} {Phys. Rev. D}\ }\textbf {\bibinfo {volume} {107}},\ \bibinfo
  {pages} {044044} (\bibinfo {year} {2023})},\ \Eprint
  {https://arxiv.org/abs/2211.08477} {arXiv:2211.08477 [gr-qc]} \BibitemShut
  {NoStop}%
\bibitem [{\citenamefont {Corelli}\ \emph
  {et~al.}(2023{\natexlab{a}})\citenamefont {Corelli}, \citenamefont
  {De~Amicis}, \citenamefont {Ikeda},\ and\ \citenamefont
  {Pani}}]{Corelli:2022pio}%
  \BibitemOpen
  \bibfield  {author} {\bibinfo {author} {\bibfnamefont {F.}~\bibnamefont
  {Corelli}}, \bibinfo {author} {\bibfnamefont {M.}~\bibnamefont {De~Amicis}},
  \bibinfo {author} {\bibfnamefont {T.}~\bibnamefont {Ikeda}},\ and\ \bibinfo
  {author} {\bibfnamefont {P.}~\bibnamefont {Pani}},\ }\bibfield  {title}
  {\bibinfo {title} {{What is the Fate of Hawking Evaporation in Gravity
  Theories with Higher Curvature Terms?}},\ }\href
  {https://doi.org/10.1103/PhysRevLett.130.091501} {\bibfield  {journal}
  {\bibinfo  {journal} {Phys. Rev. Lett.}\ }\textbf {\bibinfo {volume} {130}},\
  \bibinfo {pages} {091501} (\bibinfo {year} {2023}{\natexlab{a}})},\ \Eprint
  {https://arxiv.org/abs/2205.13006} {arXiv:2205.13006 [gr-qc]} \BibitemShut
  {NoStop}%
\bibitem [{\citenamefont {Corelli}\ \emph
  {et~al.}(2023{\natexlab{b}})\citenamefont {Corelli}, \citenamefont
  {De~Amicis}, \citenamefont {Ikeda},\ and\ \citenamefont
  {Pani}}]{Corelli:2022phw}%
  \BibitemOpen
  \bibfield  {author} {\bibinfo {author} {\bibfnamefont {F.}~\bibnamefont
  {Corelli}}, \bibinfo {author} {\bibfnamefont {M.}~\bibnamefont {De~Amicis}},
  \bibinfo {author} {\bibfnamefont {T.}~\bibnamefont {Ikeda}},\ and\ \bibinfo
  {author} {\bibfnamefont {P.}~\bibnamefont {Pani}},\ }\bibfield  {title}
  {\bibinfo {title} {{Nonperturbative gedanken experiments in
  Einstein-dilaton-Gauss-Bonnet gravity: Nonlinear transitions and tests of the
  cosmic censorship beyond general relativity}},\ }\href
  {https://doi.org/10.1103/PhysRevD.107.044061} {\bibfield  {journal} {\bibinfo
   {journal} {Phys. Rev. D}\ }\textbf {\bibinfo {volume} {107}},\ \bibinfo
  {pages} {044061} (\bibinfo {year} {2023}{\natexlab{b}})},\ \Eprint
  {https://arxiv.org/abs/2205.13007} {arXiv:2205.13007 [gr-qc]} \BibitemShut
  {NoStop}%
\bibitem [{\citenamefont {Thaalba}\ \emph {et~al.}(2023)\citenamefont
  {Thaalba}, \citenamefont {Bezares}, \citenamefont {Franchini},\ and\
  \citenamefont {Sotiriou}}]{Thaalba:2023fmq}%
  \BibitemOpen
  \bibfield  {author} {\bibinfo {author} {\bibfnamefont {F.}~\bibnamefont
  {Thaalba}}, \bibinfo {author} {\bibfnamefont {M.}~\bibnamefont {Bezares}},
  \bibinfo {author} {\bibfnamefont {N.}~\bibnamefont {Franchini}},\ and\
  \bibinfo {author} {\bibfnamefont {T.~P.}\ \bibnamefont {Sotiriou}},\
  }\bibfield  {title} {\bibinfo {title} {{Spherical collapse in
  scalar-Gauss-Bonnet gravity: taming ill-posedness with a Ricci coupling}},\
  }\href@noop {} {\  (\bibinfo {year} {2023})},\ \Eprint
  {https://arxiv.org/abs/2306.01695} {arXiv:2306.01695 [gr-qc]} \BibitemShut
  {NoStop}%
\bibitem [{\citenamefont {Ripley}(2022)}]{Ripley:2022cdh}%
  \BibitemOpen
  \bibfield  {author} {\bibinfo {author} {\bibfnamefont {J.~L.}\ \bibnamefont
  {Ripley}},\ }\bibfield  {title} {\bibinfo {title} {{Numerical relativity for
  Horndeski gravity}},\ }\href@noop {} {\  (\bibinfo {year} {2022})},\ \Eprint
  {https://arxiv.org/abs/2207.13074} {arXiv:2207.13074 [gr-qc]} \BibitemShut
  {NoStop}%
\bibitem [{\citenamefont {Okounkova}\ \emph {et~al.}(2017)\citenamefont
  {Okounkova}, \citenamefont {Stein}, \citenamefont {Scheel},\ and\
  \citenamefont {Hemberger}}]{Okounkova:2017yby}%
  \BibitemOpen
  \bibfield  {author} {\bibinfo {author} {\bibfnamefont {M.}~\bibnamefont
  {Okounkova}}, \bibinfo {author} {\bibfnamefont {L.~C.}\ \bibnamefont
  {Stein}}, \bibinfo {author} {\bibfnamefont {M.~A.}\ \bibnamefont {Scheel}},\
  and\ \bibinfo {author} {\bibfnamefont {D.~A.}\ \bibnamefont {Hemberger}},\
  }\bibfield  {title} {\bibinfo {title} {{Numerical binary black hole mergers
  in dynamical Chern-Simons gravity: Scalar field}},\ }\href
  {https://doi.org/10.1103/PhysRevD.96.044020} {\bibfield  {journal} {\bibinfo
  {journal} {Phys. Rev. D}\ }\textbf {\bibinfo {volume} {96}},\ \bibinfo
  {pages} {044020} (\bibinfo {year} {2017})},\ \Eprint
  {https://arxiv.org/abs/1705.07924} {arXiv:1705.07924 [gr-qc]} \BibitemShut
  {NoStop}%
\bibitem [{\citenamefont {Okounkova}(2019)}]{Okounkova:2019zep}%
  \BibitemOpen
  \bibfield  {author} {\bibinfo {author} {\bibfnamefont {M.}~\bibnamefont
  {Okounkova}},\ }\bibfield  {title} {\bibinfo {title} {{Stability of Rotating
  Black Holes in Einstein Dilaton Gauss-Bonnet Gravity}},\ }\href
  {https://doi.org/10.1103/PhysRevD.100.124054} {\bibfield  {journal} {\bibinfo
   {journal} {Phys. Rev. D}\ }\textbf {\bibinfo {volume} {100}},\ \bibinfo
  {pages} {124054} (\bibinfo {year} {2019})},\ \Eprint
  {https://arxiv.org/abs/1909.12251} {arXiv:1909.12251 [gr-qc]} \BibitemShut
  {NoStop}%
\bibitem [{\citenamefont {Okounkova}\ \emph {et~al.}(2020)\citenamefont
  {Okounkova}, \citenamefont {Stein}, \citenamefont {Moxon}, \citenamefont
  {Scheel},\ and\ \citenamefont {Teukolsky}}]{Okounkova:2019zjf}%
  \BibitemOpen
  \bibfield  {author} {\bibinfo {author} {\bibfnamefont {M.}~\bibnamefont
  {Okounkova}}, \bibinfo {author} {\bibfnamefont {L.~C.}\ \bibnamefont
  {Stein}}, \bibinfo {author} {\bibfnamefont {J.}~\bibnamefont {Moxon}},
  \bibinfo {author} {\bibfnamefont {M.~A.}\ \bibnamefont {Scheel}},\ and\
  \bibinfo {author} {\bibfnamefont {S.~A.}\ \bibnamefont {Teukolsky}},\
  }\bibfield  {title} {\bibinfo {title} {{Numerical relativity simulation of
  GW150914 beyond general relativity}},\ }\href
  {https://doi.org/10.1103/PhysRevD.101.104016} {\bibfield  {journal} {\bibinfo
   {journal} {Phys. Rev. D}\ }\textbf {\bibinfo {volume} {101}},\ \bibinfo
  {pages} {104016} (\bibinfo {year} {2020})},\ \Eprint
  {https://arxiv.org/abs/1911.02588} {arXiv:1911.02588 [gr-qc]} \BibitemShut
  {NoStop}%
\bibitem [{\citenamefont {Okounkova}(2020)}]{Okounkova:2020rqw}%
  \BibitemOpen
  \bibfield  {author} {\bibinfo {author} {\bibfnamefont {M.}~\bibnamefont
  {Okounkova}},\ }\bibfield  {title} {\bibinfo {title} {{Numerical relativity
  simulation of GW150914 in Einstein dilaton Gauss-Bonnet gravity}},\ }\href
  {https://doi.org/10.1103/PhysRevD.102.084046} {\bibfield  {journal} {\bibinfo
   {journal} {Phys. Rev. D}\ }\textbf {\bibinfo {volume} {102}},\ \bibinfo
  {pages} {084046} (\bibinfo {year} {2020})},\ \Eprint
  {https://arxiv.org/abs/2001.03571} {arXiv:2001.03571 [gr-qc]} \BibitemShut
  {NoStop}%
\bibitem [{\citenamefont {Witek}\ \emph {et~al.}(2019)\citenamefont {Witek},
  \citenamefont {Gualtieri}, \citenamefont {Pani},\ and\ \citenamefont
  {Sotiriou}}]{Witek:2018dmd}%
  \BibitemOpen
  \bibfield  {author} {\bibinfo {author} {\bibfnamefont {H.}~\bibnamefont
  {Witek}}, \bibinfo {author} {\bibfnamefont {L.}~\bibnamefont {Gualtieri}},
  \bibinfo {author} {\bibfnamefont {P.}~\bibnamefont {Pani}},\ and\ \bibinfo
  {author} {\bibfnamefont {T.~P.}\ \bibnamefont {Sotiriou}},\ }\bibfield
  {title} {\bibinfo {title} {{Black holes and binary mergers in scalar
  Gauss-Bonnet gravity: scalar field dynamics}},\ }\href
  {https://doi.org/10.1103/PhysRevD.99.064035} {\bibfield  {journal} {\bibinfo
  {journal} {Phys. Rev. D}\ }\textbf {\bibinfo {volume} {99}},\ \bibinfo
  {pages} {064035} (\bibinfo {year} {2019})},\ \Eprint
  {https://arxiv.org/abs/1810.05177} {arXiv:1810.05177 [gr-qc]} \BibitemShut
  {NoStop}%
\bibitem [{\citenamefont {Silva}\ \emph {et~al.}(2021)\citenamefont {Silva},
  \citenamefont {Witek}, \citenamefont {Elley},\ and\ \citenamefont
  {Yunes}}]{Silva:2020omi}%
  \BibitemOpen
  \bibfield  {author} {\bibinfo {author} {\bibfnamefont {H.~O.}\ \bibnamefont
  {Silva}}, \bibinfo {author} {\bibfnamefont {H.}~\bibnamefont {Witek}},
  \bibinfo {author} {\bibfnamefont {M.}~\bibnamefont {Elley}},\ and\ \bibinfo
  {author} {\bibfnamefont {N.}~\bibnamefont {Yunes}},\ }\bibfield  {title}
  {\bibinfo {title} {{Dynamical Descalarization in Binary Black Hole
  Mergers}},\ }\href {https://doi.org/10.1103/PhysRevLett.127.031101}
  {\bibfield  {journal} {\bibinfo  {journal} {Phys. Rev. Lett.}\ }\textbf
  {\bibinfo {volume} {127}},\ \bibinfo {pages} {031101} (\bibinfo {year}
  {2021})},\ \Eprint {https://arxiv.org/abs/2012.10436} {arXiv:2012.10436
  [gr-qc]} \BibitemShut {NoStop}%
\bibitem [{\citenamefont {Elley}\ \emph {et~al.}(2022)\citenamefont {Elley},
  \citenamefont {Silva}, \citenamefont {Witek},\ and\ \citenamefont
  {Yunes}}]{Elley:2022ept}%
  \BibitemOpen
  \bibfield  {author} {\bibinfo {author} {\bibfnamefont {M.}~\bibnamefont
  {Elley}}, \bibinfo {author} {\bibfnamefont {H.~O.}\ \bibnamefont {Silva}},
  \bibinfo {author} {\bibfnamefont {H.}~\bibnamefont {Witek}},\ and\ \bibinfo
  {author} {\bibfnamefont {N.}~\bibnamefont {Yunes}},\ }\bibfield  {title}
  {\bibinfo {title} {{Spin-induced dynamical scalarization, descalarization,
  and stealthness in scalar-Gauss-Bonnet gravity during a black hole
  coalescence}},\ }\href {https://doi.org/10.1103/PhysRevD.106.044018}
  {\bibfield  {journal} {\bibinfo  {journal} {Phys. Rev. D}\ }\textbf {\bibinfo
  {volume} {106}},\ \bibinfo {pages} {044018} (\bibinfo {year} {2022})},\
  \Eprint {https://arxiv.org/abs/2205.06240} {arXiv:2205.06240 [gr-qc]}
  \BibitemShut {NoStop}%
\bibitem [{\citenamefont {Figueras}\ and\ \citenamefont
  {Fran\c{c}a}(2020)}]{Figueras:2020dzx}%
  \BibitemOpen
  \bibfield  {author} {\bibinfo {author} {\bibfnamefont {P.}~\bibnamefont
  {Figueras}}\ and\ \bibinfo {author} {\bibfnamefont {T.}~\bibnamefont
  {Fran\c{c}a}},\ }\bibfield  {title} {\bibinfo {title} {{Gravitational
  Collapse in Cubic Horndeski Theories}},\ }\href
  {https://doi.org/10.1088/1361-6382/abb693} {\bibfield  {journal} {\bibinfo
  {journal} {Class. Quant. Grav.}\ }\textbf {\bibinfo {volume} {37}},\ \bibinfo
  {pages} {225009} (\bibinfo {year} {2020})},\ \Eprint
  {https://arxiv.org/abs/2006.09414} {arXiv:2006.09414 [gr-qc]} \BibitemShut
  {NoStop}%
\bibitem [{\citenamefont {Figueras}\ and\ \citenamefont
  {Fran\c{c}a}(2021)}]{Figueras:2021abd}%
  \BibitemOpen
  \bibfield  {author} {\bibinfo {author} {\bibfnamefont {P.}~\bibnamefont
  {Figueras}}\ and\ \bibinfo {author} {\bibfnamefont {T.}~\bibnamefont
  {Fran\c{c}a}},\ }\bibfield  {title} {\bibinfo {title} {{Black Hole Binaries
  in Cubic Horndeski Theories}},\ }\href@noop {} {\  (\bibinfo {year}
  {2021})},\ \Eprint {https://arxiv.org/abs/2112.15529} {arXiv:2112.15529
  [gr-qc]} \BibitemShut {NoStop}%
\bibitem [{\citenamefont {Bezares}\ \emph
  {et~al.}(2021{\natexlab{a}})\citenamefont {Bezares}, \citenamefont
  {Crisostomi}, \citenamefont {Palenzuela},\ and\ \citenamefont
  {Barausse}}]{Bezares:2020wkn}%
  \BibitemOpen
  \bibfield  {author} {\bibinfo {author} {\bibfnamefont {M.}~\bibnamefont
  {Bezares}}, \bibinfo {author} {\bibfnamefont {M.}~\bibnamefont {Crisostomi}},
  \bibinfo {author} {\bibfnamefont {C.}~\bibnamefont {Palenzuela}},\ and\
  \bibinfo {author} {\bibfnamefont {E.}~\bibnamefont {Barausse}},\ }\bibfield
  {title} {\bibinfo {title} {{K-dynamics: well-posed 1+1 evolutions in
  K-essence}},\ }\href {https://doi.org/10.1088/1475-7516/2021/03/072}
  {\bibfield  {journal} {\bibinfo  {journal} {JCAP}\ }\textbf {\bibinfo
  {volume} {2103}},\ \bibinfo {pages} {072}},\ \Eprint
  {https://arxiv.org/abs/2008.07546} {arXiv:2008.07546 [gr-qc]} \BibitemShut
  {NoStop}%
%%CITATION = ARXIV:2008.07546;%%
\bibitem [{\citenamefont {Bezares}\ \emph
  {et~al.}(2021{\natexlab{b}})\citenamefont {Bezares}, \citenamefont {ter
  Haar}, \citenamefont {Crisostomi}, \citenamefont {Barausse},\ and\
  \citenamefont {Palenzuela}}]{Bezares:2021yek}%
  \BibitemOpen
  \bibfield  {author} {\bibinfo {author} {\bibfnamefont {M.}~\bibnamefont
  {Bezares}}, \bibinfo {author} {\bibfnamefont {L.}~\bibnamefont {ter Haar}},
  \bibinfo {author} {\bibfnamefont {M.}~\bibnamefont {Crisostomi}}, \bibinfo
  {author} {\bibfnamefont {E.}~\bibnamefont {Barausse}},\ and\ \bibinfo
  {author} {\bibfnamefont {C.}~\bibnamefont {Palenzuela}},\ }\bibfield  {title}
  {\bibinfo {title} {{Kinetic screening in nonlinear stellar oscillations and
  gravitational collapse}},\ }\href
  {https://doi.org/10.1103/PhysRevD.104.044022} {\bibfield  {journal} {\bibinfo
   {journal} {Phys. Rev. D}\ }\textbf {\bibinfo {volume} {104}},\ \bibinfo
  {pages} {044022} (\bibinfo {year} {2021}{\natexlab{b}})},\ \Eprint
  {https://arxiv.org/abs/2105.13992} {arXiv:2105.13992 [gr-qc]} \BibitemShut
  {NoStop}%
\bibitem [{\citenamefont {Bezares}\ \emph {et~al.}(2022)\citenamefont
  {Bezares}, \citenamefont {Aguilera-Miret}, \citenamefont {ter Haar},
  \citenamefont {Crisostomi}, \citenamefont {Palenzuela},\ and\ \citenamefont
  {Barausse}}]{Bezares:2021dma}%
  \BibitemOpen
  \bibfield  {author} {\bibinfo {author} {\bibfnamefont {M.}~\bibnamefont
  {Bezares}}, \bibinfo {author} {\bibfnamefont {R.}~\bibnamefont
  {Aguilera-Miret}}, \bibinfo {author} {\bibfnamefont {L.}~\bibnamefont {ter
  Haar}}, \bibinfo {author} {\bibfnamefont {M.}~\bibnamefont {Crisostomi}},
  \bibinfo {author} {\bibfnamefont {C.}~\bibnamefont {Palenzuela}},\ and\
  \bibinfo {author} {\bibfnamefont {E.}~\bibnamefont {Barausse}},\ }\bibfield
  {title} {\bibinfo {title} {{No Evidence of Kinetic Screening in Simulations
  of Merging Binary Neutron Stars beyond General Relativity}},\ }\href
  {https://doi.org/10.1103/PhysRevLett.128.091103} {\bibfield  {journal}
  {\bibinfo  {journal} {Phys. Rev. Lett.}\ }\textbf {\bibinfo {volume} {128}},\
  \bibinfo {pages} {091103} (\bibinfo {year} {2022})},\ \Eprint
  {https://arxiv.org/abs/2107.05648} {arXiv:2107.05648 [gr-qc]} \BibitemShut
  {NoStop}%
\bibitem [{\citenamefont {Held}\ and\ \citenamefont
  {Lim}(2021)}]{Held:2021pht}%
  \BibitemOpen
  \bibfield  {author} {\bibinfo {author} {\bibfnamefont {A.}~\bibnamefont
  {Held}}\ and\ \bibinfo {author} {\bibfnamefont {H.}~\bibnamefont {Lim}},\
  }\bibfield  {title} {\bibinfo {title} {{Nonlinear dynamics of quadratic
  gravity in spherical symmetry}},\ }\href
  {https://doi.org/10.1103/PhysRevD.104.084075} {\bibfield  {journal} {\bibinfo
   {journal} {Phys. Rev. D}\ }\textbf {\bibinfo {volume} {104}},\ \bibinfo
  {pages} {084075} (\bibinfo {year} {2021})},\ \Eprint
  {https://arxiv.org/abs/2104.04010} {arXiv:2104.04010 [gr-qc]} \BibitemShut
  {NoStop}%
\bibitem [{\citenamefont {Held}\ and\ \citenamefont
  {Lim}(2023)}]{Held:2023aap}%
  \BibitemOpen
  \bibfield  {author} {\bibinfo {author} {\bibfnamefont {A.}~\bibnamefont
  {Held}}\ and\ \bibinfo {author} {\bibfnamefont {H.}~\bibnamefont {Lim}},\
  }\bibfield  {title} {\bibinfo {title} {{Nonlinear evolution of quadratic
  gravity in 3+1 dimensions}},\ }\href
  {https://doi.org/10.1103/PhysRevD.108.104025} {\bibfield  {journal} {\bibinfo
   {journal} {Phys. Rev. D}\ }\textbf {\bibinfo {volume} {108}},\ \bibinfo
  {pages} {104025} (\bibinfo {year} {2023})},\ \Eprint
  {https://arxiv.org/abs/2306.04725} {arXiv:2306.04725 [gr-qc]} \BibitemShut
  {NoStop}%
\bibitem [{\citenamefont {Rubio}\ \emph {et~al.}(2023)\citenamefont {Rubio},
  \citenamefont {Kov\'acs}, \citenamefont {Herrero-Valea}, \citenamefont
  {Bezares},\ and\ \citenamefont {Barausse}}]{Rubio:2023eva}%
  \BibitemOpen
  \bibfield  {author} {\bibinfo {author} {\bibfnamefont {M.~E.}\ \bibnamefont
  {Rubio}}, \bibinfo {author} {\bibfnamefont {A.~D.}\ \bibnamefont {Kov\'acs}},
  \bibinfo {author} {\bibfnamefont {M.}~\bibnamefont {Herrero-Valea}}, \bibinfo
  {author} {\bibfnamefont {M.}~\bibnamefont {Bezares}},\ and\ \bibinfo {author}
  {\bibfnamefont {E.}~\bibnamefont {Barausse}},\ }\bibfield  {title} {\bibinfo
  {title} {{Well-posed evolution of field theories with anisotropic scaling:
  the Lifshitz scalar field in a black hole space-time}},\ }\href
  {https://doi.org/10.1088/1475-7516/2023/11/001} {\bibfield  {journal}
  {\bibinfo  {journal} {JCAP}\ }\textbf {\bibinfo {volume} {11}},\ \bibinfo
  {pages} {001}},\ \Eprint {https://arxiv.org/abs/2307.13041} {arXiv:2307.13041
  [gr-qc]} \BibitemShut {NoStop}%
\bibitem [{\citenamefont {Kov\'acs}\ and\ \citenamefont
  {Reall}(2020{\natexlab{a}})}]{Kovacs:2020PRL}%
  \BibitemOpen
  \bibfield  {author} {\bibinfo {author} {\bibfnamefont {A.~D.}\ \bibnamefont
  {Kov\'acs}}\ and\ \bibinfo {author} {\bibfnamefont {H.~S.}\ \bibnamefont
  {Reall}},\ }\bibfield  {title} {\bibinfo {title} {{Well-Posed Formulation of
  Scalar-Tensor Effective Field Theory}},\ }\href
  {https://doi.org/10.1103/PhysRevLett.124.221101} {\bibfield  {journal}
  {\bibinfo  {journal} {Phys. Rev. Lett.}\ }\textbf {\bibinfo {volume} {124}},\
  \bibinfo {pages} {221101} (\bibinfo {year} {2020}{\natexlab{a}})},\ \Eprint
  {https://arxiv.org/abs/2003.04327} {arXiv:2003.04327 [gr-qc]} \BibitemShut
  {NoStop}%
\bibitem [{\citenamefont {Kov\'acs}\ and\ \citenamefont
  {Reall}(2020{\natexlab{b}})}]{Kovacs:2020ywu}%
  \BibitemOpen
  \bibfield  {author} {\bibinfo {author} {\bibfnamefont {A.~D.}\ \bibnamefont
  {Kov\'acs}}\ and\ \bibinfo {author} {\bibfnamefont {H.~S.}\ \bibnamefont
  {Reall}},\ }\bibfield  {title} {\bibinfo {title} {{Well-posed formulation of
  Lovelock and Horndeski theories}},\ }\href
  {https://doi.org/10.1103/PhysRevD.101.124003} {\bibfield  {journal} {\bibinfo
   {journal} {Phys. Rev. D}\ }\textbf {\bibinfo {volume} {101}},\ \bibinfo
  {pages} {124003} (\bibinfo {year} {2020}{\natexlab{b}})},\ \Eprint
  {https://arxiv.org/abs/2003.08398} {arXiv:2003.08398 [gr-qc]} \BibitemShut
  {NoStop}%
\bibitem [{\citenamefont {East}\ and\ \citenamefont
  {Ripley}(2021{\natexlab{b}})}]{East:2020hgw}%
  \BibitemOpen
  \bibfield  {author} {\bibinfo {author} {\bibfnamefont {W.~E.}\ \bibnamefont
  {East}}\ and\ \bibinfo {author} {\bibfnamefont {J.~L.}\ \bibnamefont
  {Ripley}},\ }\bibfield  {title} {\bibinfo {title} {{Evolution of
  Einstein-scalar-Gauss-Bonnet gravity using a modified harmonic
  formulation}},\ }\href {https://doi.org/10.1103/PhysRevD.103.044040}
  {\bibfield  {journal} {\bibinfo  {journal} {Phys. Rev. D}\ }\textbf {\bibinfo
  {volume} {103}},\ \bibinfo {pages} {044040} (\bibinfo {year}
  {2021}{\natexlab{b}})},\ \Eprint {https://arxiv.org/abs/2011.03547}
  {arXiv:2011.03547 [gr-qc]} \BibitemShut {NoStop}%
\bibitem [{\citenamefont {Corman}\ \emph {et~al.}(2023)\citenamefont {Corman},
  \citenamefont {Ripley},\ and\ \citenamefont {East}}]{Corman:2022xqg}%
  \BibitemOpen
  \bibfield  {author} {\bibinfo {author} {\bibfnamefont {M.}~\bibnamefont
  {Corman}}, \bibinfo {author} {\bibfnamefont {J.~L.}\ \bibnamefont {Ripley}},\
  and\ \bibinfo {author} {\bibfnamefont {W.~E.}\ \bibnamefont {East}},\
  }\bibfield  {title} {\bibinfo {title} {{Nonlinear studies of binary black
  hole mergers in Einstein-scalar-Gauss-Bonnet gravity}},\ }\href
  {https://doi.org/10.1103/PhysRevD.107.024014} {\bibfield  {journal} {\bibinfo
   {journal} {Phys. Rev. D}\ }\textbf {\bibinfo {volume} {107}},\ \bibinfo
  {pages} {024014} (\bibinfo {year} {2023})},\ \Eprint
  {https://arxiv.org/abs/2210.09235} {arXiv:2210.09235 [gr-qc]} \BibitemShut
  {NoStop}%
\bibitem [{\citenamefont {Arest\'e~Sal\'o}\ \emph {et~al.}(2022)\citenamefont
  {Arest\'e~Sal\'o}, \citenamefont {Clough},\ and\ \citenamefont
  {Figueras}}]{AresteSalo:2022hua}%
  \BibitemOpen
  \bibfield  {author} {\bibinfo {author} {\bibfnamefont {L.}~\bibnamefont
  {Arest\'e~Sal\'o}}, \bibinfo {author} {\bibfnamefont {K.}~\bibnamefont
  {Clough}},\ and\ \bibinfo {author} {\bibfnamefont {P.}~\bibnamefont
  {Figueras}},\ }\bibfield  {title} {\bibinfo {title} {{Well-Posedness of the
  Four-Derivative Scalar-Tensor Theory of Gravity in Singularity Avoiding
  Coordinates}},\ }\href {https://doi.org/10.1103/PhysRevLett.129.261104}
  {\bibfield  {journal} {\bibinfo  {journal} {Phys. Rev. Lett.}\ }\textbf
  {\bibinfo {volume} {129}},\ \bibinfo {pages} {261104} (\bibinfo {year}
  {2022})},\ \Eprint {https://arxiv.org/abs/2208.14470} {arXiv:2208.14470
  [gr-qc]} \BibitemShut {NoStop}%
\bibitem [{\citenamefont {Arest\'e~Sal\'o}\ \emph {et~al.}(2023)\citenamefont
  {Arest\'e~Sal\'o}, \citenamefont {Clough},\ and\ \citenamefont
  {Figueras}}]{AresteSalo:2023mmd}%
  \BibitemOpen
  \bibfield  {author} {\bibinfo {author} {\bibfnamefont {L.}~\bibnamefont
  {Arest\'e~Sal\'o}}, \bibinfo {author} {\bibfnamefont {K.}~\bibnamefont
  {Clough}},\ and\ \bibinfo {author} {\bibfnamefont {P.}~\bibnamefont
  {Figueras}},\ }\bibfield  {title} {\bibinfo {title} {{Puncture gauge
  formulation for Einstein-Gauss-Bonnet gravity and four-derivative
  scalar-tensor theories in d+1 spacetime dimensions}},\ }\href
  {https://doi.org/10.1103/PhysRevD.108.084018} {\bibfield  {journal} {\bibinfo
   {journal} {Phys. Rev. D}\ }\textbf {\bibinfo {volume} {108}},\ \bibinfo
  {pages} {084018} (\bibinfo {year} {2023})},\ \Eprint
  {https://arxiv.org/abs/2306.14966} {arXiv:2306.14966 [gr-qc]} \BibitemShut
  {NoStop}%
\bibitem [{\citenamefont {Cayuso}\ \emph {et~al.}(2017)\citenamefont {Cayuso},
  \citenamefont {Ortiz},\ and\ \citenamefont {Lehner}}]{Cayuso:2017iqc}%
  \BibitemOpen
  \bibfield  {author} {\bibinfo {author} {\bibfnamefont {J.}~\bibnamefont
  {Cayuso}}, \bibinfo {author} {\bibfnamefont {N.}~\bibnamefont {Ortiz}},\ and\
  \bibinfo {author} {\bibfnamefont {L.}~\bibnamefont {Lehner}},\ }\bibfield
  {title} {\bibinfo {title} {{Fixing extensions to general relativity in the
  nonlinear regime}},\ }\href {https://doi.org/10.1103/PhysRevD.96.084043}
  {\bibfield  {journal} {\bibinfo  {journal} {Phys. Rev. D}\ }\textbf {\bibinfo
  {volume} {96}},\ \bibinfo {pages} {084043} (\bibinfo {year} {2017})},\
  \Eprint {https://arxiv.org/abs/1706.07421} {arXiv:1706.07421 [gr-qc]}
  \BibitemShut {NoStop}%
\bibitem [{\citenamefont {Deppe}\ \emph {et~al.}(2024)\citenamefont {Deppe},
  \citenamefont {Throwe}, \citenamefont {Kidder}, \citenamefont {Vu},
  \citenamefont {Nelli}, \citenamefont {Armaza}, \citenamefont {Bonilla},
  \citenamefont {Hébert}, \citenamefont {Kim}, \citenamefont {Kumar},
  \citenamefont {Lovelace}, \citenamefont {Macedo}, \citenamefont {Moxon},
  \citenamefont {O'Shea}, \citenamefont {Pfeiffer}, \citenamefont {Scheel},
  \citenamefont {Teukolsky}, \citenamefont {Wittek}, \citenamefont
  {Anantpurkar}, \citenamefont {Anderson}, \citenamefont {Boyle}, \citenamefont
  {Carpenter}, \citenamefont {Ceja}, \citenamefont {Chaudhary}, \citenamefont
  {Corso}, \citenamefont {Foucart}, \citenamefont {Ghadiri}, \citenamefont
  {Giesler}, \citenamefont {Guo}, \citenamefont {Iozzo}, \citenamefont {Jones},
  \citenamefont {Lara}, \citenamefont {Legred}, \citenamefont {Li},
  \citenamefont {Ma}, \citenamefont {Melchor}, \citenamefont {Morales},
  \citenamefont {Most}, \citenamefont {Nee}, \citenamefont {Osorio},
  \citenamefont {Pajkos}, \citenamefont {Pannone}, \citenamefont {Ramirez},
  \citenamefont {Ring}, \citenamefont {Rüter}, \citenamefont {Sanchez},
  \citenamefont {Stein}, \citenamefont {Tellez}, \citenamefont {Thomas},
  \citenamefont {Vieira}, \citenamefont {Wlodarczyk}, \citenamefont {Wu},\ and\
  \citenamefont {Yoo}}]{deppe_2024_10619885}%
  \BibitemOpen
  \bibfield  {author} {\bibinfo {author} {\bibfnamefont {N.}~\bibnamefont
  {Deppe}}, \bibinfo {author} {\bibfnamefont {W.}~\bibnamefont {Throwe}},
  \bibinfo {author} {\bibfnamefont {L.~E.}\ \bibnamefont {Kidder}}, \bibinfo
  {author} {\bibfnamefont {N.~L.}\ \bibnamefont {Vu}}, \bibinfo {author}
  {\bibfnamefont {K.~C.}\ \bibnamefont {Nelli}}, \bibinfo {author}
  {\bibfnamefont {C.}~\bibnamefont {Armaza}}, \bibinfo {author} {\bibfnamefont
  {M.~S.}\ \bibnamefont {Bonilla}}, \bibinfo {author} {\bibfnamefont
  {F.}~\bibnamefont {Hébert}}, \bibinfo {author} {\bibfnamefont
  {Y.}~\bibnamefont {Kim}}, \bibinfo {author} {\bibfnamefont {P.}~\bibnamefont
  {Kumar}}, \bibinfo {author} {\bibfnamefont {G.}~\bibnamefont {Lovelace}},
  \bibinfo {author} {\bibfnamefont {A.}~\bibnamefont {Macedo}}, \bibinfo
  {author} {\bibfnamefont {J.}~\bibnamefont {Moxon}}, \bibinfo {author}
  {\bibfnamefont {E.}~\bibnamefont {O'Shea}}, \bibinfo {author} {\bibfnamefont
  {H.~P.}\ \bibnamefont {Pfeiffer}}, \bibinfo {author} {\bibfnamefont {M.~A.}\
  \bibnamefont {Scheel}}, \bibinfo {author} {\bibfnamefont {S.~A.}\
  \bibnamefont {Teukolsky}}, \bibinfo {author} {\bibfnamefont {N.~A.}\
  \bibnamefont {Wittek}}, \bibinfo {author} {\bibfnamefont {I.}~\bibnamefont
  {Anantpurkar}}, \bibinfo {author} {\bibfnamefont {C.}~\bibnamefont
  {Anderson}}, \bibinfo {author} {\bibfnamefont {M.}~\bibnamefont {Boyle}},
  \bibinfo {author} {\bibfnamefont {A.}~\bibnamefont {Carpenter}}, \bibinfo
  {author} {\bibfnamefont {A.}~\bibnamefont {Ceja}}, \bibinfo {author}
  {\bibfnamefont {H.}~\bibnamefont {Chaudhary}}, \bibinfo {author}
  {\bibfnamefont {N.}~\bibnamefont {Corso}}, \bibinfo {author} {\bibfnamefont
  {F.}~\bibnamefont {Foucart}}, \bibinfo {author} {\bibfnamefont
  {N.}~\bibnamefont {Ghadiri}}, \bibinfo {author} {\bibfnamefont
  {M.}~\bibnamefont {Giesler}}, \bibinfo {author} {\bibfnamefont {J.~S.}\
  \bibnamefont {Guo}}, \bibinfo {author} {\bibfnamefont {D.~A.~B.}\
  \bibnamefont {Iozzo}}, \bibinfo {author} {\bibfnamefont {K.~Z.}\ \bibnamefont
  {Jones}}, \bibinfo {author} {\bibfnamefont {G.}~\bibnamefont {Lara}},
  \bibinfo {author} {\bibfnamefont {I.}~\bibnamefont {Legred}}, \bibinfo
  {author} {\bibfnamefont {D.}~\bibnamefont {Li}}, \bibinfo {author}
  {\bibfnamefont {S.}~\bibnamefont {Ma}}, \bibinfo {author} {\bibfnamefont
  {D.}~\bibnamefont {Melchor}}, \bibinfo {author} {\bibfnamefont
  {M.}~\bibnamefont {Morales}}, \bibinfo {author} {\bibfnamefont {E.~R.}\
  \bibnamefont {Most}}, \bibinfo {author} {\bibfnamefont {P.~J.}\ \bibnamefont
  {Nee}}, \bibinfo {author} {\bibfnamefont {A.}~\bibnamefont {Osorio}},
  \bibinfo {author} {\bibfnamefont {M.~A.}\ \bibnamefont {Pajkos}}, \bibinfo
  {author} {\bibfnamefont {K.}~\bibnamefont {Pannone}}, \bibinfo {author}
  {\bibfnamefont {T.}~\bibnamefont {Ramirez}}, \bibinfo {author} {\bibfnamefont
  {N.}~\bibnamefont {Ring}}, \bibinfo {author} {\bibfnamefont {H.~R.}\
  \bibnamefont {Rüter}}, \bibinfo {author} {\bibfnamefont {J.}~\bibnamefont
  {Sanchez}}, \bibinfo {author} {\bibfnamefont {L.~C.}\ \bibnamefont {Stein}},
  \bibinfo {author} {\bibfnamefont {D.}~\bibnamefont {Tellez}}, \bibinfo
  {author} {\bibfnamefont {S.}~\bibnamefont {Thomas}}, \bibinfo {author}
  {\bibfnamefont {D.}~\bibnamefont {Vieira}}, \bibinfo {author} {\bibfnamefont
  {T.}~\bibnamefont {Wlodarczyk}}, \bibinfo {author} {\bibfnamefont
  {D.}~\bibnamefont {Wu}},\ and\ \bibinfo {author} {\bibfnamefont
  {J.}~\bibnamefont {Yoo}},\ }\href {https://doi.org/10.5281/zenodo.10619885}
  {\bibinfo {title} {Spectre}} (\bibinfo {year} {2024})\BibitemShut {NoStop}%
\bibitem [{\citenamefont {Allwright}\ and\ \citenamefont
  {Lehner}(2019)}]{Allwright:2018rut}%
  \BibitemOpen
  \bibfield  {author} {\bibinfo {author} {\bibfnamefont {G.}~\bibnamefont
  {Allwright}}\ and\ \bibinfo {author} {\bibfnamefont {L.}~\bibnamefont
  {Lehner}},\ }\bibfield  {title} {\bibinfo {title} {{Towards the nonlinear
  regime in extensions to GR: assessing possible options}},\ }\href
  {https://doi.org/10.1088/1361-6382/ab0ee1} {\bibfield  {journal} {\bibinfo
  {journal} {Class. Quant. Grav.}\ }\textbf {\bibinfo {volume} {36}},\ \bibinfo
  {pages} {084001} (\bibinfo {year} {2019})},\ \Eprint
  {https://arxiv.org/abs/1808.07897} {arXiv:1808.07897 [gr-qc]} \BibitemShut
  {NoStop}%
\bibitem [{\citenamefont {Cayuso}\ and\ \citenamefont
  {Lehner}(2020)}]{Cayuso:2020lca}%
  \BibitemOpen
  \bibfield  {author} {\bibinfo {author} {\bibfnamefont {R.}~\bibnamefont
  {Cayuso}}\ and\ \bibinfo {author} {\bibfnamefont {L.}~\bibnamefont
  {Lehner}},\ }\bibfield  {title} {\bibinfo {title} {{Nonlinear, noniterative
  treatment of EFT-motivated gravity}},\ }\href
  {https://doi.org/10.1103/PhysRevD.102.084008} {\bibfield  {journal} {\bibinfo
   {journal} {Phys. Rev. D}\ }\textbf {\bibinfo {volume} {102}},\ \bibinfo
  {pages} {084008} (\bibinfo {year} {2020})},\ \Eprint
  {https://arxiv.org/abs/2005.13720} {arXiv:2005.13720 [gr-qc]} \BibitemShut
  {NoStop}%
\bibitem [{\citenamefont {Lara}\ \emph {et~al.}(2022)\citenamefont {Lara},
  \citenamefont {Bezares},\ and\ \citenamefont {Barausse}}]{Lara:2021piy}%
  \BibitemOpen
  \bibfield  {author} {\bibinfo {author} {\bibfnamefont {G.}~\bibnamefont
  {Lara}}, \bibinfo {author} {\bibfnamefont {M.}~\bibnamefont {Bezares}},\ and\
  \bibinfo {author} {\bibfnamefont {E.}~\bibnamefont {Barausse}},\ }\bibfield
  {title} {\bibinfo {title} {{UV completions, fixing the equations, and
  nonlinearities in k-essence}},\ }\href
  {https://doi.org/10.1103/PhysRevD.105.064058} {\bibfield  {journal} {\bibinfo
   {journal} {Phys. Rev. D}\ }\textbf {\bibinfo {volume} {105}},\ \bibinfo
  {pages} {064058} (\bibinfo {year} {2022})},\ \Eprint
  {https://arxiv.org/abs/2112.09186} {arXiv:2112.09186 [gr-qc]} \BibitemShut
  {NoStop}%
\bibitem [{\citenamefont {Franchini}\ \emph {et~al.}(2022)\citenamefont
  {Franchini}, \citenamefont {Bezares}, \citenamefont {Barausse},\ and\
  \citenamefont {Lehner}}]{Franchini:2022ukz}%
  \BibitemOpen
  \bibfield  {author} {\bibinfo {author} {\bibfnamefont {N.}~\bibnamefont
  {Franchini}}, \bibinfo {author} {\bibfnamefont {M.}~\bibnamefont {Bezares}},
  \bibinfo {author} {\bibfnamefont {E.}~\bibnamefont {Barausse}},\ and\
  \bibinfo {author} {\bibfnamefont {L.}~\bibnamefont {Lehner}},\ }\bibfield
  {title} {\bibinfo {title} {{Fixing the dynamical evolution in
  scalar-Gauss-Bonnet gravity}},\ }\href
  {https://doi.org/10.1103/PhysRevD.106.064061} {\bibfield  {journal} {\bibinfo
   {journal} {Phys. Rev. D}\ }\textbf {\bibinfo {volume} {106}},\ \bibinfo
  {pages} {064061} (\bibinfo {year} {2022})},\ \Eprint
  {https://arxiv.org/abs/2206.00014} {arXiv:2206.00014 [gr-qc]} \BibitemShut
  {NoStop}%
\bibitem [{\citenamefont {Cayuso}\ \emph {et~al.}(2023)\citenamefont {Cayuso},
  \citenamefont {Figueras}, \citenamefont {Fran\c{c}a},\ and\ \citenamefont
  {Lehner}}]{Cayuso:2023aht}%
  \BibitemOpen
  \bibfield  {author} {\bibinfo {author} {\bibfnamefont {R.}~\bibnamefont
  {Cayuso}}, \bibinfo {author} {\bibfnamefont {P.}~\bibnamefont {Figueras}},
  \bibinfo {author} {\bibfnamefont {T.}~\bibnamefont {Fran\c{c}a}},\ and\
  \bibinfo {author} {\bibfnamefont {L.}~\bibnamefont {Lehner}},\ }\bibfield
  {title} {\bibinfo {title} {{Modelling self-consistently beyond General
  Relativity}},\ }\href@noop {} {\  (\bibinfo {year} {2023})},\ \Eprint
  {https://arxiv.org/abs/2303.07246} {arXiv:2303.07246 [gr-qc]} \BibitemShut
  {NoStop}%
\bibitem [{\citenamefont {Coates}\ and\ \citenamefont
  {Ramazano\u{g}lu}(2023)}]{Coates:2023swo}%
  \BibitemOpen
  \bibfield  {author} {\bibinfo {author} {\bibfnamefont {A.}~\bibnamefont
  {Coates}}\ and\ \bibinfo {author} {\bibfnamefont {F.~M.}\ \bibnamefont
  {Ramazano\u{g}lu}},\ }\bibfield  {title} {\bibinfo {title} {{Treatments and
  placebos for the pathologies of effective field theories}},\ }\href
  {https://doi.org/10.1103/PhysRevD.108.L101501} {\bibfield  {journal}
  {\bibinfo  {journal} {Phys. Rev. D}\ }\textbf {\bibinfo {volume} {108}},\
  \bibinfo {pages} {L101501} (\bibinfo {year} {2023})},\ \Eprint
  {https://arxiv.org/abs/2307.07743} {arXiv:2307.07743 [gr-qc]} \BibitemShut
  {NoStop}%
\bibitem [{\citenamefont {Sotiriou}\ and\ \citenamefont
  {Zhou}(2014{\natexlab{a}})}]{Sotiriou:2013qea}%
  \BibitemOpen
  \bibfield  {author} {\bibinfo {author} {\bibfnamefont {T.~P.}\ \bibnamefont
  {Sotiriou}}\ and\ \bibinfo {author} {\bibfnamefont {S.-Y.}\ \bibnamefont
  {Zhou}},\ }\bibfield  {title} {\bibinfo {title} {{Black hole hair in
  generalized scalar-tensor gravity}},\ }\href
  {https://doi.org/10.1103/PhysRevLett.112.251102} {\bibfield  {journal}
  {\bibinfo  {journal} {Phys. Rev. Lett.}\ }\textbf {\bibinfo {volume} {112}},\
  \bibinfo {pages} {251102} (\bibinfo {year} {2014}{\natexlab{a}})},\ \Eprint
  {https://arxiv.org/abs/1312.3622} {arXiv:1312.3622 [gr-qc]} \BibitemShut
  {NoStop}%
\bibitem [{\citenamefont {Sotiriou}\ and\ \citenamefont
  {Zhou}(2014{\natexlab{b}})}]{Sotiriou:2014pfa}%
  \BibitemOpen
  \bibfield  {author} {\bibinfo {author} {\bibfnamefont {T.~P.}\ \bibnamefont
  {Sotiriou}}\ and\ \bibinfo {author} {\bibfnamefont {S.-Y.}\ \bibnamefont
  {Zhou}},\ }\bibfield  {title} {\bibinfo {title} {{Black hole hair in
  generalized scalar-tensor gravity: An explicit example}},\ }\href
  {https://doi.org/10.1103/PhysRevD.90.124063} {\bibfield  {journal} {\bibinfo
  {journal} {Phys. Rev. D}\ }\textbf {\bibinfo {volume} {90}},\ \bibinfo
  {pages} {124063} (\bibinfo {year} {2014}{\natexlab{b}})},\ \Eprint
  {https://arxiv.org/abs/1408.1698} {arXiv:1408.1698 [gr-qc]} \BibitemShut
  {NoStop}%
\bibitem [{\citenamefont {Kerr}(1963)}]{Kerr:1963ud}%
  \BibitemOpen
  \bibfield  {author} {\bibinfo {author} {\bibfnamefont {R.~P.}\ \bibnamefont
  {Kerr}},\ }\bibfield  {title} {\bibinfo {title} {{Gravitational field of a
  spinning mass as an example of algebraically special metrics}},\ }\href
  {https://doi.org/10.1103/PhysRevLett.11.237} {\bibfield  {journal} {\bibinfo
  {journal} {Phys. Rev. Lett.}\ }\textbf {\bibinfo {volume} {11}},\ \bibinfo
  {pages} {237} (\bibinfo {year} {1963})}\BibitemShut {NoStop}%
\bibitem [{\citenamefont {Hui}\ and\ \citenamefont
  {Nicolis}(2013)}]{Hui:2012qt}%
  \BibitemOpen
  \bibfield  {author} {\bibinfo {author} {\bibfnamefont {L.}~\bibnamefont
  {Hui}}\ and\ \bibinfo {author} {\bibfnamefont {A.}~\bibnamefont {Nicolis}},\
  }\bibfield  {title} {\bibinfo {title} {{No-Hair Theorem for the Galileon}},\
  }\href {https://doi.org/10.1103/PhysRevLett.110.241104} {\bibfield  {journal}
  {\bibinfo  {journal} {Phys. Rev. Lett.}\ }\textbf {\bibinfo {volume} {110}},\
  \bibinfo {pages} {241104} (\bibinfo {year} {2013})},\ \Eprint
  {https://arxiv.org/abs/1202.1296} {arXiv:1202.1296 [hep-th]} \BibitemShut
  {NoStop}%
%%CITATION = ARXIV:1202.1296;%%
\bibitem [{\citenamefont {Maselli}\ \emph {et~al.}(2015)\citenamefont
  {Maselli}, \citenamefont {Silva}, \citenamefont {Minamitsuji},\ and\
  \citenamefont {Berti}}]{Maselli:2015yva}%
  \BibitemOpen
  \bibfield  {author} {\bibinfo {author} {\bibfnamefont {A.}~\bibnamefont
  {Maselli}}, \bibinfo {author} {\bibfnamefont {H.~O.}\ \bibnamefont {Silva}},
  \bibinfo {author} {\bibfnamefont {M.}~\bibnamefont {Minamitsuji}},\ and\
  \bibinfo {author} {\bibfnamefont {E.}~\bibnamefont {Berti}},\ }\bibfield
  {title} {\bibinfo {title} {{Slowly rotating black hole solutions in Horndeski
  gravity}},\ }\href {https://doi.org/10.1103/PhysRevD.92.104049} {\bibfield
  {journal} {\bibinfo  {journal} {Phys. Rev. D}\ }\textbf {\bibinfo {volume}
  {92}},\ \bibinfo {pages} {104049} (\bibinfo {year} {2015})},\ \Eprint
  {https://arxiv.org/abs/1508.03044} {arXiv:1508.03044 [gr-qc]} \BibitemShut
  {NoStop}%
\bibitem [{\citenamefont {Creminelli}\ \emph {et~al.}(2020)\citenamefont
  {Creminelli}, \citenamefont {Loayza}, \citenamefont {Serra}, \citenamefont
  {Trincherini},\ and\ \citenamefont {Trombetta}}]{Creminelli:2020lxn}%
  \BibitemOpen
  \bibfield  {author} {\bibinfo {author} {\bibfnamefont {P.}~\bibnamefont
  {Creminelli}}, \bibinfo {author} {\bibfnamefont {N.}~\bibnamefont {Loayza}},
  \bibinfo {author} {\bibfnamefont {F.}~\bibnamefont {Serra}}, \bibinfo
  {author} {\bibfnamefont {E.}~\bibnamefont {Trincherini}},\ and\ \bibinfo
  {author} {\bibfnamefont {L.~G.}\ \bibnamefont {Trombetta}},\ }\bibfield
  {title} {\bibinfo {title} {{Hairy Black-holes in Shift-symmetric Theories}},\
  }\href {https://doi.org/10.1007/JHEP08(2020)045} {\bibfield  {journal}
  {\bibinfo  {journal} {JHEP}\ }\textbf {\bibinfo {volume} {08}},\ \bibinfo
  {pages} {045}},\ \Eprint {https://arxiv.org/abs/2004.02893} {arXiv:2004.02893
  [hep-th]} \BibitemShut {NoStop}%
\bibitem [{\citenamefont {Capuano}\ \emph {et~al.}(2023)\citenamefont
  {Capuano}, \citenamefont {Santoni},\ and\ \citenamefont
  {Barausse}}]{Capuano:2023yyh}%
  \BibitemOpen
  \bibfield  {author} {\bibinfo {author} {\bibfnamefont {L.}~\bibnamefont
  {Capuano}}, \bibinfo {author} {\bibfnamefont {L.}~\bibnamefont {Santoni}},\
  and\ \bibinfo {author} {\bibfnamefont {E.}~\bibnamefont {Barausse}},\
  }\bibfield  {title} {\bibinfo {title} {{Black hole hairs in scalar-tensor
  gravity and the lack thereof}},\ }\href
  {https://doi.org/10.1103/PhysRevD.108.064058} {\bibfield  {journal} {\bibinfo
   {journal} {Phys. Rev. D}\ }\textbf {\bibinfo {volume} {108}},\ \bibinfo
  {pages} {064058} (\bibinfo {year} {2023})},\ \Eprint
  {https://arxiv.org/abs/2304.12750} {arXiv:2304.12750 [gr-qc]} \BibitemShut
  {NoStop}%
\bibitem [{\citenamefont {Doneva}\ \emph {et~al.}(2022)\citenamefont {Doneva},
  \citenamefont {Ramazano\u{g}lu}, \citenamefont {Silva}, \citenamefont
  {Sotiriou},\ and\ \citenamefont {Yazadjiev}}]{Doneva:2022ewd}%
  \BibitemOpen
  \bibfield  {author} {\bibinfo {author} {\bibfnamefont {D.~D.}\ \bibnamefont
  {Doneva}}, \bibinfo {author} {\bibfnamefont {F.~M.}\ \bibnamefont
  {Ramazano\u{g}lu}}, \bibinfo {author} {\bibfnamefont {H.~O.}\ \bibnamefont
  {Silva}}, \bibinfo {author} {\bibfnamefont {T.~P.}\ \bibnamefont
  {Sotiriou}},\ and\ \bibinfo {author} {\bibfnamefont {S.~S.}\ \bibnamefont
  {Yazadjiev}},\ }\bibfield  {title} {\bibinfo {title} {{Scalarization}},\
  }\href@noop {} {\  (\bibinfo {year} {2022})},\ \Eprint
  {https://arxiv.org/abs/2211.01766} {arXiv:2211.01766 [gr-qc]} \BibitemShut
  {NoStop}%
\bibitem [{\citenamefont {Perkins}\ \emph {et~al.}(2021)\citenamefont
  {Perkins}, \citenamefont {Nair}, \citenamefont {Silva},\ and\ \citenamefont
  {Yunes}}]{Perkins:2021mhb}%
  \BibitemOpen
  \bibfield  {author} {\bibinfo {author} {\bibfnamefont {S.~E.}\ \bibnamefont
  {Perkins}}, \bibinfo {author} {\bibfnamefont {R.}~\bibnamefont {Nair}},
  \bibinfo {author} {\bibfnamefont {H.~O.}\ \bibnamefont {Silva}},\ and\
  \bibinfo {author} {\bibfnamefont {N.}~\bibnamefont {Yunes}},\ }\bibfield
  {title} {\bibinfo {title} {{Improved gravitational-wave constraints on
  higher-order curvature theories of gravity}},\ }\href
  {https://doi.org/10.1103/PhysRevD.104.024060} {\bibfield  {journal} {\bibinfo
   {journal} {Phys. Rev. D}\ }\textbf {\bibinfo {volume} {104}},\ \bibinfo
  {pages} {024060} (\bibinfo {year} {2021})},\ \Eprint
  {https://arxiv.org/abs/2104.11189} {arXiv:2104.11189 [gr-qc]} \BibitemShut
  {NoStop}%
\bibitem [{\citenamefont {Lyu}\ \emph {et~al.}(2022)\citenamefont {Lyu},
  \citenamefont {Jiang},\ and\ \citenamefont {Yagi}}]{Lyu:2022gdr}%
  \BibitemOpen
  \bibfield  {author} {\bibinfo {author} {\bibfnamefont {Z.}~\bibnamefont
  {Lyu}}, \bibinfo {author} {\bibfnamefont {N.}~\bibnamefont {Jiang}},\ and\
  \bibinfo {author} {\bibfnamefont {K.}~\bibnamefont {Yagi}},\ }\bibfield
  {title} {\bibinfo {title} {{Constraints on Einstein-dilation-Gauss-Bonnet
  gravity from black hole-neutron star gravitational wave events}},\ }\href
  {https://doi.org/10.1103/PhysRevD.105.064001} {\bibfield  {journal} {\bibinfo
   {journal} {Phys. Rev. D}\ }\textbf {\bibinfo {volume} {105}},\ \bibinfo
  {pages} {064001} (\bibinfo {year} {2022})},\ \bibinfo {note} {[Erratum:
  Phys.Rev.D 106, 069901 (2022), Erratum: Phys.Rev.D 106, 069901 (2022)]},\
  \Eprint {https://arxiv.org/abs/2201.02543} {arXiv:2201.02543 [gr-qc]}
  \BibitemShut {NoStop}%
\bibitem [{\citenamefont {Herrero-Valea}(2022)}]{Herrero-Valea:2021dry}%
  \BibitemOpen
  \bibfield  {author} {\bibinfo {author} {\bibfnamefont {M.}~\bibnamefont
  {Herrero-Valea}},\ }\bibfield  {title} {\bibinfo {title} {{The shape of
  scalar Gauss-Bonnet gravity}},\ }\href
  {https://doi.org/10.1007/JHEP03(2022)075} {\bibfield  {journal} {\bibinfo
  {journal} {JHEP}\ }\textbf {\bibinfo {volume} {03}},\ \bibinfo {pages}
  {075}},\ \Eprint {https://arxiv.org/abs/2106.08344} {arXiv:2106.08344
  [gr-qc]} \BibitemShut {NoStop}%
\bibitem [{\citenamefont {Silva}\ \emph {et~al.}(2019)\citenamefont {Silva},
  \citenamefont {Macedo}, \citenamefont {Sotiriou}, \citenamefont {Gualtieri},
  \citenamefont {Sakstein},\ and\ \citenamefont {Berti}}]{Silva:2018qhn}%
  \BibitemOpen
  \bibfield  {author} {\bibinfo {author} {\bibfnamefont {H.~O.}\ \bibnamefont
  {Silva}}, \bibinfo {author} {\bibfnamefont {C.~F.~B.}\ \bibnamefont
  {Macedo}}, \bibinfo {author} {\bibfnamefont {T.~P.}\ \bibnamefont
  {Sotiriou}}, \bibinfo {author} {\bibfnamefont {L.}~\bibnamefont {Gualtieri}},
  \bibinfo {author} {\bibfnamefont {J.}~\bibnamefont {Sakstein}},\ and\
  \bibinfo {author} {\bibfnamefont {E.}~\bibnamefont {Berti}},\ }\bibfield
  {title} {\bibinfo {title} {{Stability of scalarized black hole solutions in
  scalar-Gauss-Bonnet gravity}},\ }\href
  {https://doi.org/10.1103/PhysRevD.99.064011} {\bibfield  {journal} {\bibinfo
  {journal} {Phys. Rev.}\ }\textbf {\bibinfo {volume} {D99}},\ \bibinfo {pages}
  {064011} (\bibinfo {year} {2019})},\ \Eprint
  {https://arxiv.org/abs/1812.05590} {arXiv:1812.05590 [gr-qc]} \BibitemShut
  {NoStop}%
%%CITATION = ARXIV:1812.05590;%%
\bibitem [{\citenamefont {Silva}\ \emph {et~al.}(2018)\citenamefont {Silva},
  \citenamefont {Sakstein}, \citenamefont {Gualtieri}, \citenamefont
  {Sotiriou},\ and\ \citenamefont {Berti}}]{Silva:2017uqg}%
  \BibitemOpen
  \bibfield  {author} {\bibinfo {author} {\bibfnamefont {H.~O.}\ \bibnamefont
  {Silva}}, \bibinfo {author} {\bibfnamefont {J.}~\bibnamefont {Sakstein}},
  \bibinfo {author} {\bibfnamefont {L.}~\bibnamefont {Gualtieri}}, \bibinfo
  {author} {\bibfnamefont {T.~P.}\ \bibnamefont {Sotiriou}},\ and\ \bibinfo
  {author} {\bibfnamefont {E.}~\bibnamefont {Berti}},\ }\bibfield  {title}
  {\bibinfo {title} {{Spontaneous scalarization of black holes and compact
  stars from a Gauss-Bonnet coupling}},\ }\href
  {https://doi.org/10.1103/PhysRevLett.120.131104} {\bibfield  {journal}
  {\bibinfo  {journal} {Phys. Rev. Lett.}\ }\textbf {\bibinfo {volume} {120}},\
  \bibinfo {pages} {131104} (\bibinfo {year} {2018})},\ \Eprint
  {https://arxiv.org/abs/1711.02080} {arXiv:1711.02080 [gr-qc]} \BibitemShut
  {NoStop}%
\bibitem [{\citenamefont {Dima}\ \emph {et~al.}(2020)\citenamefont {Dima},
  \citenamefont {Barausse}, \citenamefont {Franchini},\ and\ \citenamefont
  {Sotiriou}}]{Dima:2020yac}%
  \BibitemOpen
  \bibfield  {author} {\bibinfo {author} {\bibfnamefont {A.}~\bibnamefont
  {Dima}}, \bibinfo {author} {\bibfnamefont {E.}~\bibnamefont {Barausse}},
  \bibinfo {author} {\bibfnamefont {N.}~\bibnamefont {Franchini}},\ and\
  \bibinfo {author} {\bibfnamefont {T.~P.}\ \bibnamefont {Sotiriou}},\
  }\bibfield  {title} {\bibinfo {title} {{Spin-induced black hole spontaneous
  scalarization}},\ }\href@noop {} {\  (\bibinfo {year} {2020})},\ \Eprint
  {https://arxiv.org/abs/2006.03095} {arXiv:2006.03095 [gr-qc]} \BibitemShut
  {NoStop}%
\bibitem [{\citenamefont {Damour}\ and\ \citenamefont
  {Esposito-Farese}(1996)}]{Damour:1996ke}%
  \BibitemOpen
  \bibfield  {author} {\bibinfo {author} {\bibfnamefont {T.}~\bibnamefont
  {Damour}}\ and\ \bibinfo {author} {\bibfnamefont {G.}~\bibnamefont
  {Esposito-Farese}},\ }\bibfield  {title} {\bibinfo {title} {{Tensor - scalar
  gravity and binary pulsar experiments}},\ }\href
  {https://doi.org/10.1103/PhysRevD.54.1474} {\bibfield  {journal} {\bibinfo
  {journal} {Phys. Rev. D}\ }\textbf {\bibinfo {volume} {54}},\ \bibinfo
  {pages} {1474} (\bibinfo {year} {1996})},\ \Eprint
  {https://arxiv.org/abs/gr-qc/9602056} {arXiv:gr-qc/9602056} \BibitemShut
  {NoStop}%
\bibitem [{\citenamefont {Doneva}\ and\ \citenamefont
  {Yazadjiev}(2018)}]{Doneva:2017bvd}%
  \BibitemOpen
  \bibfield  {author} {\bibinfo {author} {\bibfnamefont {D.~D.}\ \bibnamefont
  {Doneva}}\ and\ \bibinfo {author} {\bibfnamefont {S.~S.}\ \bibnamefont
  {Yazadjiev}},\ }\bibfield  {title} {\bibinfo {title} {{New Gauss-Bonnet Black
  Holes with Curvature-Induced Scalarization in Extended Scalar-Tensor
  Theories}},\ }\href {https://doi.org/10.1103/PhysRevLett.120.131103}
  {\bibfield  {journal} {\bibinfo  {journal} {Phys. Rev. Lett.}\ }\textbf
  {\bibinfo {volume} {120}},\ \bibinfo {pages} {131103} (\bibinfo {year}
  {2018})},\ \Eprint {https://arxiv.org/abs/1711.01187} {arXiv:1711.01187
  [gr-qc]} \BibitemShut {NoStop}%
\bibitem [{\citenamefont {Antoniou}\ \emph {et~al.}(2018)\citenamefont
  {Antoniou}, \citenamefont {Bakopoulos},\ and\ \citenamefont
  {Kanti}}]{Antoniou:2017acq}%
  \BibitemOpen
  \bibfield  {author} {\bibinfo {author} {\bibfnamefont {G.}~\bibnamefont
  {Antoniou}}, \bibinfo {author} {\bibfnamefont {A.}~\bibnamefont
  {Bakopoulos}},\ and\ \bibinfo {author} {\bibfnamefont {P.}~\bibnamefont
  {Kanti}},\ }\bibfield  {title} {\bibinfo {title} {{Evasion of No-Hair
  Theorems and Novel Black-Hole Solutions in Gauss-Bonnet Theories}},\ }\href
  {https://doi.org/10.1103/PhysRevLett.120.131102} {\bibfield  {journal}
  {\bibinfo  {journal} {Phys. Rev. Lett.}\ }\textbf {\bibinfo {volume} {120}},\
  \bibinfo {pages} {131102} (\bibinfo {year} {2018})},\ \Eprint
  {https://arxiv.org/abs/1711.03390} {arXiv:1711.03390 [hep-th]} \BibitemShut
  {NoStop}%
\bibitem [{\citenamefont {Minamitsuji}\ and\ \citenamefont
  {Ikeda}(2019)}]{Minamitsuji:2018xde}%
  \BibitemOpen
  \bibfield  {author} {\bibinfo {author} {\bibfnamefont {M.}~\bibnamefont
  {Minamitsuji}}\ and\ \bibinfo {author} {\bibfnamefont {T.}~\bibnamefont
  {Ikeda}},\ }\bibfield  {title} {\bibinfo {title} {{Scalarized black holes in
  the presence of the coupling to Gauss-Bonnet gravity}},\ }\href
  {https://doi.org/10.1103/PhysRevD.99.044017} {\bibfield  {journal} {\bibinfo
  {journal} {Phys. Rev. D}\ }\textbf {\bibinfo {volume} {99}},\ \bibinfo
  {pages} {044017} (\bibinfo {year} {2019})},\ \Eprint
  {https://arxiv.org/abs/1812.03551} {arXiv:1812.03551 [gr-qc]} \BibitemShut
  {NoStop}%
\bibitem [{\citenamefont {Herdeiro}\ \emph {et~al.}(2021)\citenamefont
  {Herdeiro}, \citenamefont {Radu}, \citenamefont {Silva}, \citenamefont
  {Sotiriou},\ and\ \citenamefont {Yunes}}]{Herdeiro:2020wei}%
  \BibitemOpen
  \bibfield  {author} {\bibinfo {author} {\bibfnamefont {C.~A.~R.}\
  \bibnamefont {Herdeiro}}, \bibinfo {author} {\bibfnamefont {E.}~\bibnamefont
  {Radu}}, \bibinfo {author} {\bibfnamefont {H.~O.}\ \bibnamefont {Silva}},
  \bibinfo {author} {\bibfnamefont {T.~P.}\ \bibnamefont {Sotiriou}},\ and\
  \bibinfo {author} {\bibfnamefont {N.}~\bibnamefont {Yunes}},\ }\bibfield
  {title} {\bibinfo {title} {{Spin-induced scalarized black holes}},\ }\href
  {https://doi.org/10.1103/PhysRevLett.126.011103} {\bibfield  {journal}
  {\bibinfo  {journal} {Phys. Rev. Lett.}\ }\textbf {\bibinfo {volume} {126}},\
  \bibinfo {pages} {011103} (\bibinfo {year} {2021})},\ \Eprint
  {https://arxiv.org/abs/2009.03904} {arXiv:2009.03904 [gr-qc]} \BibitemShut
  {NoStop}%
\bibitem [{\citenamefont {Wittek}\ \emph {et~al.}(2023)\citenamefont {Wittek}
  \emph {et~al.}}]{Wittek:2023nyi}%
  \BibitemOpen
  \bibfield  {author} {\bibinfo {author} {\bibfnamefont {N.~A.}\ \bibnamefont
  {Wittek}} \emph {et~al.},\ }\bibfield  {title} {\bibinfo {title} {{Worldtube
  excision method for intermediate-mass-ratio inspirals: scalar-field model in
  3+1 dimensions}},\ }\href@noop {} {\  (\bibinfo {year} {2023})},\ \Eprint
  {https://arxiv.org/abs/2304.05329} {arXiv:2304.05329 [gr-qc]} \BibitemShut
  {NoStop}%
\bibitem [{\citenamefont {Lindblom}\ \emph {et~al.}(2006)\citenamefont
  {Lindblom}, \citenamefont {Scheel}, \citenamefont {Kidder}, \citenamefont
  {Owen},\ and\ \citenamefont {Rinne}}]{Lindblom:2005qh}%
  \BibitemOpen
  \bibfield  {author} {\bibinfo {author} {\bibfnamefont {L.}~\bibnamefont
  {Lindblom}}, \bibinfo {author} {\bibfnamefont {M.~A.}\ \bibnamefont
  {Scheel}}, \bibinfo {author} {\bibfnamefont {L.~E.}\ \bibnamefont {Kidder}},
  \bibinfo {author} {\bibfnamefont {R.}~\bibnamefont {Owen}},\ and\ \bibinfo
  {author} {\bibfnamefont {O.}~\bibnamefont {Rinne}},\ }\bibfield  {title}
  {\bibinfo {title} {{A New generalized harmonic evolution system}},\ }\href
  {https://doi.org/10.1088/0264-9381/23/16/S09} {\bibfield  {journal} {\bibinfo
   {journal} {Class. Quant. Grav.}\ }\textbf {\bibinfo {volume} {23}},\
  \bibinfo {pages} {S447} (\bibinfo {year} {2006})},\ \Eprint
  {https://arxiv.org/abs/gr-qc/0512093} {arXiv:gr-qc/0512093} \BibitemShut
  {NoStop}%
\bibitem [{\citenamefont {Throwe}\ and\ \citenamefont
  {Teukolsky}(2020)}]{throwe2020highorder}%
  \BibitemOpen
  \bibfield  {author} {\bibinfo {author} {\bibfnamefont {W.}~\bibnamefont
  {Throwe}}\ and\ \bibinfo {author} {\bibfnamefont {S.~A.}\ \bibnamefont
  {Teukolsky}},\ }\href@noop {} {\bibinfo {title} {A high-order, conservative
  integrator with local time-stepping}} (\bibinfo {year} {2020}),\ \Eprint
  {https://arxiv.org/abs/1811.02499} {arXiv:1811.02499 [math.NA]} \BibitemShut
  {NoStop}%
\bibitem [{\citenamefont {Bayliss}\ and\ \citenamefont
  {Turkel}(1980)}]{bayliss1980radiation}%
  \BibitemOpen
  \bibfield  {author} {\bibinfo {author} {\bibfnamefont {A.}~\bibnamefont
  {Bayliss}}\ and\ \bibinfo {author} {\bibfnamefont {E.}~\bibnamefont
  {Turkel}},\ }\bibfield  {title} {\bibinfo {title} {Radiation boundary
  conditions for wave-like equations},\ }\href@noop {} {\bibfield  {journal}
  {\bibinfo  {journal} {Communications on Pure and applied Mathematics}\
  }\textbf {\bibinfo {volume} {33}},\ \bibinfo {pages} {707} (\bibinfo {year}
  {1980})}\BibitemShut {NoStop}%
\bibitem [{SXS()}]{SXSinprep}%
  \BibitemOpen
  \href@noop {} {\bibinfo {title} {{SXS Collaboration, in
  preparation}}}\BibitemShut {NoStop}%
\bibitem [{\citenamefont {Scheel}\ \emph {et~al.}(2006)\citenamefont {Scheel},
  \citenamefont {Pfeiffer}, \citenamefont {Lindblom}, \citenamefont {Kidder},
  \citenamefont {Rinne},\ and\ \citenamefont {Teukolsky}}]{Scheel:2006gg}%
  \BibitemOpen
  \bibfield  {author} {\bibinfo {author} {\bibfnamefont {M.~A.}\ \bibnamefont
  {Scheel}}, \bibinfo {author} {\bibfnamefont {H.~P.}\ \bibnamefont
  {Pfeiffer}}, \bibinfo {author} {\bibfnamefont {L.}~\bibnamefont {Lindblom}},
  \bibinfo {author} {\bibfnamefont {L.~E.}\ \bibnamefont {Kidder}}, \bibinfo
  {author} {\bibfnamefont {O.}~\bibnamefont {Rinne}},\ and\ \bibinfo {author}
  {\bibfnamefont {S.~A.}\ \bibnamefont {Teukolsky}},\ }\bibfield  {title}
  {\bibinfo {title} {{Solving Einstein's equations with dual coordinate
  frames}},\ }\href {https://doi.org/10.1103/PhysRevD.74.104006} {\bibfield
  {journal} {\bibinfo  {journal} {Phys. Rev. D}\ }\textbf {\bibinfo {volume}
  {74}},\ \bibinfo {pages} {104006} (\bibinfo {year} {2006})},\ \Eprint
  {https://arxiv.org/abs/gr-qc/0607056} {arXiv:gr-qc/0607056} \BibitemShut
  {NoStop}%
\bibitem [{\citenamefont {Lindblom}\ and\ \citenamefont
  {Szilagyi}(2009)}]{Lindblom:2009tu}%
  \BibitemOpen
  \bibfield  {author} {\bibinfo {author} {\bibfnamefont {L.}~\bibnamefont
  {Lindblom}}\ and\ \bibinfo {author} {\bibfnamefont {B.}~\bibnamefont
  {Szilagyi}},\ }\bibfield  {title} {\bibinfo {title} {{An Improved Gauge
  Driver for the GH Einstein System}},\ }\href
  {https://doi.org/10.1103/PhysRevD.80.084019} {\bibfield  {journal} {\bibinfo
  {journal} {Phys. Rev. D}\ }\textbf {\bibinfo {volume} {80}},\ \bibinfo
  {pages} {084019} (\bibinfo {year} {2009})},\ \Eprint
  {https://arxiv.org/abs/0904.4873} {arXiv:0904.4873 [gr-qc]} \BibitemShut
  {NoStop}%
\bibitem [{\citenamefont {Choptuik}\ and\ \citenamefont
  {Pretorius}(2010)}]{Choptuik:2009ww}%
  \BibitemOpen
  \bibfield  {author} {\bibinfo {author} {\bibfnamefont {M.~W.}\ \bibnamefont
  {Choptuik}}\ and\ \bibinfo {author} {\bibfnamefont {F.}~\bibnamefont
  {Pretorius}},\ }\bibfield  {title} {\bibinfo {title} {{Ultra Relativistic
  Particle Collisions}},\ }\href
  {https://doi.org/10.1103/PhysRevLett.104.111101} {\bibfield  {journal}
  {\bibinfo  {journal} {Phys. Rev. Lett.}\ }\textbf {\bibinfo {volume} {104}},\
  \bibinfo {pages} {111101} (\bibinfo {year} {2010})},\ \Eprint
  {https://arxiv.org/abs/0908.1780} {arXiv:0908.1780 [gr-qc]} \BibitemShut
  {NoStop}%
\bibitem [{\citenamefont {Szilagyi}\ \emph {et~al.}(2009)\citenamefont
  {Szilagyi}, \citenamefont {Lindblom},\ and\ \citenamefont
  {Scheel}}]{Szilagyi:2009qz}%
  \BibitemOpen
  \bibfield  {author} {\bibinfo {author} {\bibfnamefont {B.}~\bibnamefont
  {Szilagyi}}, \bibinfo {author} {\bibfnamefont {L.}~\bibnamefont {Lindblom}},\
  and\ \bibinfo {author} {\bibfnamefont {M.~A.}\ \bibnamefont {Scheel}},\
  }\bibfield  {title} {\bibinfo {title} {{Simulations of Binary Black Hole
  Mergers Using Spectral Methods}},\ }\href
  {https://doi.org/10.1103/PhysRevD.80.124010} {\bibfield  {journal} {\bibinfo
  {journal} {Phys. Rev. D}\ }\textbf {\bibinfo {volume} {80}},\ \bibinfo
  {pages} {124010} (\bibinfo {year} {2009})},\ \Eprint
  {https://arxiv.org/abs/0909.3557} {arXiv:0909.3557 [gr-qc]} \BibitemShut
  {NoStop}%
\bibitem [{\citenamefont {Rinne}\ \emph {et~al.}(2007)\citenamefont {Rinne},
  \citenamefont {Lindblom},\ and\ \citenamefont {Scheel}}]{Rinne:2007ui}%
  \BibitemOpen
  \bibfield  {author} {\bibinfo {author} {\bibfnamefont {O.}~\bibnamefont
  {Rinne}}, \bibinfo {author} {\bibfnamefont {L.}~\bibnamefont {Lindblom}},\
  and\ \bibinfo {author} {\bibfnamefont {M.~A.}\ \bibnamefont {Scheel}},\
  }\bibfield  {title} {\bibinfo {title} {{Testing outer boundary treatments for
  the Einstein equations}},\ }\href
  {https://doi.org/10.1088/0264-9381/24/16/006} {\bibfield  {journal} {\bibinfo
   {journal} {Class. Quant. Grav.}\ }\textbf {\bibinfo {volume} {24}},\
  \bibinfo {pages} {4053} (\bibinfo {year} {2007})},\ \Eprint
  {https://arxiv.org/abs/0704.0782} {arXiv:0704.0782 [gr-qc]} \BibitemShut
  {NoStop}%
\bibitem [{\citenamefont {Scheel}\ \emph {et~al.}(2004)\citenamefont {Scheel},
  \citenamefont {Erickcek}, \citenamefont {Burko}, \citenamefont {Kidder},
  \citenamefont {Pfeiffer},\ and\ \citenamefont {Teukolsky}}]{Scheel:2003vs}%
  \BibitemOpen
  \bibfield  {author} {\bibinfo {author} {\bibfnamefont {M.~A.}\ \bibnamefont
  {Scheel}}, \bibinfo {author} {\bibfnamefont {A.~L.}\ \bibnamefont
  {Erickcek}}, \bibinfo {author} {\bibfnamefont {L.~M.}\ \bibnamefont {Burko}},
  \bibinfo {author} {\bibfnamefont {L.~E.}\ \bibnamefont {Kidder}}, \bibinfo
  {author} {\bibfnamefont {H.~P.}\ \bibnamefont {Pfeiffer}},\ and\ \bibinfo
  {author} {\bibfnamefont {S.~A.}\ \bibnamefont {Teukolsky}},\ }\bibfield
  {title} {\bibinfo {title} {{3-D simulations of linearized scalar fields in
  Kerr space-time}},\ }\href {https://doi.org/10.1103/PhysRevD.69.104006}
  {\bibfield  {journal} {\bibinfo  {journal} {Phys. Rev. D}\ }\textbf {\bibinfo
  {volume} {69}},\ \bibinfo {pages} {104006} (\bibinfo {year} {2004})},\
  \Eprint {https://arxiv.org/abs/gr-qc/0305027} {arXiv:gr-qc/0305027}
  \BibitemShut {NoStop}%
\end{thebibliography}%

\end{document}